\documentclass[fleqn,usenatbib]{mnras}

\usepackage{newtxtext,newtxmath}

\usepackage[T1]{fontenc}

\DeclareRobustCommand{\VAN}[3]{#2}
\let\VANthebibliography\thebibliography
\def\thebibliography{\DeclareRobustCommand{\VAN}[3]{##3}\VANthebibliography}

\usepackage{graphicx}	
\usepackage{tabularx}
\usepackage{amsmath}	
\usepackage{indentfirst}
\usepackage{appendix}
\usepackage{makecell}
\usepackage{enumerate}
\usepackage{enumitem}
\setlist[enumerate]{labelwidth=0pt, labelsep=0.5em}
\usepackage{multirow}
\usepackage{multicol}
\usepackage{gensymb}
\usepackage{longtable}
\usepackage{pdflscape}
\usepackage{float}
\usepackage{lineno}
\usepackage{scrextend}
\usepackage{orcidlink}

\newcommand\sersic{S\'ersic}
\newcommand\OIII{[\text{O}\,\textsc{iii}]}
\newcommand\OII{[\text{O}\,\textsc{ii}]}

\DeclareUnicodeCharacter{2212}{-}
\DeclareUnicodeCharacter{00B4}{'}
\DeclareUnicodeCharacter{00A8}{"}

\title[Spectroscopic confirmation of dual quasars]{Spectroscopic confirmation of dual and offset quasars from the Subaru HSC-SSP program}

\author[S. Tang et al.]{
Shenli Tang,$^{\orcidlink{0000-0002-2185-5679}\,}$$^{1}$\thanks{E-mail: st1c23@soton.ac.uk}
John Silverman,$^{\orcidlink{0000-0002-0000-6977}\,}$$^{2,3,4,5}$
Xavier Prochaska,$^{\orcidlink{0000-0002-7738-6875}\,}$$^{2,6}$ 
Manda Banerji,$^{\orcidlink{0000-0002-0639-5141}\,}$$^{1}$
Xuheng Ding,$^{\orcidlink{0000-0001-8917-2148}\,}$$^{7}$\newauthor
Masafusa Onoue,$^{\orcidlink{0000-0003-2984-6803}\,}$$^{8}$
and Knud Jahnke$^{\orcidlink{0000-0003-3804-2137}}$$^{9}$
\\
$^{1}$School of Physics \& Astronomy, University of Southampton, Highfield Campus, Southampton SO17 1BJ, UK\\
$^{2}$Kavli Institute for the Physics and Mathematics of the Universe (WPI), The University of Tokyo, Kashiwa, Chiba 277-8583, Japan\\
$^{3}$Department of Astronomy, School of Science, The University of Tokyo, 7-3-1 Hongo, Bunkyo, Tokyo 113-0033, Japan\\
$^{4}$Center for Data-Driven Discovery, Kavli IPMU (WPI), UTIAS, The University of Tokyo, Kashiwa, Chiba 277-8583, Japan\\
$^{5}$Center for Astrophysical Sciences, Department of Physics \& Astronomy, Johns Hopkins University, Baltimore, MD 21218, USA\\ 
$^{6}$Department of Astronomy and Astrophysics, University of California, Santa Cruz, 1156 High Street, Santa Cruz, CA 95064, USA\\
$^{7}$School of Physics and Technology, Wuhan University, Wuhan 430072, China\\
$^{8}$Waseda Institute for Advanced Study (WIAS), Waseda University, 1-21-1, Nishi-Waseda, Shinjuku, Tokyo 169-0051, Japan\\
$^{9}$Max-Planck-Institut für Astronomie, Königstuhl 17, 69117 Heidelberg, Germany}

\date{Accepted XXX. Received YYY; in original form ZZZ}

\pubyear{2025}

\begin{document}
\label{firstpage}
\pagerange{\pageref{firstpage}--\pageref{lastpage}}
\maketitle

\begin{abstract}
We present a spectroscopic follow-up program targeting closely-separated dual quasar candidates selected from imaging of SDSS quasars with the Subaru Hyper Suprime-Cam Subaru Strategic Program (HSC-SSP). Using two-dimensional image decomposition, our selection identifies PSF-like companions within 0$\farcs$6–4\arcsec separation ($\lesssim$ 30 kpc) around the SDSS quasar. We newly confirm six broad-line dual quasars and eleven offset quasars (quasar-galaxy pairs), spanning $1.5 < z < 3.3$ for the duals and predominantly $z < 0.6$ for the offset systems. No obvious lensed quasars were discovered from this program. We obtained 99 spectra of these candidates from NTT/EFOSC2, Gemini/GMOS-N, Keck/NIRES, and Subaru/FOCAS. From the spectra, we measure the emission-line properties of these dual black holes (BH). At $z>1.5$, the confirmed duals exhibit high black hole mass ($M_{\rm BH}$ $=10^{8.5}$–$10^{10} M_{\odot}$) with high bolometric luminosities ($L_{\rm bol}$ $=10^{45.5}$–$10^{47.5}$ erg s$^{−1}$), yet accrete at moderate Eddington ratios ($\lambda_{\rm Edd}=$0.01–0.4). From the spectroscopically-confirmed samples, we estimate the dual fraction of SDSS quasars with separations of $0.6\arcsec$–$4\arcsec$ to be 0.2\%-1.2\% at $z<0.8$, 0.08\%-0.24\% at $0.8<z<1.5$, and 0.06\% at $1.5<z<3.3$. These values are broadly consistent with other recent optical studies, but lower than theoretical expectations of a rising dual fraction at cosmic noon. However, we note that these fractions, especially at high $z$, still need a more accurate assessment of selection and observation effects.
\end{abstract}




\section{Introduction}
Dual supermassive black holes (SMBHs; $M_{\rm BH} \gtrsim 10^6~M_\odot$) are an inevitable outcome of galaxy mergers in the hierarchical growth of structure \citep{begelman1980massive}. Gas inflows triggered by mergers can fuel the SMBHs, igniting them as quasars \citep{di2005energy, hopkins2006unified, hopkins2010massive}. When both SMBHs are simultaneously accreting, they appear as a dual quasar system on kiloparsec scales. The frequency and properties of such systems are therefore the key to understanding the statistical importance and physical mechanisms of mergers in triggering quasars, which is still under debate \citep[e.g.,][]{silverman2011impact, ellison2011galaxy, treister2012major, villforth2014morphologies, villarroel2017agn, marian2019major, marian2020significant, zhao2022relation, tang2023morphological}. 
Dual quasars also trace the small-scale clustering of quasars \citep{hennawi2006binary, eftekharzadeh2017clustering} and represent progenitors of parsec-scale SMBH binaries, which are expected to generate the nanohertz gravitational-wave background now reported by multiple pulsar timing array (PTA) collaborations \citep{ agazie2023nanograv} and of future space-based detections with LISA \citep{amaro2017laser, colpi2017gravitational}.
\par
Of particular interest are duals with physical separations $\lesssim\!30$ kpc, corresponding to the later stages of mergers prior to coalescence. Simulations show that tidal torques and shocks reduce gas angular momentum, boosting both star formation and SMBH accretion and increase the likelihood that both SMBHs are active \citep{ hopkins2010massive,van2012observability, capelo2015growth, capelo2017survey}. Cosmological simulations broadly indicate that the dual-AGN fraction increases toward $z\!\sim\!2$–3, with comoving number densities of order $10^{-4}$–$10^{-3}\,{\rm Mpc}^{-3}$ for $L_{\rm bol}\!>\!10^{43}\,{\rm erg\,s^{-1}}$ and separations $\lesssim\!30$ kpc, although predictions vary with model assumptions and selection criteria \citep{steinborn2016origin, rosas2019abundances, volonteri2022dual, saeedzadeh2024dual}. Wide-field surveys are therefore essential to identify such systems in statistically meaningful numbers.
\par
Early systematic searches were driven by SDSS imaging and spectroscopy \citep[e.g.,][]{hennawi2006binary, hennawi2010binary, inada2008sloan, inada2012sloan, kayo2012very, more2016sdss}, demonstrating an excess of quasar pairs on $\lesssim$100 kpc scales \citep{eftekharzadeh2017clustering}. However, SDSS’s fiber collision issue, seeing-limited imaging, and spectroscopic depth bias such samples toward relatively wide separations (typically a few arcseconds) and bright quasars ($i\!\lesssim\!20$). More recently, \textit{Gaia} has enabled all-sky searches at sub-arcsecond scales. “Varstrometry” exploits small-scale astrometric jitter from unresolved pairs with independently varying light curves \citep{hwang2020varstrometry, shen2021hidden, chen2022varstrometry}, while the Gaia Multi-Peak (GMP) method identifies multiple peaks in the along-scan light profile \citep{mannucci2022unveiling, mannucci2023gmp, ciurlo2023new, scialpi2024muse}. While powerful, these approaches are limited to bright quasars ($G \lesssim 20.5$) and are prone to contamination from host galaxies, making them most effective at $z \gtrsim 1$.
\par
Building on the method of \citet{silverman2020dual} as described in Section~\ref{subsection:selection}, we initiated an image-based dual-quasar search within the Subaru Hyper Suprime-Cam Subaru Strategic Program (HSC-SSP) footprint. The HSC Wide layer reaches $i\!\simeq\!26$ (5$\sigma$) with median seeing $\sim\!0\farcs6$ \citep{aihara2022third}, enabling discovery of pairs several magnitudes fainter than the \textit{Gaia} selections. Our candidates are typically detected as one quasar with SDSS but two separate sources with HSC, with angular separations $\gtrsim\!0\farcs6$ (seeing limited), placing our parameter space squarely between SDSS-resolved pairs and \textit{Gaia} sub-arcsecond candidates (see \citealt{tang2025alma} for a literature summary and our HSC-selected sample). In a companion work, we have also tuned the pipeline to select out more closely-separated pairs with $<\!0\farcs6$ in HSC \citep{wang2025hst}. Importantly, this image-based approach can be scaled to next-generation wide-field surveys such as the Rubin Observatory Legacy Survey of Space and Time (LSST).
\par
In \citet{silverman2020dual} and \citet{tang2021optical}, we reported spectroscopic follow-up of 32 HSC-selected candidates and confirmed six dual quasars. Here we present additional spectroscopy for 90 HSC-selected candidates, confirming six dual quasars and eleven quasar–galaxy pairs—17 physical pairs in total.
\par
The paper is organized as follows. Section~\ref{sec:methods} describes sample selection, observing setups, data reduction, and analysis. Section~\ref{sec:results} summarizes our classification using spectroscopy and HSC imaging, and presents emission-line and black-hole property measurements for the duals. Section~\ref{sec:discussion} discusses success rates, selection biases, and caveats. We adopt a flat $\Lambda$CDM cosmology with $\Omega_\Lambda\!=\!0.7$, $\Omega_{\rm m}\!=\!0.3$, and $H_0\!=\!70~{\rm km\,s^{-1}\,Mpc^{-1}}$.


\section{Methods} \label{sec:methods}
\subsection{Candidate selection} \label{subsection:selection}
\begin{figure*}
	\includegraphics[width=0.95\textwidth]{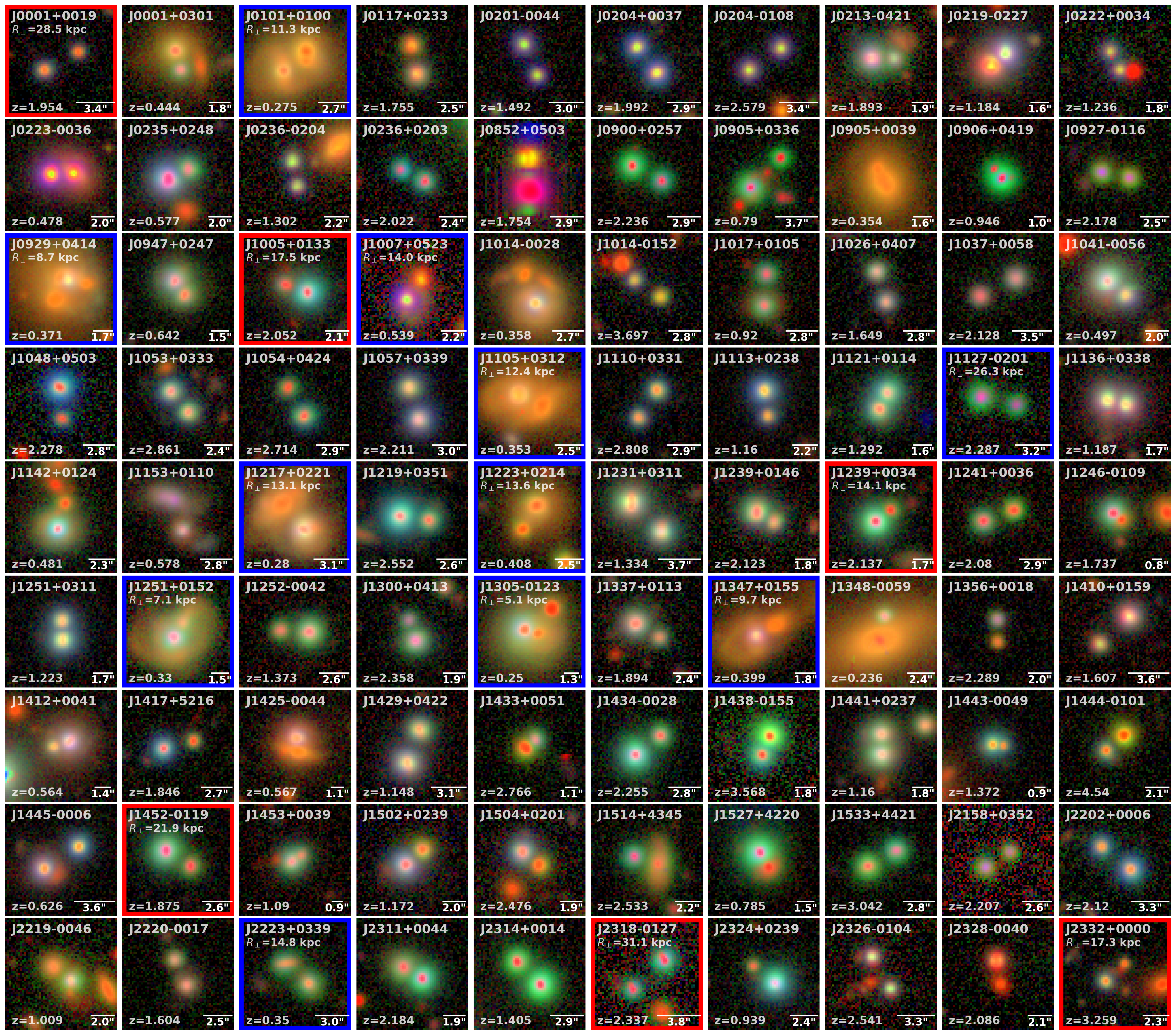}
    \caption{HSC colored images of all 90 candidates that are observed in this work. The side length of each cutout stamp is 10\arcsec. North is to the up, and east to the left side. The spectroscopic redshifts of the SDSS quasars are labeled at the bottom left. The angular separations between the two sources of each pair are plotted as a scale bar at the bottom right. The confirmed six dual quasars and eleven offset quasars are framed in red and blue, respectively. For these physical pairs, their projected physical separations ($R_\perp$) are noted at the top left. }
    \label{fig:all_HSC_img}
\end{figure*}

\begin{table*}
    \centering
    \caption{Basic information of the observed dual quasar candidates. Column (2): Spectroscopic redshift of the known SDSS quasar in each pair. Columns (3-6): PSF positions of the known SDSS quasar (qso\_RA and qso\_DEC) and the companion source (com\_RA and com\_DEC) in each pair. Columns (7) and (8): The separation between the two nuclei in arcsec and projected physical distance (kpc) assuming them to be at the same redshift. Columns (9-10): $g$ band PSF magnitudes (after decomposing the host galaxy if detected) of the known SDSS quasar (qso\_g) and the companion (com\_g) source. The full table, including 90 sources and photometry of the other four bands, is available via the online version.}
    \label{tab:basic}
    \begin{tabular}{l|ccccccccccc}
    \hline
    & Name & specz & qso\_RA & qso\_DEC & com\_RA & com\_DEC & Sep & $R_{\perp}$ & qso\_g & com\_g & ...\\	
    & (J2000) &  & (degree) & (degree) & (degree) & (degree) & (") & (kpc) & (AB) & (AB) & ...\\
    & (1) & (2) & (3) & (4) & (5) & (6) & (7) & (8) & (9) &  (10) & ... \\
    \hline
    1 & 000129.98+001911.3 & 1.9599 & 0.37492 & 0.31985 & 0.37409 & 0.32030 & 3.4 & 28.52 & 21.80 & 22.74 & ...\\
    2 & 000157.44+030147.6 & 0.4440 & 0.48934 & 3.02992 & 0.48921 & 3.02945 & 1.8 & 10.29 & 20.67 & 21.07 & ...\\
    3 & 010123.64+010045.3 & 0.2749 & 15.34851 & 1.01260 & 15.34795 & 1.01309 & 2.7 & 11.31 & 20.75 & 21.34 & ...\\
    4 & 011728.83+023309.9 & 1.7552 & 19.37019 & 2.55275 & 19.37032 & 2.55344 & 2.5 & 21.14 & 21.02 & 22.20 & ... \\
    5 & 020130.46-004448.9 & 1.4915 & 30.37693 & -0.74691 & 30.37660 & -0.74768 & 3.0 & 25.38 & 21.29 & 21.78 & ... \\
    6 & 020425.44+003741.6 & 1.9921 & 31.10602 & 0.62821 & 31.10553 & 0.62757 & 2.9 & 24.28 & 20.72 & 20.35 & ... \\
    7 & 020452.73-010822.4 & 2.5790 & 31.21977 & -1.13957 & 31.21899 & -1.13903 & 3.4 & 27.25 & 21.37 & 21.43 & ... \\
    8 & 021322.99-042134.3 & 1.9053 & 33.34584 & -4.35956 & 33.34530 & -4.35956 & 1.9 & 15.98 & 19.21 & 21.16 & ... \\
    9 & 021907.35-022715.2 & 1.1836 & 34.78065 & -2.45426 & 34.78098 & -2.45456 & 1.6 & 13.24 & 18.61 & 19.25 & ... \\
    10 & 022225.92+003401.7 & 1.2360 & 35.60796 & 0.56717 & 35.60772 & 0.56672 & 1.8 & 14.99 & 21.74 & 21.61 & ... \\
    ... & ... & ... & ... & ... & ... & ... & ... & ... & ... & ... & ... \\
    \hline
\end{tabular}
\end{table*}

The parent sample of this work was constructed following the procedure in \citet{silverman2020dual}. Briefly, we matched the SDSS DR14 quasar catalog \citep[DR14Q;][]{paris2018sloan} against the Subaru HSC-SSP footprint. The DR14Q contains 526,265 spectroscopically confirmed quasars selected by color and magnitude criteria up to $z \sim 5$ \citep{richards2001colors, richards2002spectroscopic, myers2015sdss}. For HSC imaging, we used the s20a\_wide release, included in Public Data Release 3 (PDR3; \citealt{aihara2022third}), which covers 1,264 deg$^2$ in the $i$ band, of which 881 deg$^2$ reaches the full 5$\sigma$ depth of $i \simeq 26.2$ mag. This overlap yielded 59,482 SDSS quasars imaged in HSC and with $i$-band CModel photometry.
\par
We then applied the two-dimensional image decomposition tool \texttt{GaLight} \citep{ding2022galight} to the HSC $i$-band data to identify dual-quasar candidates. A candidate is defined as an SDSS quasar with an additional unresolved source detected ($i$-band PSF magnitude brighter than 24.5) within a projected radius of $0\farcs6$–4\arcsec. The lower bound is set by the typical HSC seeing ($\sim0\farcs6$), while the upper bound complements SDSS-based searches, which were largely limited to pairs with wider separations. This procedure yielded 1007 candidates, of which 30 were located at survey edges and lacked full multi-band coverage. Removing these leaves 977 candidate pairs. We then visually inspected and rejected 89 sources contaminated by artifacts, saturated stars, and extended foreground galaxies. We further rejected 5 confirmed lens systems from literature \citep{inada2008sloan, inada2012sloan, jackson2012new}. Our final s20a\_wide parent sample list consists of 883 dual quasar candidates. We did not reject the objects that have galaxy-like photometry to identify potential type 1-type 2 pairs and offset quasars (more discussions about such pairs will be presented in Araujo et al. in prep). Here we report spectroscopic follow-up observations of 90 candidates drawn from the s20a\_wide sample.
\par
Figure~\ref{fig:all_HSC_img} presents $gri$-composite HSC cutouts (10\arcsec per side) of the 90 observed systems. The SDSS J2000 designation of the known quasar is shown at the top of each panel, with its spectroscopic redshift noted at the bottom left, and the angular separation plotted as a scale bar at bottom right. The confirmed dual quasars and offset quasars are marked with red and blue frames, respectively. For these physical pairs, we note their projected separation ($R_\perp$ in kpc) in the panel.
\par
Table~\ref{tab:basic} lists basic information for the observed candidates. Column (2) gives the spectroscopic redshift of the known SDSS quasar from DR14Q. For close pairs, the standard HSC pipeline forced photometry \citep{huang2017character} can suffer from deblending failures. Therefore, refined photometric measurements from \texttt{GaLight} are adopted. Columns (3–6) provide the PSF-fitted coordinates of the quasar (qso\_RA, qso\_DEC) and its companion (com\_RA, com\_DEC), measured using \texttt{GaLight}. Each system was modeled as two PSFs plus, where required, additional \sersic\ profiles for extended components. The PSFs were generated from the HSC pipeline using \texttt{PSFEx} \citep{bertin2013psfex, bosch2018hyper}. We assume consistent source positions across all five bands, with reported coordinates measured from the $i$-band images. Based on these positions, we compute angular separations (Sep, arcsec) and projected physical separations ($R_\perp$, kpc). Columns (9–10) list the PSF magnitudes of the quasar and companion in the $g$ band.  The full table, including photometry in all five bands for all 90 candidates, is available in the online version.


\subsection{Observations}
Long-slit spectroscopy was carried out with four ground-based facilities. For each target, we determined the relative positions of the two sources from the HSC $i$-band images and rotated the slit position angle (PA) to simultaneously cover both.

\subsubsection{NTT/EFOSC2}

We observed 60 targets using the ESO Faint Object Spectrograph and Camera version 2 \citep[EFOSC2, ][]{buzzoni1984eso} on the 3.58-m NTT at La Silla. Observations were conducted during three programs in 2021–2022: 105.20MF (P107, PI: K. Jahnke), 108.22F0 (PI: M. Onoue), and 109.23BF (PI: M. Onoue). In total, six nights of data were successfully obtained. The instrumental setups and weather conditions are summarized in Table~\ref{tab:observation_NTT}. We used a fixed $1\arcsec$ slit with either Grism \#5 (5200–9350~\AA) or \#6 (3860–8070~\AA), both offering 2.06~\AA\ pixel$^{-1}$ and 
$\sim$15.5~\AA\ FWHM resolution. The CG375 order-blocking filter was applied in runs using Grism \#6 (P108, P109). Seeing varied between 1\arcsec and occasionally 2\arcsec. The wavelengths are calibrated with the Ar/He lamps, and fluxes are calibrated with standard stars EG 21, EG 274, and LTT 1788 \citep{oke1990faint}. 
\par
In Table~\ref{tab:observation_NTT}, the rows of the companions that are found at the same redshift of the known SDSS quasar (velocity offset $<$ 500 km~s$^{-1}$) are highlighted with bold fonts. Otherwise, the companion is a projected source. The superscripts ``G", ``K", and ``S" on top of the source names indicate additional observations with Gemini (Table \ref{tab:observation_Gemini}), Keck (Table \ref{tab:observation_Keck}), and Subaru (Table \ref{tab:observation_Subaru}) on the same source. They are noted with superscript ``N" (refer to NTT) in those tables correspondingly. The classification of these sources is based on the combined spectrum. The above mentioned three tables also follow the same format as this table.

\begin{table*}
    \caption{Observation setups, weather conditions, and classifications of the candidates observed with NTT/EFOSC2. The slit width is fixed to $1\arcsec$. The classified type of the companion source of the pair is listed in column (8). The basis of classification is noted in column (9), as detailed in Section~\ref{subsec:classification}. The superscripts ``G", ``K", and ``S" on top of the source names indicate additional observations with Gemini, Keck, and Subaru on the same source. The bold lines highlight the confirmed physical pairs.}
    \label{tab:observation_NTT}
    \begin{tabular}{lcccccccc}
    \hline
    Name & Grism+Filter & PA & Date & Seeing & Airmass & Exposure & Type & Note\\
    (J2000) &  & (degree) & (dd.mm.yy) & (") &  & (s) &  & \\
    (1) & (2) & (3) & (4) & (5) & (6) & (7) & (8) & (9)\\
    \hline
    \textbf{000129.98+001911.3} & \textbf{Gr\#6+CG375} & \textbf{118} & \textbf{04.09.22} & \textbf{1.9} & \textbf{1.2} & $\mathbf{1000\times3}$ & \textbf{quasar (z=1.946)} & \textbf{(i-b), (ii-a), (iii-a), (iv-a2, b3)}\\
    000157.44+030147.6 & Gr\#6+CG375 & 118 & 05.09.22 & 0.9 & 1.5 & $1000\times3$ & A star & (v-a)\\
    \textbf{010123.64+010045.3} & \textbf{Gr\#5} & \textbf{131} & \textbf{05.09.22} & \textbf{0.8} & \textbf{1.5} & $\mathbf{800\times3}$ & \textbf{LRG (z=0.275)} & \textbf{(v-cd)}\\
    011728.83+023309.9 & Gr\#6+CG375 & 11 & 04.09.22 & 1.7 & 1.2 & $1000\times3$ & G star & -\\
    020130.46-004448.9 & Gr\#6+CG375 & 23 & 05.09.22 & 0.8 & 1.5 & $1000\times4$ & O star & (v-a)\\
    020425.44+003741.6 & Gr\#6+CG375 & 38 & 04.09.22 & 1.3 & 1.2 & $800\times3$ & O star & (v-a)\\
    020452.73-010822.4 & Gr\#6+CG375 & 125 & 05.09.22 & 1.0 & 1.3 & $1200\times3$ & O star & (v-a)\\
    022225.92+003401.7 & Gr\#6+CG375 & 29 & 05.09.22 & 1.7 & 1.2 & $1000\times3$ & O star & (v-a)\\
    022325.00-003612.6 & Gr\#6+CG375 & 92 & 05.09.22 & 1.8 & 1.1 & $600\times3$ & LRG (z=0) & (v-bcd)\\
    023546.83+024814.3 & Gr\#6+CG375 & 118 & 05.09.22 & 1.8 & 1.2 & $1000\times3$ & F star & (v-ad)\\
    023606.55-020410.3 & Gr\#6+CG375 & 11 & 04.09.22 & 1.4 & 1.1 & $1000\times3$ & A star & (v-a)\\
    023629.70+020327.5 & Gr\#6+CG375 & 64 & 04.09.22 & 1.7 & 1.2 & $1000\times3$ & A star & (v-d)\\
    085244.38+050325.6 & Gr\#5 \& Gr\#6 & 2 & 14.04.21 & 0.9 & 1.2 & $600\times6$ & K star & (v-bc)\\
    090033.55+025727.9 & Gr\#6+CG375 & 63 & 07.02.22 & 0.7 & 1.4 & $1000\times3$ & K star & -\\
    090500.09+033609.1 & Gr\#6+CG375 & 135 & 07.02.22 & 0.75 & 1.3 & $1200\times3$ & K star & -\\
    090550.30+003948.1 & Gr\#5 & 31 & 09.04.22 & 0.8 & 1.2 & $600\times3$ & unclassified & spatially unresolved\\
    092725.37-011652.9 & Gr\#6+CG375 & 79 & 07.02.22 & 0.5 & 1.1 & $1000\times3$ & F/A star & (v-a)\\
    094713.15+024743.6 & Gr\#6+CG375 & 35 & 07.02.22 & 0.55 & 1.2 & $1000\times3$ & F/A star & -\\
    \textbf{100547.80+013348.4} & \textbf{Gr\#6+CG375} & \textbf{71} & \textbf{07.02.22} & \textbf{0.6} & \textbf{1.2} & $\mathbf{1200\times3}$ & \textbf{quasar (z=2.055)} & \textbf{(i-b), (ii-a), (iii-a), (iv-a2, b3)}\\
    101419.40-002834.2 & Gr\#6 & 57 & 15.04.21 & 0.9 & 1.2 & $800\times1$ & QG (z=0.34?) & (v-a), slit misaligned, promising\\
    101444.48-015244.2 & Gr\#6 & 20 & 14.04.21 & 0.95 & 1.1 & $1200\times3$ & A star & (v-f)\\
    102614.83+040730.6 & Gr\#6+CG375 & 17 & 09.04.22 & 0.7 & 1.3 & $800\times3$ & quasar (z=1.18?) & (i-b), (ii-b), (iii-a)\\
    103737.45+005836.6 & Gr\#6+CG375 & 116 & 09.04.22 & 0.7 & 1.2 & $1000\times3$ & quasar (z=1.708) & (i-b), (ii-b), (iii-a)\\
    104122.84-005618.4 & Gr\#6 & 53 & 13.04.21 & 0.8 & 1.1 & $600\times4$ & F/A star & (v-ad)\\
    104802.22+050301.0 & Gr\#6+CG375 & 5 & 07.02.22 & 0.65 & 1.3 & $800\times3$ & quasar (z=2.019) & (i-a), (ii-b), (iii-a)\\
    105302.94+033355.8 & Gr\#6 & 40 & 13.04.21 & 0.85 & 1.2 & $1000\times3$ & A star & (v-ad)\\
    105409.83+042435.6 & Gr\#6+CG375 & 116 & 09.04.22 & 0.8 & 1.2 & $800\times3$ & F star & -\\
    105716.18+033930.0 & Gr\#6 & 17 & 14.04.21 & 0.95 & 1.2 & $600\times3$ & F star & -\\
    \textbf{110556.18+031243.1} & \textbf{Gr\#5} & \textbf{59} & \textbf{09.04.22} & \textbf{0.8} & \textbf{1.4} & $\mathbf{300\times3}$ & \textbf{SFG/QG (z=0.356)} & \textbf{(i-b), (ii-a), (iii-c), (v-a)}\\
    111006.87+033107.1 & Gr\#6 & 146 & 14.04.21 & 0.9 & 1.2 & $800\times3$ & F star & -\\
    112144.76+011421.9 & Gr\#6 & 154 & 13.04.21 & 0.9 & 1.2 & $600\times3$ & F star & (v-a)\\
    \textbf{112753.39-020142.7}$^K$ & \textbf{Gr\#6} & \textbf{78} & \textbf{14.04.21} & \textbf{0.9} & \textbf{1.2} & $\mathbf{1200\times3}$ & \textbf{SFG (z=2.292)} & \textbf{(i-b), (ii-a), (iii-a), (iv-a2, b3)}\\
    113613.37+033840.9 & Gr\#6 & 77 & 13.04.21 & 1.0 & 1.3 & $600\times3$ & K star & -\\
    114222.52+012443.8 & Gr\#6 & 164 & 15.04.21 & 0.9 & 1.2 & $800\times3$ & unclassified & faint but likely LRG\\
    \textbf{121745.76+022109.5} & \textbf{Gr\#5} & \textbf{40} & \textbf{15.04.21} & \textbf{0.9} & \textbf{1.2} & $\mathbf{800\times3}$ &\textbf{SFG (z=0.281)} & \textbf{(i-b), (ii-a), (iii-c), (v-bc)}\\ 
    121912.94+035142.2$^K$ & Gr\#6+CG375 & 82 & 07.02.22 & 0.7 & 1.2 & $800\times3$ & F star & (v-a)\\
    123101.66+031118.2$^G$ & Gr\#6 & 47 & 14.04.21 & 0.9 & 1.3 & $400\times3$ & F/A star & (v-abcd)\\
    123916.97+014654.2 & Gr\#6 & 62 & 13.04.21 & 0.9 & 1.3 & $1000\times3$ & G star & -\\
    124153.74+003648.0 & Gr\#6 & 110 & 14.04.21 & 0.9 & 1.3 & $1200\times3$ & G star & -\\
    125141.90+031140.9 & Gr\#6 & 1 & 13.04.21 & 1.2 & 1.2 & $800\times3$ & F star & (v-a)\\
    125258.75-004236.1 & Gr\#5 & 87 & 09.04.22 & 0.7 & 1.2 & $1000\times3$ & G star & (v-a)\\
    133709.29+011351.1 & Gr\#6+CG375 & 60 & 09.04.22 & 1.0 & 1.2 & $1200\times3$ & G star & -\\
    \textbf{134722.75+015504.9} & \textbf{Gr\#5} & \textbf{114} & \textbf{09.04.22} & \textbf{1.2} & \textbf{1.3} & $\mathbf{600\times3}$ & \textbf{LRG (z=0.399)} & \textbf{(v-ab)}\\
    134826.86-005943.9 & Gr\#5 & 169 & 15.04.21 & 0.9 & 1.2 & $800\times3$ & F star & (v-d)\\
    135619.16+001813.3 & Gr\#6 & 1 & 15.04.21 & 0.9 & 1.3 & $800\times3$ & unclassified & too faint\\
    141235.36+004104.1 & Gr\#6 & 108 & 15.04.21 & 0.9 & 1.5 & $800\times3$ & unclassified & too faint and blended\\
    142944.25+042202.6 & Gr\#6+CG375 & 160 & 09.04.22 & 1.3 & 1.4 & $1000\times2$ & K star & (v-ad)\\
    143452.74-002828.0 & Gr\#6+CG375 & 128 & 09.04.22 & 1.1 & 1.4 & $1000\times2$ & K star & (v-c)\\
    143822.97-015557.9 & Gr\#6 & 158 & 13.04.21 & 0.75 & 1.2 & $1000\times3$ & F/A star & -\\
    144145.09+023743.0 & Gr\#6 & 1 & 13.04.21 & 0.8 & 1.3 & $600\times3$ & F star & (v-a)\\
    144407.63-010152.7 & Gr\#5 & 130 & 09.04.22 & 1.1 & 1.6 & $1000\times2$ & unclassified & too faint and blended\\
    144551.13-000650.3 & Gr\#6 & 122 & 14.04.21 & 0.9 & 1.4 & $800\times3$ & F star & (v-d)\\
    150257.02+023917.7 & Gr\#6 & 131 & 14.04.21 & 0.9 & 1.5 & $800\times3$ & LRG (z=0.085) & (v-a)\\
    150414.02+020112.0 & Gr\#6 & 52 & 14.04.21 & 0.9 & 1.5 & $800\times3$ & LRG (z=0.325) & (v-b)\\
    215824.40+035217.5$^S$ & Gr\#6+CG375 & 122 & 04.09.22 & 2.0 & 1.4 & $1000\times4$ & A star & (v-a)\\
    220239.30+000650.0 & Gr\#6+CG375 & 52 & 05.09.22 & 1.0 & 1.4 & $1000\times3$ & A star & (v-d)\\
    222050.95-001744.5$^S$ & Gr\#6+CG375 & 25 & 04.09.22 & 1.7 & 1.2 & $1000\times3$ & F star & (v-d)\\
    231402.87+001430.2$^S$ & Gr\#6+CG375 & 45 & 04.09.22 & 1.7 & 1.2 & $400\times3$ & F star & (v-ad)\\
    \textbf{231854.69-012725.9$^S$} & \textbf{Gr\#6+CG375} & \textbf{133} & \textbf{05.09.22} & \textbf{1.5} & \textbf{1.5} & $\mathbf{1000\times3}$ & \textbf{quasar (z=2.346)} & \textbf{(i-a), (ii-a), (iii-a), (iv-a2, b2)}\\
    232610.23-010427.8 & Gr\#6+CG375 & 29 & 04.09.22 & 1.9 & 1.2 & $800\times3$ & O star & -\\
    \hline
    \end{tabular}
\end{table*}

\subsubsection{Gemini/GMOS-N}
We observed 30 candidates with the Gemini Multi-Object Spectrograph on Gemini North (GMOS-N; \citealt{hook2004gemini}) in queue mode during three programs: GN-2020A-Q-108, GN-2020B-Q-116, and GN-2021A-Q-121 (PI: J. Silverman). Observational details are listed in Table~\ref{tab:observation_Gemini}. A slit width of $0\farcs75$ was used in 2020A, while $1\arcsec$ slits were used in 2020B and 2021A.
\par
We employed the B600 grating with the CG455 filter and the R831 grating with the OG515 filter. The central wavelength setting (in $\mu$m) is reported in Table~\ref{tab:observation_Gemini}. The B600 grating provides a dispersion of 0.45~\AA\ pixel$^{-1}$ and spectral resolution $R=\lambda/\Delta\lambda \approx 1688$, while R831 offers 0.34~\AA\ pixel$^{-1}$ and $R \approx 4396$. Under queue scheduling, observations were executed in median conditions, corresponding to a typical seeing of $\sim0\farcs5$ at Maunakea \citep{lyman2020forecasting}. CuAr lamps were used for wavelength calibration, and the spectrophotometric standards HZ~44 and Feige~66 \citep{oke1990faint} were observed for flux calibration.

\begin{table*}
    \centering
    \caption{Observation setups, weather conditions, and classifications for the candidates observed with Gemini/GMOS-N. The slit width is $0.75\arcsec$ in 2020A and $1\arcsec$ thereafter. Observations were carried out in queue mode; when multiple nights contributed to a source, only the principal date is listed.}
    \label{tab:observation_Gemini}
    \begin{tabular}{llccccccc}
    \hline
    Name & Grism+Filter & PA & Date & Airmass & Exposure & Type & Note\\
    (J2000) &  & (degree) & (dd.mm.yy) &  & (s) & (spec) & \\
    (1) & (2) & (3) & (4) & (5) & (6) & (7) & (8)\\
    \hline
    021907.35-022715.2 & B600:0.580+CG455 & 131.7 & 16.08.20 & 1.4 & $120\times2$ & K star & (v-bc)\\
    021322.99-042134.3 & B600:0.580+CG455 & 89.5 & 16.08.20 & 1.3 & $847\times3$ & quasar (z=0.97) &  (i-b), (ii-b), (iii-a)\\ 
    090613.75+041923.1 & R831:0.860+OG515 & 40.7 & 22.12.20 & 1.1 & $847\times7$ & unclassified & too faint\\
    \textbf{092919.16+041414.2} & \textbf{B600:0.680+CG455} & \textbf{143.2} & \textbf{16.02.21} & \textbf{1.1} & $\mathbf{300\times3}$ & \textbf{QG (z=0.369)} & \textbf{(v-bcf)}\\ 
    \textbf{092919.16+041414.2} & \textbf{B600:0.680+CG455} & \textbf{74.7} & \textbf{16.02.21} & \textbf{1.1} & $\mathbf{300\times3}$ & \textbf{QG (z=0.369)} & \textbf{(v-bf)}\\ 
    \textbf{100701.56+052315.4} & \textbf{B600:0.680+CG455} & \textbf{144.8} & \textbf{16.02.21} & \textbf{1.1} & $\mathbf{600\times3}$ & \textbf{QG (z=0.539)} & \textbf{(v-bf)} \\ 
    101719.42+010534.6 & B600:0.630+CG455 & 176.1 & 22.02.21 & 1.1 & $600\times3$ & unclassified & featureless but promising\\
    111347.70+023830.3 & B600:0.630+CG455 & 8 & 20.04.21 & 1.5 & $600\times3$ & quasar (z=1.596) & (i-a), (ii-b), (iii-a)\\ 
    115302.69+011027.8 & R831:0.800+OG515 & 73 & 14.04.21 & 1.1 & $600\times3$ & SFG (z=0.075) & (i-a), (ii-b), (iii-c)\\ 
    \textbf{122349.29+021449.1} & \textbf{B600:0.680+CG455} & \textbf{153.8} & \textbf{22.03.21} & \textbf{1.1} & $\mathbf{600\times3}$ & \textbf{QG (z=0.408)} & (v-ab)\\ 
    123101.66+031118.2$^N$ & B600:0.630+CG455 & 46.6 & 21.03.21 & 1.1 & $300\times2$ & F/A star & (v-abcd)\\ 
    \textbf{123939.06+003439.8$^K$} & \textbf{R831:0.860+OG515} & \textbf{127.4} & 15.01.21 & 1.1 & $847\times3$ & \textbf{quasar (z=2.134)} & \textbf{(i-a), (ii-a), (iii-a), (iv-a2, b3)}\\ 
    124604.03-010954.6 & B600:0.660+CG455 & 51.9 & 20.05.20 & 1.2 & $847\times3$ & LRG (z=0.31?) & (v-a)\\ 
    \textbf{125156.49+015249.7} & \textbf{B600:0.630+CG455} & \textbf{149.2} & \textbf{30.04.21} & \textbf{1.1} & $\mathbf{600\times3}$ & \textbf{SFG/QG (z=0.33)} & \textbf{(i-a), (ii-a), (iii-c), (v-bef)}\\
    \textbf{130550.51-012331.5} & \textbf{B600:0.630+CG455} & \textbf{75.9} & \textbf{15.04.21} & \textbf{1.1} & $\mathbf{600\times3}$ & \textbf{SFG (z=0.251)} & (i-a), (ii-a), (iii-c)\\ 
    141003.70+015901.8 & B600:0.630+CG455 & 132.8 & 24.03.21 & 1.1 & $600\times3$ & G star & (v-c)\\
    141712.26+521655.8 & R831:0.800+OG515 & 103.8 & 15.04.21 & 1.2 & $600\times6$ & unclassified & common absorptions, promising\\
    142526.50-004421.5 & R831:0.860+OG515 & 5.4 & 23.05.20 & 1.1 & $847\times1$ & QG (z=0.52) & - \\
    143316.38+005145.5 & B600:0.660+CG455 & 131.6 & 20.05.20 & 1.2 & $847\times3$ & M1 star & (v-e)\\
    144308.16-004913.4 & B600:0.660+CG455 & 84.2 & 20.05.20 & 1.2 & $847\times4$ & unclassified & featureless but promising\\
    \textbf{145201.59-011945.3} & \textbf{R831:0.800+OG515} & \textbf{58.5} & \textbf{17.04.21} & \textbf{1.3} & $\mathbf{600\times3}$ & \textbf{quasar (z=1.876)} & \textbf{(i-b), (ii-a), (iii-a), (iv-a1, b3)}\\
    145347.46+003927.0 & B600:0.660+CG455 & 131.8 & 21.05.20 & 1.2 & $847\times3$ & unclassified & featureless but promising\\
    151434.38+434507.1 & B600:0.630+CG455 & 73.5 & 24.03.21 & 1.1 & $300\times3$ & SFG (z=0.33) & (i-a), (ii-b), (iii-c)\\ 
    152750.12+422035.5 & R831:0.800+OG515 & 30.5 & 30.04.21 & 1.1 & $180\times3$ & M1 star & (v-e)\\
    153342.05+442106.5 & B600:0.630+CG455 & 118.4 & 24.03.21 & 1.1 & $500\times3$ & quasar (z=1.325) & (i-b), (ii-b), (iii-a)\\
    221926.32-004612.9 & B600:0.580+CG455 & 55.1 & 13.08.20 & 1.3 & $847\times4$ & SFG (z=0.333) & (i-a), (ii-b), (iii-c)\\ 
    \textbf{222327.52+033902.0} & \textbf{B600:0.630+CG455} & \textbf{49.2} & \textbf{14.05.21} & \textbf{1.6} & $\mathbf{600\times3}$ & \textbf{SFG (z=0.35)} & (i-a), (ii-a), (iii-c)\\ 
    231148.54+004426.0 & B600:0.580+CG455 & 59.9 & 13.08.20 & 1.1 & $847\times2$ & SFG (z=0.194) & (i-a), (ii-b), (iii-c)\\ 
    232443.06+023922.7 & B600:0.630+CG455 & 50.2 & 14.06.21 & 1.3 & $600\times6$ & F star & -\\
    232821.40-004045.2 & R831:0.800+OG515 & 9.5 & 02.08.21 & 1.1 & $600\times6$ & M1 star & (v-e)\\
    \textbf{233225.33+000042.5} & \textbf{B600:0.630+CG455} & \textbf{135.4} & \textbf{04.07.21} & \textbf{1.3} & $\mathbf{600\times12}$ & \textbf{quasar (z=3.271)} & (i-b), (ii-a), (iii-a), (iv-a2, b3)\\ 
    \hline
    \end{tabular}
\end{table*}

\subsubsection{Keck II/NIRES}
We observed four targets with the Near-Infrared Echellette Spectrometer (NIRES; \citealt{wilson2004mass}) on the 10-meter Keck~II telescope. The observations were conducted during half a night on 2022 February 22 (Program ID: U197, PI: X. Prochaska). The observing setups and weather conditions are summarized in Table~\ref{tab:observation_Keck}. These targets were selected for follow-up after the NTT and Gemini runs to obtain complementary near-infrared spectroscopy. Adaptive optics was not employed because the natural seeing was sufficient to resolve the two sources. The slit width of NIRES is fixed at $0.55\arcsec$. NIRES provides simultaneous coverage of $0.94$–$2.45~\mu$m across five echelle orders, with a mean spectral resolution of $R\approx2700$. We used an ABBA dithering sequence to enable accurate subtraction of the near-infrared sky background. Wavelength calibration was performed using night-sky OH lines, and the standard star Feige~110 was observed for flux calibration.

\begin{table*}
    \centering
    \caption{Observation setups, weather conditions, and classifications for the candidates observed with Keck/NIRES, the width of the slit is fixed to $0.55\arcsec$.}
    \label{tab:observation_Keck}
    \begin{tabular}{lcccccccc}
    \hline
    Name & PA & Date & Seeing & Airmass & Exposure & Type & Note\\
    (J2000) & (degree) & (dd.mm.yy) & (\arcsec) &  & (s) & (spec) & \\
    (1) & (2) & (3) & (4) & (5) & (6) & (7) & (8)\\
    \hline
    \textbf{112753.39-020142.7$^N$} & \textbf{77.6} & \textbf{10.02.22} & \textbf{1.0} & \textbf{1.2} & $\mathbf{300\times4}$ & \textbf{SFG (z=2.292)} & \textbf{(i-b), (ii-a), (iii-a), (iv-a2, b3)}\\
    121912.94+035142.2$^N$ & 81.9 & 10.02.22 & 0.8 & 1.4 & $300\times4$ & F star & (v-a)\\
    \textbf{123939.06+003439.8$^G$} & \textbf{127.4} & \textbf{10.02.22} & \textbf{1.4} & \textbf{1.1} & $\mathbf{300\times4}$ & \textbf{quasar (z=2.134)} & \textbf{(i-a), (ii-a), (iii-a), (iv-a2, b3)}\\
    130047.68+041320.4 & 18.7 & 10.02.22 & 1.0 & 1.2 & $300\times8$ & unclassified & too faint, unable to extract spectrum\\
    \hline
    \end{tabular}
\end{table*}

\subsubsection{Subaru/FOCAS}
We observed four targets with the Faint Object Camera and Spectrograph (FOCAS; \citealt{Kashikawa2002focas}) on the 8.2-meter Subaru telescope. The observations were conducted during half a night on 2023 September 5 as filler time in Program S23B-116 (PI: S. Tang). The observing setups and weather conditions are summarized in Table~\ref{tab:observation_Subaru}. These targets were selected as follow-ups to the NTT sample to achieve higher signal-to-noise ratios and to detect additional diagnostic emission lines. We used the VPH850 grating in combination with the SO58 order-sorting filter, which provides a wavelength coverage of 5800–10350~\AA\ at a spectral resolution of $R\approx1500$. A slit of type \texttt{SCFCSLLC08} with $0.5\arcsec$ width was used for all targets. Wavelength calibration was performed using ThAr arc lamps, and the standard star BD+28D4211 was observed for flux calibration.
\begin{table*}
    \centering
    \caption{Observation setups, weather conditions, and classifications for the candidates observed with Subaru/FOCAS, the width of the slit is fixed to $0.55\arcsec$. All of the sources are also observed with NTT.}
    \label{tab:observation_Subaru}
    \begin{tabular}{lcccccccc}
    \hline
    Name & PA & Date & Seeing & Airmass & Exposure & Type & Note\\
    (J2000) & (degree) & (dd.mm.yy) & (\arcsec) &  & (s) & (spec) & \\
    (1) & (2) & (3) & (4) & (5) & (6) & (7) & (8)\\
    \hline
    215824.40+035217.5$^N$ & 122 & 05.09.24 & 0.4 & 1.1 & $1200\times3$ & F star & -\\
    222050.95-001744.5$^N$ & 25 & 05.09.24 & 0.3 & 1.1 & $1000\times2$ & F star & (v-d)\\
    231402.87+001430.2$^N$ & 45 & 05.09.24 & 0.3 & 1.1 & $600\times2$ & F star & (v-ad)\\
    \textbf{231854.69-012725.9$^N$} & \textbf{133} & \textbf{05.09.24} & \textbf{0.4} & \textbf{1.1} & $\mathbf{1200\times2}$ & \textbf{quasar (z=2.346)} & \textbf{(i-a), (ii-a), (iii-a), (iv-a2, b2)} \\
    \hline
    \end{tabular}
\end{table*}

\subsection{Data analysis} \label{subsec:data_analysis}
\subsubsection{Spectrum reduction} \label{subsubsec:spec_reduce}
All spectroscopic data in this work were reduced using the Python Spectroscopic Data Reduction Pipeline \citep[\texttt{PypeIt,}][]{j_xavier_prochaska_2023_7779402}, a semi-automated package for astronomical spectroscopy. The general procedures follow the documentation of \texttt{PypeIt} \citep[\url{https://pypeit.readthedocs.io/en/latest/}][]{prochaska2020pypeit}. 
\par
In brief, the pipeline first trims the overscan region and applies a bad-pixel mask appropriate for each instrument. For Gemini/GMOS data, which are read out in three CCDs with four amplifiers each, \texttt{PypeIt} mosaics the twelve amplifier outputs into a single frame, an essential step for optimizing the wavelength solution. Flat frames are then used to trace the slit edges (and the echelle orders for Keck/NIRES), while arc frames provide a two-dimensional wavelength solution across the detector. In the case of NIRES, which lacks arc lamps, the sky emission lines in the science exposures are used directly for wavelength calibration. Subsequently, the sky background is modeled and subtracted with an improved algorithm from \citet{kelson2003optimal}, yielding reduced two-dimensional spectra.
\par
We then extracted one-dimensional spectra for each component of every pair. Because \texttt{PypeIt}'s automated extraction is not optimized for blended sources with separations of $\lesssim$1\arcsec, we implemented a custom three-step method: (1) from the reduced 2D spectrum, we binned the image along the wavelength axis to enhance S/N and fit a double-Gaussian profile at each row; (2) the Gaussian centers were then fit with a second-order polynomial to trace the spatial positions of the two spectra; (3) using these traces, we refit double Gaussians at each wavelength position on the original 2D frame, and extracted the Gaussian areas as the photon counts. The resulting spectra were flux-calibrated using the sensitivity functions derived from standard stars within \texttt{PypeIt}, and rescaled to match HSC photometry to correct for slit losses (see Appendix~\ref{sec:psf_vs_cmodel} for a comparison between our measurements of the PSF magnitudes and the HSC Pipeline-generated CModel magnitudes). 

\subsubsection{Spectrum fitting} \label{subsec:spec_fitting}
For the confirmed quasar--quasar and quasar--galaxy pairs, we fit the one-dimensional spectra with the \texttt{PyQSOFit} package \citep{guo2018pyqsofit,guo2019constraining,shen2019sloan}. When a source was observed with multiple telescopes, we first combined the spectra by resampling to a common wavelength array and weighted averaging in the overlapping wavelengths before fitting. All fitting was performed in the rest frame. For quasars at $z<0.5$, we applied the prior-informed PCA method in \texttt{PyQSOFit} to subtract host-galaxy emission \citep{ren2024prior}. 
\par
For pure quasar spectra, the code first fits a pseudo-continuum consisting of a power-law plus a third-order polynomial plus an Fe\,{\sc ii} template in line-free windows: 6000–6250~\AA\ and 6800–7000~\AA\ around H$\alpha$, 4435–4630~\AA\ and 5100–5535~\AA\ around H$\beta$, and 2200–2700~\AA\ and 2900–3090~\AA\ around Mg\,{\sc ii}. Regarding the Fe\,{\sc ii} emission, the \cite{boroson1992emission} model is applied to the UV part between 1200 and 3500~\AA, and the \cite{vestergaard2001empirical} model is applied to the optical part between 3686 and 7484~\AA. More details about the compilation of the continuum and iron emission were discussed in \cite{shen2019sloan}. After subtracting the pseudo-continuum, the emission-line-only spectra were modeled with single or multiple Gaussian components, including the H$\alpha$ composite (H$\alpha$ broad and narrow, [N\,{\sc ii}]~$\lambda\lambda$6549,6585, [S\,{\sc ii}]~$\lambda\lambda$6718,6732), H$\beta$ composite (H$\beta$ broad and narrow, core and wing components of [O\,{\sc iii}]~$\lambda\lambda$4959,5007), H$\gamma$ composite (H$\gamma$ broad and narrow, [O\,{\sc iii}]~$\lambda$4364), Mg\,{\sc ii}, C\,{\sc iii}], C\,{\sc iv}, Ly$\alpha$, and N\,{\sc v}~$\lambda$1240. 
\par
The fitting returns the emission-line properties, including central wavelength, velocity dispersion, FWHM, equivalent width (EW), peak position, integrated flux, signal-to-noise ratio, and continuum parameters. The parameter uncertainties are estimated with the MCMC method. These fitting results are then employed to classify the types of sources and measure the BH property.

\subsubsection{BH property measurements}
For the broad-line AGNs with FWHM $>$ 2000 km s$^{-1}$, we derived their black hole (BH) masses using the single-epoch virial method, which is calibrated against reverberation mapping \citep{kaspi2000reverberation}: 
\begin{equation}
M_{\mathrm{BH}}=10^{\alpha} L^{\beta} \left(\frac{\mathrm{FWHM}}{1000~\mathrm{km}~ \mathrm{s}^{-1}}\right)^{\gamma} M_{\odot},
\label{eq:virial}
\end{equation}
where $L$ is the relevant continuum or line luminosity. In this work we adopt the calibrations of \citet{shen2011catalog} for H$\alpha$, \citet{vestergaard2006determining} for H$\beta$ and C\,{\sc iv}, and \citet{schulze2017near} for Mg\,{\sc ii}. The corresponding parameters are listed in Table~\ref{tab:viral}. The total uncertainties include both measurement errors (from the \texttt{PyQSOFit} MCMC fits) and the intrinsic systematic uncertainty of the virial method, $\sim$0.4 dex \citep{shen2011catalog,schulze2018fmos}. We note that the C\,{\sc iv} BH estimator is known to be systematically biased compared to the Balmer estimators due to its outflow-related origin \citep{netzer2015revisiting, coatman2016c}. The blueshift of the C\,{\sc iv} line might be used to mitigate this bias \citep{coatman2017correcting}, which, however, requires a robust measurement of the systematic redshift of the quasar. For our newly confirmed dual quasars with C\,{\sc iv} detections (Table~\ref{tab:emission_lines}), the spectrum coverages are all limited to rest wavelengths $<$3000~\AA, and thus lack a robust redshift estimator. We therefore did not apply any correction to the C\,{\sc iv}-based BH masses.
\par
Bolometric luminosities were estimated from the monochromatic continuum luminosities as $L_{\rm bol} = 3.81\times L_{1350}$, $5.15\times L_{3000}$, and $9.26\times L_{5100}$, respectively \citep{richards2006spectral}. The Eddington luminosity and Eddington ratio are then:
\begin{equation}
\begin{aligned}
    L_{\rm Edd}&=1.26\times10^{38}M_{\rm BH} \\
    \lambda_{\rm Edd}&=L_{\rm bol}/L_{\rm Edd}
\end{aligned}
\end{equation}

\begin{table}
\caption{Parameters of the virial BH mass calibration (Eq.~\ref{eq:virial}). Luminosities are defined as follows: $L_{1350,44}$ is the monochromatic luminosity at 1350~\AA\ in units of $10^{44}\,\mathrm{erg}\,\mathrm{s}^{-1}$, and similarly for $L_{3000,44}$, $L_{5100,44}$, and $L_{\mathrm{H\alpha},42}$ (in units of $10^{42}\,\mathrm{erg}\,\mathrm{s}^{-1}$).}
\begin{tabular}{ccccc}
\hline
line & L & $\alpha$ & $\beta$ & $\gamma$ \\
\hline
C\,{\sc iv} & $L_{1350,44}$ & 6.66 & 0.5 & 2 \\
Mg\,{\sc ii} & $L_{3000,44}$ & 6.74 & 0.62 & 2 \\
H$\beta$ &  $L_{5100,44}$ & 6.91 & 0.5 & 2 \\
H$\alpha$ &  $L_{\mathrm{H\alpha},42}$ & 6.71 & 0.48 & 2.12 \\
\hline
\label{tab:viral}
\end{tabular}
\end{table}

\section{Results} \label{sec:results}
\subsection{Source type classification} \label{subsec:classification}
We classified the spectral types of the companion sources in each pair primarily by cross-correlation with the SDSS spectral templates\footnote{\url{https://classic.sdss.org/dr5/algorithms/spectemplates/}}. The abbreviations and the corresponding SDSS templates are as follows:  
(1) quasar (SDSS templates No.30, 33);
(2) LRG (template 29): Luminous Red Galaxy;  
(3) QG (template 24): Quiescent (``early-type") Galaxy;  
(4) SFG (templates 25–28): Star-Forming (``late-type") Galaxy;  
(5) Stars: classified by the corresponding stellar templates (templates 1–23). 
We note that these templates are sufficient for the classification purpose of this paper, and that more objective and complete galaxy templates can be found in, e.g., \cite{fraix2021unsupervised}.
The final classifications of our sources are listed in the ``Type" columns of Tables~\ref{tab:observation_NTT}, \ref{tab:observation_Gemini}, \ref{tab:observation_Keck}, and \ref{tab:observation_Subaru}. 
Rows corresponding to physical pairs (dual quasars and quasar–galaxy pairs at the same redshift) are highlighted in bold fonts.
The basis of classification is noted in the final columns of the tables. In particular, we considered the following conditions:  
\begin{enumerate}
    \item \textbf{Emission line detection}
    \begin{enumerate}
        \item More than two emission lines detected above $3\sigma$.  
        \item Only one emission line detected above $3\sigma$.  
        \item Only tentative emission-line detections below $3\sigma$.  
    \end{enumerate}
    \item \textbf{Emission line position} 
    \begin{enumerate}
        \item Velocity offset from the known SDSS quasar $<$ 500 km s$^{-1}$.  
        \item Velocity offset $>$ 500 km s$^{-1}$.  
    \end{enumerate}
    \item \textbf{Emission line width} 
    \begin{enumerate}
        \item FWHM $>$ 2000 km s$^{-1}$.  
        \item $1200 < \mathrm{FWHM} < 2000$ km s$^{-1}$.  
        \item FWHM $<$ 1200 km s$^{-1}$.  
    \end{enumerate}
    \item \textbf{Lensing} 
    \begin{enumerate}
        \item Comparison of the two spectra:  
        \begin{enumerate}
            \item Nearly identical (standard deviation of the flux ratios $<$50\% of their median).  
            \item Non-identical (standard deviation $>$50\% of the median).  
        \end{enumerate}
        \item HSC imaging:  
        \begin{enumerate}
            \item A third source detected between the two objects.  
            \item A third source detected within $3\arcsec$ of the system.  
            \item No third source detected within $3\arcsec$.  
        \end{enumerate}
    \end{enumerate}
    \item \textbf{Absorption line detection}:  
    \begin{enumerate}
        \item Ca\,{\sc ii} H\&K $\lambda\lambda$3934.8, 3969.6~\AA.  
        \item Mg\,{\sc i} $\lambda$5176.7~\AA.  
        \item Na\,{\sc i} $\lambda$5895.6~\AA (``D" line).  
        \item Balmer absorption lines.  
        \item TiO molecular bands.  
    \end{enumerate}
\end{enumerate}
Our classification strategy is as follows. If at least one emission line is detected above $3\sigma$ (Condition~i), we measure the velocity offset relative to the SDSS quasar (Condition~ii). Previous works \citep[e.g.,][]{hennawi2006binary,hennawi2010binary} used 2000 km s$^{-1}$ as the threshold, accounting for both the expected peculiar velocities of bound quasar pairs ($\lesssim 500$ km s$^{-1}$) and SDSS redshift uncertainties ($\sim$1500 km s$^{-1}$). Since none of our observed systems have velocity offsets between 500-2000 km s$^{-1}$, the choice of this threshold within this range will not affect our results. Here we adopt 500 km s$^{-1}$ as the criterion. If the offset exceeds this threshold (Condition~ii-b), we classify the system as a projected pair.
\par
Meanwhile, we use the emission-line width to distinguish between quasars and galaxies. Sources with FWHM $>$2000 km s$^{-1}$ are classified as quasars; those with FWHM $<$1200 km s$^{-1}$ as star-forming galaxies (SFGs); and those in between as composites \citep{hao2005active,shen2011catalog}. The 1200 km s$^{-1}$ threshold is a very conservative assumption as the typical FWHM of SFGs is 200-300 km s$^{-1}$. Nevertheless, none of our sources has FWHM in between 1200-2000 km s$^{-1}$. There are two offset quasars with their companions having FWHM $>$500 km s$^{-1}$: J112753.39-020142.7 and J121745.76+022109.5, which we will discuss in detail in Section \ref{subsec:offset_quasars}. We do not adopt the BPT diagrams in our main classification procedure, as the spectra of most of our offset pairs do not cover both axes. It will be discussed in detail if available for specific cases.
\par
If the companion satisfies Condition~(i), Condition~(ii-a), and Condition~(iii-a/b),\footnote{Because the primary source is an SDSS quasar, if the companion is a narrow-line source (Condition~iii-c), it automatically rules out lensing.} we further assess whether the system could be a lensed quasar. In the most straightforward case, if the spectral ratio is non-identical across the wavelength range (Condition~iv-a2) and no third source is detected in the HSC image (Condition~iv-b3), we can confidently reject lensing and classify the system as a dual quasar. For more ambiguous cases, we provide object-specific discussions in Section~\ref{sec:individuals}.  
\par
If no emission lines are detected, the source is most likely a quiescent galaxy (QG) or a star (but can also be quasars merely covered with weak emission lines). In such cases, we attempt to match the continuum shape to SDSS templates and identify redshifts from strong absorption features, including the Fraunhofer lines (Ca\,{\sc ii} H\&K, Mg\,{\sc i} $\lambda$5176.7, Na\,{\sc i} $\lambda$5895.6~\AA), Balmer absorption, and TiO bands\footnote{Line list from \url{https://classic.sdss.org/dr6/algorithms/linestable.php}}. If the spectrum is featureless but has sufficient S/N, we match the continuum slope to SDSS templates to assign a classification. Unclassified cases are generally due to blending or insufficient S/N, as described in the ``Note" column of the tables. 
\par
In total, we have classified six systems as broad-line quasar pairs and eleven systems as offset quasars (quasar-galaxy pairs). The redshifts of these pairs are between 0.2 and 3.3, the angular separations are between $1\farcs3$ and $3\farcs8$, corresponding to projected separations between 5 and 31 kpc. The maximum flux ratio between the two sources of each pair is $\sim$15, as estimated from the spectrum. The faintest case has $i$-band magnitude of 22.6.

\subsection{Physical properties of the dual quasars} \label{subsec:phy_props}
We present the broad emission-line properties of the confirmed dual quasars in Table~\ref{tab:emission_lines}, measured with \texttt{PyQSOFit} (Section~\ref{subsec:spec_fitting}). Within each pair, the newly confirmed companion source is highlighted in bold. For the broad component of each line, we report the signal-to-noise ratio (SNR, in $\sigma$), full width at half maximum (FWHM, in km\,s$^{-1}$), and equivalent width (EW, in \text{\AA}). Lines not covered by our observations are denoted with ``--", while covered but undetected lines (SNR $< 1\sigma$) are marked with ``$\cdots$".
\par
From the continuum and emission-line fitting results, we estimate the black hole (BH) properties of the dual quasars as described in Section~\ref{subsec:data_analysis}. Table~\ref{tab:BH_props} lists the bolometric luminosity ($L_{\rm bol}$), BH mass ($M_{\rm BH}$), and Eddington ratio ($\lambda_{\rm Edd}$) for our six newly confirmed dual quasars. For the pair of J1239+0034, multiple estimators of $L_{\rm bol}$ and $M_{\rm BH}$ are available, we adopt $~L_{\rm bol}^{5100 \text{\AA}}$ and $M_{\rm BH}^{\rm H\alpha}$ when computing $\lambda_{\rm Edd}$. For the pair of J2318–0127, although Mg\,{\sc ii} is detected in the spectrum, $L_{3000}$ falls at the edge of the coverage, thus $L^{3000}_{\rm bol}$ and $M_{\rm BH}^{\rm \text{Mg \sc{ii}}}$ were not measured for this case.
\par
\begin{figure*}
    \includegraphics[width=0.9\textwidth]{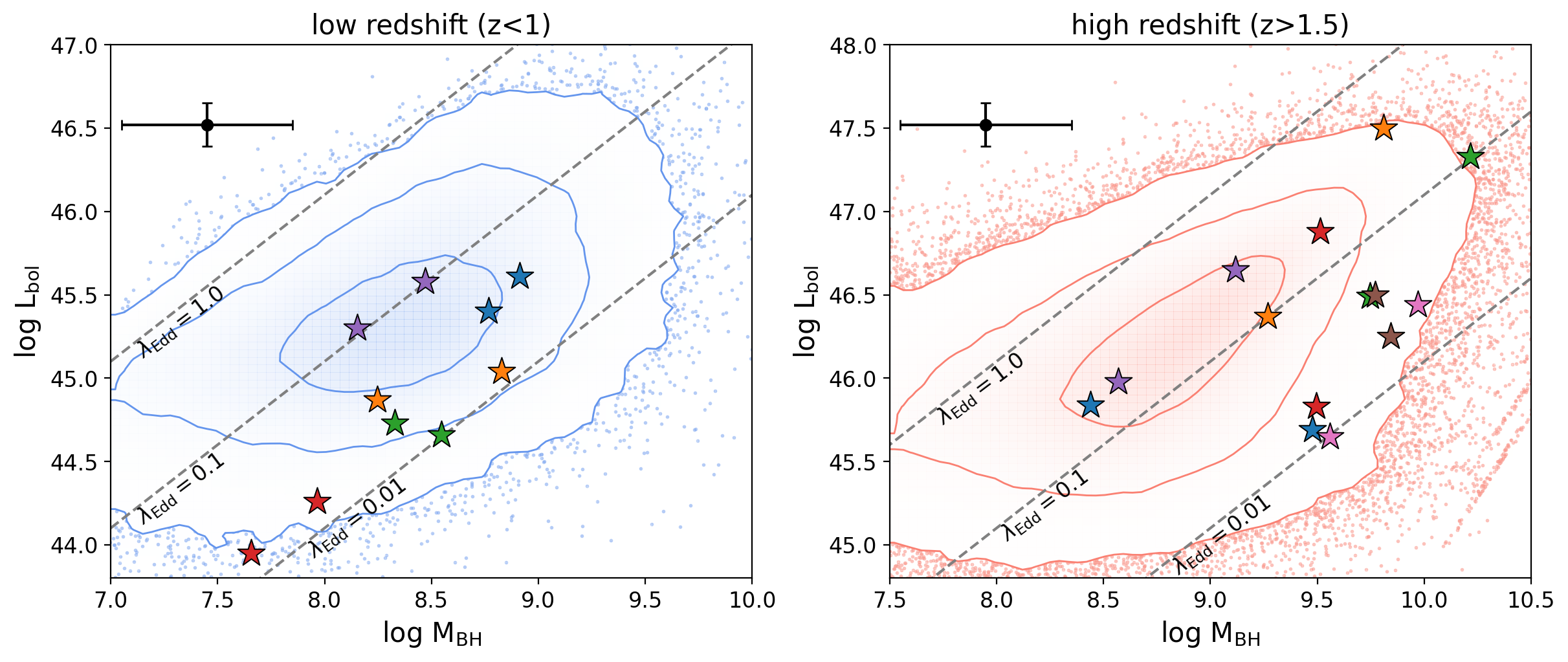}
    \caption{Bolometric luminosities vs. black hole masses for confirmed dual quasars by our project (colored star marks), including six pairs from our previous works. \textbf{Left panel:} $z<1$ systems; \textbf{Right panel:} $z>1.5$ systems. In each panel, the same colored star marks indicate every two members of each pair. Representative uncertainties in $M_{\rm BH}$ and $L_{\rm bol}$ are shown in the top-left corner of each panel. Underlying contours show the 1$\sigma$, 2$\sigma$, and 3$\sigma$ distributions of SDSS DR14 quasars in the corresponding redshift range. Dashed gray lines indicate constant Eddington ratios ($\lambda_{\rm Edd}=1.0, 0.1, 0.01$).}
    \label{fig:BH_bol_edd}
\end{figure*}

Figure~\ref{fig:BH_bol_edd} compares the bolometric luminosities and black hole masses of our confirmed dual quasars \citep[including six pairs from][]{silverman2020dual, tang2021optical} with the general SDSS quasar population \citep[][shown as colored contours]{rakshit2020spectral}. The left panel shows the $z<1$ systems, while the right panel presents the $z>1.5$ systems. In each panel, the two quasars within a given pair are plotted with the same colored star symbols. Representative uncertainties for $M_{\rm BH}$ and $L_{\rm bol}$ are shown as error bars in the upper left. These uncertainties are dominated by systematics: $\sim$0.4 dex for $M_{\rm BH}$ from the virial method \citep{shen2011catalog}, and $\sim$0.13 dex for $L_{\rm bol}$ from optical/UV bolometric corrections \citep{runnoe2012updating}. Statistical fitting uncertainties are negligible in most cases, as indicated by the \texttt{PyQSOFit} MCMC analyses.
\par
All the dual quasars newly confirmed in this work are at $z>1.5$. As shown in Figure~\ref{fig:BH_bol_edd}, they occupy the relatively high BH mass ($\log M_{\rm BH}/M_\odot \sim 8.5$--10) and wide range of bolometric luminosity ($\log L_{\rm bol}/{\rm erg\,s^{-1}} \sim 45.5$--47.5) region of the entire $z>1.5$ SDSS quasar population. The dashed lines indicate constant Eddington ratios ($\lambda_{\rm Edd}=1.0, 0.1, 0.01$). Most of our systems accrete at moderate to low Eddington ratios ($\lambda_{\rm Edd}\sim 0.01$--0.4), consistent with the broader single quasar population of similar mass \citep[e.g.,][]{jones2016intrinsic}. 
\par
All the low-redshift pairs in Figure~\ref{fig:BH_bol_edd} were reported in our earlier work, from which we found that BH masses and host stellar masses of dual quasars are slightly elevated above the local $M_{\rm BH}$--$M_\ast$ scaling relation \citep{tang2021optical}. However, we cannot robustly measure host stellar masses for our new $z>1.5$ dual quasars from HSC imaging and SED fitting alone. At these redshifts, space-based imaging is generally required to resolve quasar host galaxies \citep{ding2020mass}. Alternatively, stellar masses can be approximated by subtracting the gas mass from the dynamical mass in submillimeter interferometric observations, such as those provided by ALMA \citep[e.g.,][]{bischetti2017wissh, molyneux2025evidence}.
\par
Crucially, neither our sample nor previous studies show compelling evidence that galaxy mergers systematically drive both BHs to near-Eddington accretion simultaneously. Instead, confirmed dual quasars exhibit $\lambda_{\rm Edd}$ distributions indistinguishable from those of single quasars, suggesting that while mergers can trigger concurrent nuclear activity, the detailed fueling and duty cycle of each SMBH are controlled primarily by the local gas supply and feedback processes rather than by a wholesale boost in accretion efficiency \citep{kawaguchi2020mass}. We also note that these dual quasars probably have not reached their peak of activity among the mergers, given that their physical separations are still above the kpc level. Numerical simulations suggested increasing AGN activities and accretion rate with decreasing separations between the BHs down to the sub-kpc scale \citep{capelo2015growth}. The dual quasars provide observational constraints on the initial conditions of sub-kpc SMBH pairs. Follow-up observations probing the ISM environment of the duals will shed light on the BH growth mechanism in such systems.
    
\begin{landscape}
\begin{table}
    \caption{Broad emission line measurements of the newly-confirmed dual quasars. For each source, the broad component of the listed lines is reported with signal-to-noise ratio (SNR in $\sigma$), full width at half maximum (FWHM in km~s$^{-1}$), and equivalent width (EW in \AA). Lines not covered by our spectra are indicated as ``--", and lines covered but undetected (SNR $<1\sigma$) are shown as ``$\cdots$". Newly confirmed companion sources in each pair are highlighted in bold fonts.}
    \label{tab:emission_lines}
    \begin{tabularx}{\linewidth}{lc|c|X|X|c|X|X|c|X|X|c|X|X|c|X|X|}
    \hline
    \multirow{3}{*}{Name} & \multirow{3}{*}{redshift} & \multicolumn{3}{c|}{C\,{\sc iv}$\lambda$1549} & \multicolumn{3}{c|}{C\,{\sc iii}]$\lambda$1909} & \multicolumn{3}{c|}{Mg\,{\sc ii}$\lambda$2799} & \multicolumn{3}{c|}{H$\beta$} & \multicolumn{3}{c|}{H$\alpha$}\\
    \cline{3-17}
     &  & SNR & FWHM & EW & SNR & FWHM & EW & SNR & FWHM & EW & SNR & FWHM & EW & SNR & FWHM & EW\\
     &  & ($\sigma$) & (km/s) & (\text{\AA}) & ($\sigma$) & (km/s) & (\text{\AA}) & ($\sigma$) & (km/s) & (\text{\AA}) & ($\sigma$) & (km/s) & (\text{\AA}) & ($\sigma$) & (km/s) & (\text{\AA})\\
    \hline
    \textbf{0001+0019\_N} & 1.9457 & $6.0\pm1.0$ & $2874\pm670$ & $53\pm10$ & $2.1\pm0.6$ & $5639\pm5643$ & $36\pm14$ & -- & -- & -- & -- & -- & -- & -- & -- & -- \\
    0001+0019\_S & 1.9474 & $5.7\pm0.6$ & $10568\pm1059$ & $128\pm11$ & $2.6\pm0.5$ & $6457\pm1200$ & $22\pm4$ & -- & -- & -- & -- & -- & -- & -- & -- & -- \\ \hline
    \textbf{1005+0133\_N} & 2.0551 & $5.3\pm1.0$ & $5124\pm2257$ & $21\pm4$ & $\cdots$ & $\cdots$ & $\cdots$ & -- & -- & -- & -- & -- & -- & -- & -- & -- \\
    1005+0133\_S & 2.0525 & $13.2\pm0.2$ & $4256\pm89$ & $30\pm0$ & $4.9\pm0.1$ & $8105\pm409$ & $30\pm1$ & -- & -- & -- & -- & -- & -- & -- & -- & -- \\ \hline
    \textbf{1239+0034\_N} & 2.1338 & -- & -- & -- & -- & -- & -- & $2.3\pm0.0$ & $6712\pm199$ & $29\pm1$ & $2.6\pm0.1$ & $16950\pm875$ & $821\pm35$ & $5.0\pm0.3$ & $8093\pm350$ & $599\pm22$ \\ 
    1239+0034\_S & 2.1371 & -- & -- & -- & -- & -- & -- & $6.7\pm0.0$ & $4095\pm11$ & $24\pm0$ & $6.7\pm0.2$ & $10383\pm90$ & $65\pm1$ & $16.5\pm2.3$ & $5734\pm443$ & $251\pm15$ \\ \hline
    1452--0119\_N & 1.8753 & -- & -- & -- & -- & -- & -- & $14.3\pm0.1$ & $3857\pm51$ & $44\pm1$ & -- & -- & -- & -- & -- & -- \\
    \textbf{1452--0119\_S} & 1.8755 & -- & -- & -- & -- & -- & -- & $6.9\pm0.4$ & $3339\pm218$ & $33\pm2$ & -- & -- & -- & -- & -- & -- \\ \hline
    2318--0127\_N & 2.3369 & $5.6\pm0.2$ & $8312\pm494$ & $82\pm4$ & $9.4\pm0.2$ & $8389\pm128$ & $35\pm0$ & $7.8\pm0.1$ & $7684\pm115$ & $41\pm1$ & -- & -- & -- & -- & -- & -- \\
    \textbf{2318--0127\_S} & 2.3461 & $4.0\pm0.5$ & $10796\pm1369$ & $68\pm7$ & $6.5\pm0.2$ & $10141\pm270$ & $31\pm1$ & $4.5\pm0.1$ & $10118\pm260$ & $41\pm1$ & -- & -- & -- & -- & -- & -- \\ \hline
    \textbf{2332+0000\_N} & 3.2706 & $2.0\pm0.1$ & $11944\pm961$ & $38\pm2$ & -- & -- & -- & -- & -- & -- & -- & -- & -- & -- & -- & --  \\
    2332+0000\_S & 3.2649 & $6.4\pm0.1$ & $10926\pm147$ & $58\pm0$ & -- & -- & -- & -- & -- & -- & -- & -- & -- & -- & -- & --  \\
    \hline
    \end{tabularx}
\end{table}
\end{landscape}

\begin{table*}
\caption{BH properties of the confirmed dual quasars in this work.\\
Columns (2--4) list bolometric luminosities ($L_{\rm bol}$) estimated from monochromatic luminosities at 1350 \AA, 3000 \AA, and 5100 \AA. \\
Columns (5--8) give BH masses ($M_{\rm BH}$) derived from the virial method 
using broad emission lines. \\
Column (9) reports the Eddington ratio ($\lambda_{\rm Edd}$). 
Where multiple measurements exist, averages are adopted for $\lambda_{\rm Edd}$.}
\label{tab:BH_props}
\begin{tabular}{cccccccccc}
\hline
Name & $\log~L_{\rm bol}^{1350 \text{\AA}}$ & $\log~L_{\rm bol}^{3000 \text{\AA}}$ & $\log~L_{\rm bol}^{5100 \text{\AA}}$ & $\log~M_{\rm BH}^{\rm \text{C \sc{iv}}}$ & $\log~M_{\rm BH}^{\rm \text{Mg \sc{ii}}}$ & $\log~M_{\rm BH}^{\rm H\beta}$ & $\log~M_{\rm BH}^{\rm H\alpha}$ & $\log~\lambda_{\rm Edd}$\\
 & ($\mathrm{erg~s^{-1}}$) & ($\mathrm{erg~s^{-1}}$) & ($\mathrm{erg~s^{-1}}$) & ($\mathrm{M_{\odot}}$) & ($\mathrm{M_{\odot}}$) & ($\mathrm{M_{\odot}}$) & ($\mathrm{M_{\odot}}$) \\
(1) & (2) & (3) & (4) & (5) & (6) & (7) & (8) & (9)\\
\hline
0001+0019\_N & $45.84$ & -- & -- & $8.44\pm0.45$ & -- & -- & -- & $-0.71$ \\ 
0001+0019\_S & $45.69$ & -- & -- & $9.48\pm0.41$ & -- & -- & -- & $-1.90$ \\ 
\hline
1005+0133\_N & $46.37$ & -- & -- & $9.27\pm0.55$ & -- & -- & -- & $-1.01$ \\ 
1005+0133\_S & $47.50$ & -- & -- & $9.81\pm0.40$ & -- & -- & -- & $-0.42$ \\ 
\hline
1239+0034\_N & -- & $45.77$ & $45.87$ & -- & $9.05\pm0.40$ & $9.82\pm0.40$ & $9.49\pm0.40$ & $-1.78$\\
1239+0034\_S & -- & $46.82$ & $46.94$ & -- & $9.27\pm0.40$ & $9.93\pm0.40$ & $9.51\pm0.41$ & $-0.75$\\
\hline
1452–0119\_N & -- & $46.65$ & -- & -- & $9.12\pm0.40$ & -- & -- & $-0.58$\\
1452–0119\_S & -- & $45.98$ & -- & -- & $8.57\pm0.40$ & -- & -- & $-0.70$\\
\hline
2318–0127\_N & $46.50$ & -- & -- & $9.77\pm0.40$ & -- & -- & -- & $-1.38$\\
2318–0127\_S & $46.25$ & -- & -- & $9.84\pm0.41$ & -- & -- & -- & $-1.71$\\
\hline
2332+0000\_N & $45.65$ & -- & -- & $9.56\pm0.41$ & -- & -- & -- & $-2.02$\\
2332+0000\_S & $46.44$ & -- & -- & $9.97\pm0.40$ & -- & -- & -- & $-1.64$\\
\hline
\end{tabular}
\end{table*}


    
\section{Notes on individual systems} \label{sec:individuals}
\begin{figure*}
    \includegraphics[width=0.9\textwidth]{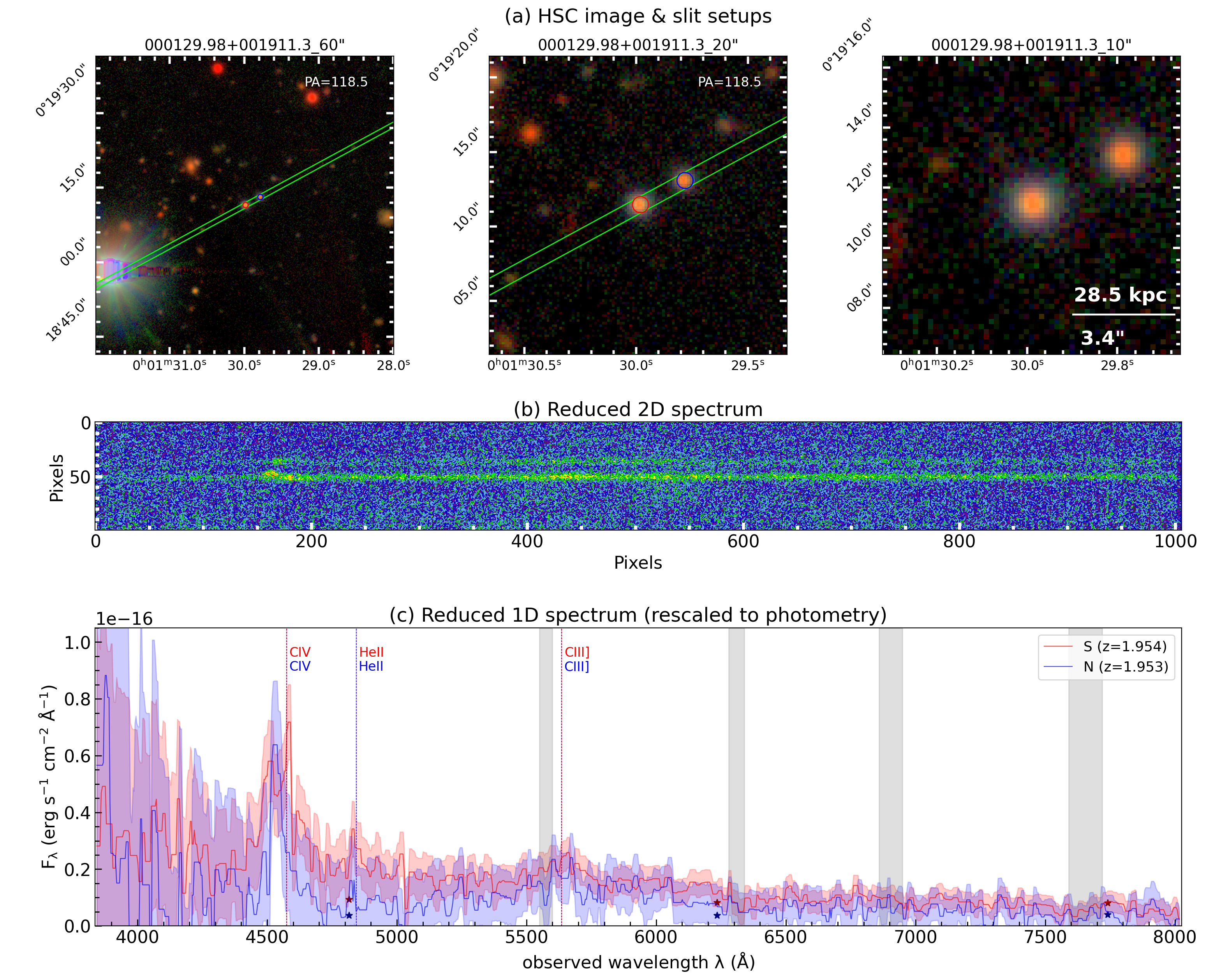}
    \par
    (d) PyQSOFit results
    \vspace*{-1em}
    \begin{multicols}{2}
    {\includegraphics[width=0.5\textwidth]{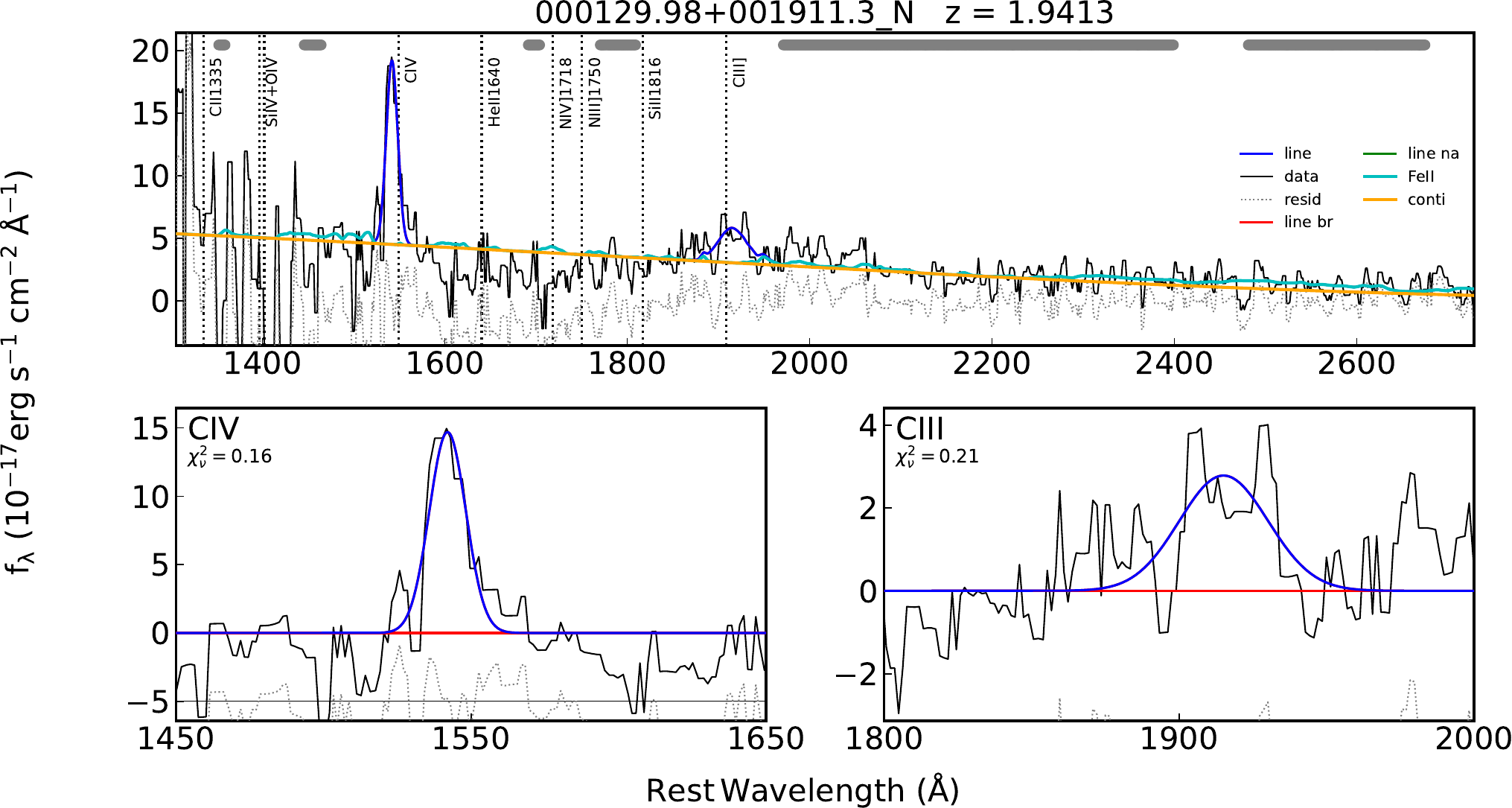}}\par
    {\includegraphics[width=0.5\textwidth]{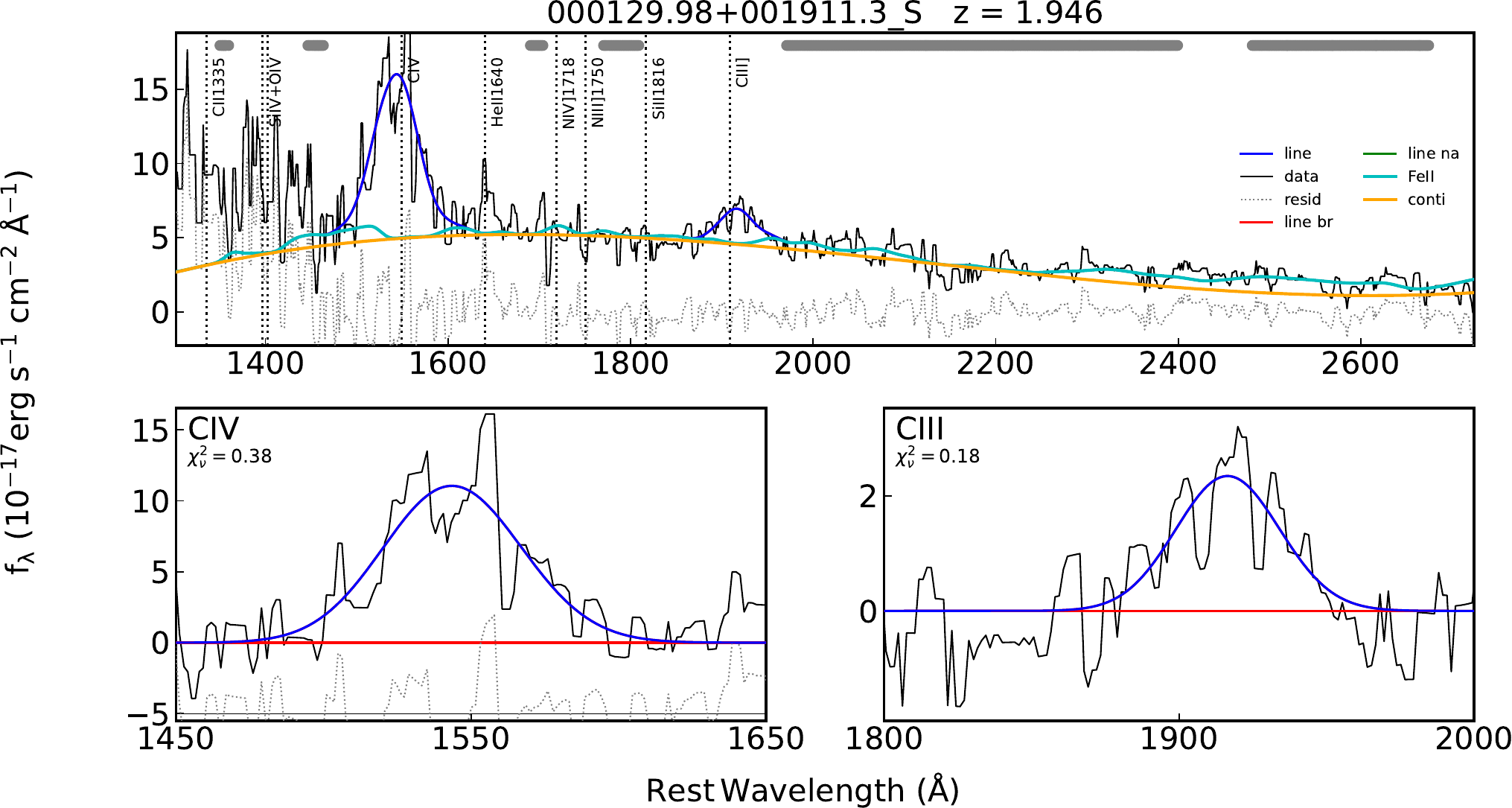}}\par
    \end{multicols}
    \caption{Example of a standardized ``Discovery Panel" for J000129.98+001911.3. 
    Panel (a): HSC three-color images with side lengths of 60\arcsec, 20\arcsec, and 10\arcsec, including slit orientation and source separation. 
    Panel (b): \texttt{PypeIt}-reduced 2D spectrum. The spectrum at the center (50 pixels) corresponds to the SDSS quasar (source circled in red in panel a).
    Panel (c): 1D spectra of the SDSS quasar (red) and the companion source (blue), rescaled to HSC photometry. 
    Panel (d): PyQSOFit results for both 1D spectra.
    A detailed description of the Discovery Panel format is provided in Section~\ref{sec:individuals}. 
    The full figure set for all systems analyzed in this work is available as supplementary material online.}
    \label{fig:J000129.98+001911.3}
\end{figure*}

For each observed system, we present a standardized ``Discovery Panel'' to summarize the imaging and spectroscopic information. Figure~\ref{fig:J000129.98+001911.3} illustrates the format using J000129.98+001911.3 as an example. From top to bottom, each panel contains:
\begin{itemize}
    \item \textbf{HSC imaging:} $gri$ color-composite cutouts with side lengths of 60\arcsec, 20\arcsec, and 10\arcsec. On the 60\arcsec\ and 20\arcsec\ views, the spectroscopic slit is shown as a green rectangle, with its position angle labeled in the upper-right corner. The 10\arcsec\ cutout includes both angular and physical scale bars, indicating the separation between the two sources.
    \item \textbf{2D spectrum:} The reduced two-dimensional spectrum from \texttt{PypeIt} (Section~\ref{subsubsec:spec_reduce}), displayed with wavelength along the horizontal axis and spatial direction along the slit on the vertical axis (both in pixels). The spectrum at the center of the cutout always corresponds to the SDSS quasar, which is also the source at the center of the image cutouts in panel (a), circled in red. The format varies by instrument: a single row for NTT/EFOSC2 and Subaru/FOCAS, three rows for Gemini/GMOS-N (corresponding to its three CCD chips), and five rows for Keck/NIRES (corresponding to its five echelle orders, ordered blue to red from top to bottom). The spectra are normalized with Z-scale and linear stretch (typically $v_{\rm min}=$ median$-1.5\sigma$, $v_{\rm max}=$ median$+3\sigma$), with minor adjustments applied in a few cases to enhance line visibility. 
    \item \textbf{1D spectra:} Extracted spectra of the known SDSS quasar (red) and the companion source (blue). We label the spectra with the relative location of the sources in the image (mostly by ``N"orth or ``S"outh, rarely by ``L"eft or ``R"ight). Gray shaded regions mark wavelengths affected by strong sky emission or absorption. The PyQSOFit-estimated redshifts \footnote{They are generally similar to the SDSS redshifts as given in Table\ref{tab:basic}.} of both sources are shown in the legend, along with vertical dashed lines marking the expected positions of key emission or absorption features. The spectra are flux-calibrated and rescaled to HSC CModel photometry to correct for slit loss; the photometric flux points are shown as star symbols. In high-contrast cases, one spectrum may be rescaled to facilitate comparison (e.g., ``N$\times 2$" indicates the companion spectrum has been multiplied by two).
    \item \textbf{Line fitting results:} For the confirmed dual quasars and quasar-SFG pairs, we show the best-fit results from PyQSOFit for the known SDSS quasar (left) and the companion source (right). The techniques of PyQSOFit are detailed in Section~\ref{subsec:spec_fitting}. The top panels show the fitting of the entire spectra, composed of the continuum (orange curve), iron template (cyan curve), and emission lines (green for narrow lines, and red for broad lines). The wavelength ranges which are used for the continuum fitting are marked with the gray bars. The bottom panels show zoom-in views of the emission lines, after subtracting the continuum and iron emission. For the quasar-QG/LRG pairs, we show the fitting results of the absorption lines of the companion sources instead.
\end{itemize}

The complete set of Discovery Panels for all observed systems is available in the online supplementary material associated with this article. The rest of those for the confirmed dual and offset quasars are presented in Appendix~\ref{sec:discovery_panels}. We encourage readers to use these figures as visual guides to the classifications presented below, and as resources for potential follow-up observations.

\subsection{Dual quasars} \label{subsec:qso_pairs}
In this section, we present six dual quasars with at least one broad emission line detected in both sources at the same redshift.
\subsubsection{J000129.98+001911.3} \label{subsubsection:J000129.98+001911.3}
This system was observed with NTT/EFOSC2 using the Gr\#6+CG375 setup with three exposures of 1000~s each (Figure~\ref{fig:J000129.98+001911.3}). The separation between the two sources is $3.4\arcsec$, corresponding to a projected distance of 28.5~kpc. The southern component (J0001+0019S) is the known SDSS quasar. From its spectrum, we measure $z=1.9474$. 
\par
In the spectrum of the northern source (J0001+0019N), we detect a $6.0\sigma$ broad emission line with ${\rm FWHM}=2874$~km~s$^{-1}$ at 4542~\AA. Interpreting this feature as C\,{\sc iv} $\lambda1549$ yields a redshift of $z=1.9457$, which corresponds to a line-of-sight velocity offset of $64$~km~s$^{-1}$ relative to the SDSS quasar. A tentative $2.1\sigma$ feature is also seen near the expected position of C\,{\sc iii}] $\lambda1909$, though with an unconstrained width (Table~\ref{tab:emission_lines}). An alternative interpretation of the $6.0\sigma$ line as Mg\,{\sc ii} $\lambda2799$ would imply a lower redshift of $z=0.68$. However, in that scenario, additional strong lines ([Ne\,{\sc v}], [O\,{\sc ii}], [Ne\,{\sc iii}], H$\gamma$) should fall within our spectral coverage, none of which are detected. We therefore conclude that both components are most likely at the same redshift ($z\simeq1.95$). 
\par
Next, we consider the lensing hypothesis. The flux ratio between the spectra of J0001+0019S and J0001+0019N is non-identical, varying between $\sim$1 and 3 across the observed wavelength range. Moreover, the C\,{\sc iv} line profile of J0001+0019S is significantly broader than that of J0001+0019N, and shows a double-peaked feature which is unseen in the latter. Finally, no additional galaxies are detected within a $4\arcsec$ radius in the HSC imaging. Taken together, these factors argue against the lensing interpretation and favor the classification of this system as a dual quasar system. 



\subsubsection{J100547.80+013348.4} \label{subsubsection:J100547.80+013348.4}
This system was observed with NTT/EFOSC2 using the Gr\#6+CG375 setup with three exposures of 1200~s each (Figure~\ref{fig:J100547.80+013348.4}). The two components are separated by $2.1\arcsec$, corresponding to a projected distance of 17.5~kpc. The southern source (J1005+0133S) is the known SDSS quasar, for which we measure $z=2.0551$. 
\par
In the spectrum of the northern source (J1005+0133N), we detect a $5.3\sigma$ broad emission line with ${\rm FWHM}=5124$~km~s$^{-1}$ at 4730~\AA. This feature is best identified as C\,{\sc iv} $\lambda1549$, yielding $z=2.0525$, only $89$~km~s$^{-1}$ offset from the SDSS quasar. While C\,{\sc iii}] $\lambda1909$ is not detected, we identify a $\sim3\sigma$ emission feature at the expected wavelength of He\,{\sc ii} $\lambda1640$, consistent with the same redshift. 
\par
We also assess the lensing scenario. J1005+0133S appears slightly bluer than J1005+0133N, and the overall spectral flux ratio is $\sim$8 across the observed wavelength range. Moreover, C\,{\sc iii}] is clearly detected in J1005+0133S but absent in J1005+0133N, indicating intrinsic spectral differences between the two sources. Finally, the HSC imaging reveals no additional galaxies within a $10\arcsec$ radius. Taken together, these arguments strongly disfavor a lensing interpretation. We therefore classify J1005+0133 as a dual quasar system.


\subsubsection{J123939.06+003439.8} \label{subsubsection:J123939.06+003439.8}
This system was first observed with Gemini/GMOS-N using the R831+OG515 grating centered at 860~nm, with three exposures of 847~s each  (Figure~\ref{fig:J123939.06+003439.8}), and subsequently followed up with Keck/NIRES using four exposures of 300~s each (Figure~\ref{fig:J123939.06+003439.8_K}). The two components are separated by $1.7\arcsec$, corresponding to a projected distance of 14.1~kpc. The southern source (J1239+0034S) is the known SDSS quasar, for which we measure $z=2.1371$. 
\par
In the GMOS spectrum of the northern source (J1239+0034N), we detect a tentative $2.3\sigma$ broad emission line at 5975~\AA\ with ${\rm FWHM}=6712$~km~s$^{-1}$, most consistent with Mg\,{\sc ii} $\lambda2799$. To verify this identification, we obtained follow-up NIRES spectroscopy. The NIRES data reveal a $2.6\sigma$ broad H$\beta$ emission line (${\rm FWHM}=16950$~km~s$^{-1}$) and a strong $5.0\sigma$ broad H$\alpha$ emission line (${\rm FWHM}=8093$~km~s$^{-1}$) at the corresponding redshift. From these lines, we determine $z=2.1338$ for J1239+0034N, implying a velocity offset of $106$~km~s$^{-1}$ relative to the SDSS quasar. 
\par
The continua of the two sources are nearly identical in color, with a flux ratio of $\sim$10. However, their emission-line properties differ: H$\beta$ is significantly stronger in J1239+0034S than in J1239+0034N relative to H$\alpha$, indicating intrinsic differences between the quasars. In the HSC imaging, a faint galaxy ($i\simeq22$, $y\simeq22$) lies $4.7\arcsec$ from J1239+0034S at PA=$-138^\circ$. Given its faintness and projected separation, as well as the spectral differences between the two quasars, it is unlikely to act as a lensing galaxy. We therefore classify J1239+0034 as a dual quasar system. 

\subsubsection{J145201.59-011945.3} \label{subsubsection:J145201.59-011945.3}
This system was observed with Gemini/GMOS-N using the R831+OG515 grating centered at 800~nm, with three exposures of 600~s each (Figure~\ref{fig:J145201.59-011945.3}). The two components are separated by $2.6\arcsec$, corresponding to a projected distance of 21.9~kpc. The northern source (J1452$-$0119N) is the known SDSS quasar, for which we measure $z=1.8753$. 
\par
In the spectrum of the southern source (J1452$-$0119S), we detect a $6.9\sigma$ broad emission line at 8044~\AA\ with ${\rm FWHM}=3339$~km~s$^{-1}$. This feature is best identified as Mg\,{\sc ii} $\lambda2799$, from which we determined z=1.8755, with almost no line-of-sight velocity offset from the SDSS quasar. No additional emission lines are detected between 7000–9000~\AA, but the iron emission template provides a good fit under this redshift assumption. 
\par
The two sources exhibit nearly identical optical colors, with an overall flux ratio of $\sim$5. The HSC imaging shows no additional galaxy brighter than $i=23$ within $5\arcsec$ of the pair. Given the consistent redshift interpretation and the absence of a potential lensing galaxy, we classify J1452$-$0119 as a dual quasar system. 


\subsubsection{J231854.69-012725.9} \label{subsubsection:J231854.69-012725.9}
This system was observed with NTT/EFOSC2 using the Gr\#6+CG375 setup with 3$\times$1000~s exposure (Figure~\ref{fig:J231854.69-012725.9}) and subsequently followed up with Subaru/FOCAS with 2$\times$1200~s exposure (Figure~\ref{fig:J231854.69-012725.9_S}). The two components are separated by $3.8\arcsec$, corresponding to a projected distance of 31.1~kpc. The northern source (J2318$-$0127N) is the known SDSS quasar, for which we measure $z=2.3369$. 
\par
In the combined EFOSC2+FOCAS spectrum of the southern source (J2318$-$0127S), we detect broad C\,{\sc iv}, C\,{\sc iii}], and Mg\,{\sc ii} emission lines at the same wavelengths as in J2318$-$0127N, each with significance $\gtrsim4\sigma$. From these lines, we estimate $z=2.3461$, corresponding to a velocity offset of $239$~km~s$^{-1}$ relative to the SDSS quasar. 
\par
The two quasars exhibit nearly identical optical colors with an overall flux ratio of $\sim$2. In the HSC imaging, a third red source lies $\sim4\arcsec$ south of the primary quasar. However, this object is classified as a point source in all five HSC bands (based on the extendedness flag in the forced photometry catalog), and is therefore most likely a foreground star rather than a lensing galaxy. Moreover, the spectrum of J2318$-$0127S shows narrow absorption features superimposed on the C\,{\sc iv} and C\,{\sc iii}] emission lines that are not present in J2318$-$0127N. These intrinsic differences further disfavor a lensing interpretation. We therefore classify J2318$-$0127 as a dual quasar system.

\subsubsection{J233225.33+000042.5} \label{subsubsection:J233225.33+000042.5}
This system was observed with Gemini/GMOS-N using the B600+CG455 setup centered at 630~nm, with $600$\,s\,$\times\,12$ exposures (Figure~\ref{fig:J233225.33+000042.5}). The two sources are separated by $2\farcs3$, corresponding to a projected distance of 17.3~kpc. The southern source (J2332+0000S) is the known SDSS quasar, for which we measure $z=3.2649$. Its spectrum shows a strong Ly$\alpha$ emission line. In the northern source (J2332+0000N), Ly$\alpha$ is detected at $\sim$3$\sigma$ at a consistent wavelength, and a tentative C\,{\sc iv} line is seen at $\sim$2$\sigma$. From these features we estimate $z=3.2706$ for J2332+0000N, corresponding to a velocity offset of 79~km~s$^{-1}$ from the SDSS quasar. 
\par
The continuum flux ratio of the two quasars is $\sim$5, while their Ly$\alpha$ flux ratio is significantly higher ($\sim$14). In the HSC imaging, the only additional bright source in the vicinity is a red point-like object located $\sim$5\arcsec from J2332+0000S, consistent with a foreground star rather than a lensing galaxy. Combined with the spectral differences between the two sources, we therefore classify this system as a dual quasar rather than a lensed quasar. 
\par
Interestingly, both quasars in this system exhibit narrow C\,{\sc iv} absorption lines superimposed on their C\,{\sc iv} emission. Narrow, associated absorbers of this kind are common in quasar spectra \citep[e.g.,][]{vestergaard2003occurrence,hamann2011high}, and may originate from gas inflows, circumnuclear material, or the larger-scale circumgalactic medium (CGM). At face value, the absorbers in J2332+0000 appear slightly redshifted relative to the C\,{\sc iv} emission peaks, which could be interpreted as evidence for inflow. The close similarity of the absorption profiles in both quasars suggests that the absorber may spans at least 17~kpc and may trace shared halo gas. Coherent metal-line absorbers are reported with up to 200 kpc from the center \citep[e.g.,][]{landoni2016circumgalactic}, supporting an interpretation in terms of enriched CGM associated with the merging system. 
\par
In summary, J2332+0000 is a dual quasar system at $z\sim3.27$, and its shared narrow absorption features make it a particularly compelling laboratory for studying the interplay between quasar activity and the CGM in merging environments. 

\subsection{Offset quasars} \label{subsec:offset_quasars}
In this section, we describe the offset quasars where the companion to the SDSS quasar is identified as a paired galaxy without significant broad emission line detection. The companion spectra are classified either as star-forming galaxy (SFGs), based on the presence of narrow emission lines, or as quiescent galaxy (QGs) and luminous red galaxy (LRG), when dominated by absorption features.

\subsubsection{J010123.64+010045.3} \label{subsubsection:J010123.64+010045.3}
This system, at $z=0.275$, was observed with NTT/EFOSC2 using the Gr\#5 grism for three exposures of 800\,s each (Figure~\ref{fig:J010123.64+010045.3}). The angular separation between the two sources is 2.7\arcsec, corresponding to a projected physical distance of 11.3\,kpc. The southern source of the pair is the known SDSS quasar.  
\par
The companion source is brighter in the continuum, with a quasar-to-companion flux ratio of $\sim0.5$. Its spectrum shows significant Mg\,{\sc i}~$\lambda5176.7$ and Na~$\lambda5895.6$ absorption features at the same redshift as the quasar (bottom panels of Figure~\ref{fig:J010123.64+010045.3}). Furthermore, the continuum shape does not match any stellar template in the SDSS library, while well reproduced by the Luminous Red Galaxy (LRG) template redshifted to $z=0.275$. We therefore classify this system as a quasar–LRG pair.  

\subsubsection{J092919.16+041414.2} \label{subsubsection:J092919.16+041414.2}
This system, at $z=0.37$, was observed with Gemini/GMOS-N using the B600+CG455 grating centered at 680\,nm. It is at least a triple system, consisting of a quasar flanked by two elliptical galaxies and connected by a curved optical feature. We refer to the quasar as J0929+0414A, the brighter southern galaxy as J0929+0414B, and the fainter western galaxy as J0929+0414C. The projected separations are 2.1\arcsec\ (10.8\,kpc) between A and B, and 1.6\arcsec\ (8.2\,kpc) between A and C.  
\par
We obtained two sets of spectra, each with $3\times300$\,s exposures, at position angles (PA) of 143.2\degree\ (Figure~\ref{fig:J092919.16+041414.2A}) and 74.7\degree\ (Figure~\ref{fig:J092919.16+041414.2B}) to cover the quasar with either galaxy. Because of the complex morphology, HSC photometry rescaling is not applied. From the flux-calibrated spectra, the continuum flux ratios are $\sim$2 (A/B) and $\sim$5 (A/C). Both J0929+0414B and J0929+0414C show absorption features typical of old stellar populations at the same redshift as the quasar: Ca H\&K $\lambda\lambda$3934.8, 3969.6, the G-band $\lambda$4305.6, and Mg\,{\sc i} $\lambda$5176.7. The G-band absorption is also evident in the quasar spectrum, confirming its association with the same galaxy group. The two galaxies have nearly identical spectra, but given that they share the quasar’s redshift, they cannot be lens images. We therefore classify this system as a quasar–QG–QG triple system.  
\par
Intriguingly, the HSC imaging reveals an arc-like structure northeast of the quasar. The J0929+0414A spectrum obtained at PA=74.7\degree\ is $\sim$60\% brighter than that at PA=143.2\degree, likely due to partial slit coverage of this arc. The morphology resembles shock fronts seen in nearby interacting galaxies \citep[e.g.,][]{rodriguez2014study}, suggestive of tidal debris or merger-driven gas compression. Alternatively, the feature could represent a lensed background galaxy, which cannot be ruled out without deeper spectroscopy or higher-resolution imaging.  
\par
Comparable systems have been recently discovered at higher redshift, such as the remarkable double-ringed merging galaxy at $z=1.14$ identified in the COSMOS field with JWST \citep{li2025cosmic, van2025galaxy}, where the central AGN appears offset within the merger structure. J0929+0414 may serve as a rare low-redshift analog to test scenarios of merger-driven AGN triggering and the role of gas inflows in shaping SMBH growth. Motivated by its appearance in HSC images, we assign this system the nickname of the ``clown" galaxy.

\subsubsection{J100701.56+052315.4} \label{subsubsection:J100701.56+052315.4}
This system, at $z=0.539$, was observed with Gemini/GMOS-N using the B600+CG455 grating centered at 680\,nm, with $3\times600$\,s exposures (Figure~\ref{fig:J100701.56+052315.4}). The two sources are separated by 2.2\arcsec, corresponding to 14.0\,kpc. The southern source is the known SDSS quasar. The continuum flux ratio between the quasar and its companion decreases from $\sim$18 at 5300\,\AA\ to $\sim$3 at 8000\,\AA, indicating that the companion is relatively red. The spectrum of the companion shows clear Ca\,{\sc ii} H\&K absorption and weak G-band absorption at the same redshift as the quasar, while no Balmer, [O\,{\sc ii}], or [O\,{\sc iii}] emission lines are detected. We therefore classify this system as a quasar–QG pair.  

\subsubsection{J110556.18+031243.1} \label{subsubsection:J110556.18+031243.1}
This system, at $z=0.354$, was observed with NTT/EFOSC2 using Gr\#5 with $3\times300$\,s exposures (Figure~\ref{fig:J110556.18+031243.1}). The projected separation between the two sources is 2.4\arcsec\ (11.9\,kpc). The northern source is the known SDSS quasar. The continuum flux ratio between the quasar and its companion is $\sim$1.5, although the 2D spectrum of the companion appears fainter than expected, likely due to a slight slit mis-centering. The companion spectrum exhibits narrow H$\alpha$ and [N\,{\sc ii}] emission lines with ${\rm FWHM}\approx15$\,\AA\ ($\sim$685\,km\,s$^{-1}$) at $z=0.356$. Ca\,{\sc ii} H\&K absorption lines are also found at this redshift. Other expected features, including H$\gamma$, H$\beta$, and [O\,{\sc iii}], are covered by our spectral setup but remain undetected. Therefore, this galaxy is likely made up of a mixed population of young and old stars. We classify this system as a quasar–SFG/QG pair. 

\subsubsection{J112753.39-020142.7} \label{subsubsection:J112753.39-020142.7}
This system was observed with NTT/EFOSC2 using Gr\#6+CG375 with $3\times1200$\,s exposures (Figure~\ref{fig:J112753.39-020142.7}) and subsequently followed up using Keck/NIRES with $4\times300$\,s exposures (Figure~\ref{fig:J112753.39-020142.7_K}). The two sources are separated by 3.2\arcsec, corresponding to a projected distance of 26.3\,kpc. The northern source is the known SDSS quasar, for which we measure $z=2.2867$. The two objects show nearly identical colors in HSC imaging, with a continuum flux ratio of $\sim$3.  
\par
In the NIRES spectrum of the southern source (J1127–0201S), we detect a strong ($\sim6\sigma$) emission line at $\sim2.187\,\mu$m with ${\rm FWHM}=13.6$\,\AA\ ($\sim$619\,km\,s$^{-1}$). Interpreting this feature as H$\alpha$ yields $z=2.2924$, corresponding to a velocity offset of 161\,km\,s$^{-1}$ relative to the SDSS quasar. A tentative broad C\,{\sc iii}] emission feature is also present in the NTT spectrum at the same redshift. However, other key lines expected at this redshift, including C\,{\sc iv}, H$\beta$, and [O\,{\sc iii}], are not detected despite being within the observed wavelength coverage.  
\par
We also considered alternative identifications of the 2.187\,$\mu$m feature. If interpreted as H$\beta$, the implied redshift would be $z\sim3.5$, which should place Ly$\alpha$ in the optical spectrum obtained with NTT. However, no Ly$\alpha$ emission is observed at the expected wavelength, disfavoring this scenario.  
\par
The line width of the putative H$\alpha$ emission ($\sim600$\,km\,s$^{-1}$) is broader than typically seen in pure star-forming galaxies, where nebular recombination lines generally exhibit ${\rm FWHM}\lesssim200$–300\,km\,s$^{-1}$ \citep[e.g.,][]{schreiber2018sins}. Also given the tentative broad C\,{\sc iii}] detection, this suggests that the companion could host a narrow-line AGN (type~2) or a composite SFG+AGN. We therefore classify this system as a quasar–SFG pair, with the companion possibly hosting a narrow-line AGN. 

\subsubsection{J121745.76+022109.5} \label{subsubsection:J121745.76+022109.5}
This system is at $z=0.28$, observed using NTT/EFOSC2 Gr\#5 with $3\times800$\,s exposures (Figure~\ref{fig:J121745.76+022109.5}). The two sources are separated by 3.0\arcsec, corresponding to a projected distance of 12.7\,kpc. The southern source is the known SDSS quasar. The continuum flux ratio between the quasar and its companion decreases from $\sim$20 at 5500\,\AA\ to $\sim$10 at 9000\,\AA, indicating that the companion becomes relatively redder toward longer wavelengths.
\par
The companion spectrum shows clear narrow H$\alpha$ and [N\,{\sc ii}] emission lines, with ${\rm FWHM}\simeq12$\,\AA\ ($\sim$529\,km\,s$^{-1}$) at $z=0.281$. The [O\,{\sc iii}] and H$\beta$ lines are also within the spectral coverage, but only tentatively detected with peak fluxes of $F_\lambda\simeq2\times10^{-18}\,{\rm erg\,s^{-1}\,cm^{-2}\,\text{\AA}^{-1}}$. Weak Mg\,{\sc i} $\lambda$5176.7 and Na\,{\sc i} $\lambda$5895.6 absorptions are also found at corresponding wavelengths, suggesting a mixed old and young stellar population of this galaxy.
\par
Both the [N\,{\sc ii}]/H$\alpha$ and [O\,{\sc iii}]/H$\beta$ ratios of the companion are close to unity. This places the source in the composite region of the standard BPT diagram \citep{baldwin1981classification}. The relatively narrow line widths ($\sim$500\,km\,s$^{-1}$) are broader than expected for pure H\,{\sc ii} regions but consistent with ionized gas in galaxies hosting weak or obscured AGN \citep[e.g.,][]{fernandes2010alternative}.  
\par
We therefore classify this system as a quasar–SFG/QG pair, with the companion likely being a composite system where both star formation and AGN contribute to the ionization. 

\subsubsection{J122349.29+021449.1} \label{subsubsection:J122349.29+021449.1}
This system is at $z=0.408$, observed using Gemini/GMOS-N B600+CG455 with $3\times600$\,s exposures, centered at 680\,nm (Figure~\ref{fig:J122349.29+021449.1}). The separation between the two sources is 2.4\arcsec, corresponding to a projected distance of 13.1\,kpc. The northern source is the known SDSS quasar. The quasar and the companion have similar colors in HSC imaging. Because the photometry of this system is unreliable, we did not apply photometric rescaling to the spectrum. From the Gemini data, we measure a continuum flux ratio of $\sim$3 between the quasar and the companion, although the true ratio is likely higher given that the slit was slightly offset from the companion.  
\par
The companion spectrum shows Ca\,{\sc ii} H\&K and G band absorption features at the same redshift as the quasar (Figure~\ref{fig:J122349.29+021449.1}). The continuum slope matches with the SDSS galaxy template at this redshift. No additional emission features are securely detected. We therefore classify this system as a quasar–QG pair.  
\par
Interestingly, the quasar itself exhibits strong spectral variability. The original SDSS spectrum taken at MJD=52283 shows prominent broad Balmer emission lines. In contrast, the later Gemini/GMOS-N spectrum (MJD=59295) reveals only narrow emission lines, with the broad components having disappeared. This dramatic transition identifies the quasar as a changing-look AGN, with the accretion state changes probably driven by disk instabilities and/or variations in accretion rate \citep[e.g.,][]{ricci2023changing}.  

\subsubsection{J125156.49+015249.7} \label{subsubsection:J125156.49+015249.7}
This system is at z=0.33, observed with Gemini/GMOS-N B600+CG455 centered at 630 nm with 600s$\times$3 exposures (Figure~\ref{fig:J125156.49+015249.7}). The separation between the two sources is 1.5\arcsec, which corresponds to 7.1 kpc. The southern source of the pair is the known SDSS quasar. The continuum flux ratio between the quasar and the companion source is $\sim$7. The companion spectrum shows solid detections of the $\OII$ and $\OIII$ lines at the same redshift as the quasar, while the position of the H$\beta$ line rather shows a weak absorption. The blue side of the spectrum shows Ca\,{\sc ii} H\&K doublet absorption and the other Fraunhofer-like absorption features. The tidal features in the HSC image indicates that this system is in a late merger stage. Considering these evidences, we consider two possible scenarios for the companion source: (1) it is under a transition stage from star-formation galaxy to post-starburst galaxy. It has little or no ongoing star formation but with some residual ionized gas ($\OII$ and $\OIII$) — possibly excited by evolved stars or shocks. (2) it is a quiescent galaxy, and the gas is ionized by an obscured AGN. Therefore, we classify this system a bona fide physical pair, with the ionization source of the companion to be confirmed. Therefore, we suggest the companion source a quiescent or post-starburst galaxy, and the $\OIII$ line indicates a potential obscured AGN residing in it.

\subsubsection{J130550.51-012331.5} \label{subsubsection:J130550.51-012331.5}
This system is at z=0.251, observed with Gemini/GMOS-N B600+CG455 centered at 630 nm with 600s$\times$3 exposures (Figure~\ref{fig:J130550.51-012331.5}). The separation between the two sources is 1.3\arcsec, which corresponds to 5.1 kpc. The northern source of the pair is the known SDSS quasar. The continuum ratio between the quasar and the companion source reduces from $\sim$27 at $\sim$4800 \text{\AA} to $\sim$6 at $\sim$7500 \text{\AA}. An intriguing feature of this system is the extremely extended $\OIII$ emission line. As shown in the 2D spectrum, the $\OIII$ line-emitting region spatially extends beyond the two nuclei, reaching the outskirts of the galaxies. This feature is more extended from the quasar to the companion side than to the opposite side. In addition to the $\OIII$ doublet, the H$\beta$ line is also extended on both sides, but less obviously. The physical mechanisms attributed to this kind of extended emission line region (EELR) are still uncertain. Recent IFU studies reported geometric connections between EELRs and radio emissions, indicating connections with the AGN feedback effect \citep{balmaverde2022murales}. When following the reduction steps (Section \ref{subsubsec:spec_reduce}), we extract the spectrum along the spatial axis at the position of the $\OIII$ $\lambda$5007 \text{\AA} line peak, which can be fit with two separate Gaussian profiles. Therefore, we suggest that the companion source also contributes to the EELR, and this system is a quasar-SFG pair.
\par
Furthermore, the SDSS quasar in the pair is also a potential changing-look AGN. In our observation (MJD=59319), the broad H$\beta$ line still exists, with FWHM=7431 \text{\AA}, EW=66 \text{\AA}, and SNR=6.5. However, the narrow H$\beta$ becomes much weaker than in the SDSS spectrum (observed at MJD=51692).

\subsubsection{J134722.75+015504.9} \label{subsubsection:J134722.75+015504.9}
This system lies at $z=0.399$ and was observed with NTT/EFOSC2 using Gr\#5 with three $\times$600s exposures (Figure~\ref{fig:J134722.75+015504.9}). The projected separation between the two sources is 1.9\arcsec\ (10.2 kpc). The southern source is the known SDSS quasar. The continuum flux ratio between the quasar and the companion decreases from $\sim$15 at $\sim$5500\,\AA\ to $\sim$8 at $\sim$9000\,\AA. The companion spectrum shows tentative absorption features, including Ca\,{\sc ii} H\&K and Mg\,{\sc i}, at a redshift consistent with the quasar, but no emission lines are detected. HSC imaging reveals tidal features connecting the quasar and the companion, supporting an interacting system. We therefore classify J1347+0155 as a quasar–QG pair.

\subsubsection{J222327.52+033902.0} \label{subsubsection:J222327.52+033902.0}
This system lies at $z=0.35$ and was observed with Gemini/GMOS-N using the B600+CG455 grating centered at 630 nm, with three $\times$600s exposures (Figure~\ref{fig:J222327.52+033902.0}). The projected separation is 2.7\arcsec\ (13.3 kpc). The southern source is the known SDSS quasar. The continuum flux ratio between the quasar and its companion is $\sim$3. The companion spectrum shows narrow $\OII$, H$\gamma$, H$\beta$, and $\OIII$ emission lines at the same redshift as the quasar. We measure FWHM values of 286 km\,s$^{-1}$ for [O\,{\sc iii}] and H$\beta$, and 467 km\,s$^{-1}$ for [O\,{\sc ii}]. The line ratio $\OIII$/H$\beta$ is 2.6.
\par
Because the spectrum does not cover H$\alpha$ and [N\,{\sc ii}], we cannot apply the classical BPT diagram \citep{baldwin1981classification}. Alternatively, we use the $R_{23}$ \citep[($\OII~\lambda$3727 + $\OIII~\lambda\lambda$4959, 5007) / H$\beta$, ][]{pagel1979composition, kewley2002using} and $O_{32}$ ratio \citep[$\OIII~\lambda\lambda$4959, 5007) / $\OII~\lambda$3727, ][]{kewley2002using} diagnostics, which are sensitive to the ionization parameter and gas-phase metallicity. Assuming the emission originates purely from star formation, we find $\log R_{23} = 0.92$ and $\log O_{32} = -0.13$. These values correspond to an intermediate metallicity (12+log(O/H) $\sim$ 8.4) typical of SDSS star-forming galaxies \citep{nakajima2014ionization, curti2016new}. Using the diagnostic diagrams of \citet{kewley2002using}, we estimate an ionization parameter of $\log(q_{\rm ion}/{\rm cm\,s^{-1}}) \sim 7.4$, intermediate between local ($z<0.3$) SDSS galaxies and $z\sim 1$ galaxies \citep{nakajima2014ionization}. The ionization state of the companion is thus consistent with star formation activity. We therefore classify J2223+0339 as a quasar-SFG pair.

\section{Discussion} \label{sec:discussion}
\begin{figure*}
    \includegraphics[width=0.75\textwidth]{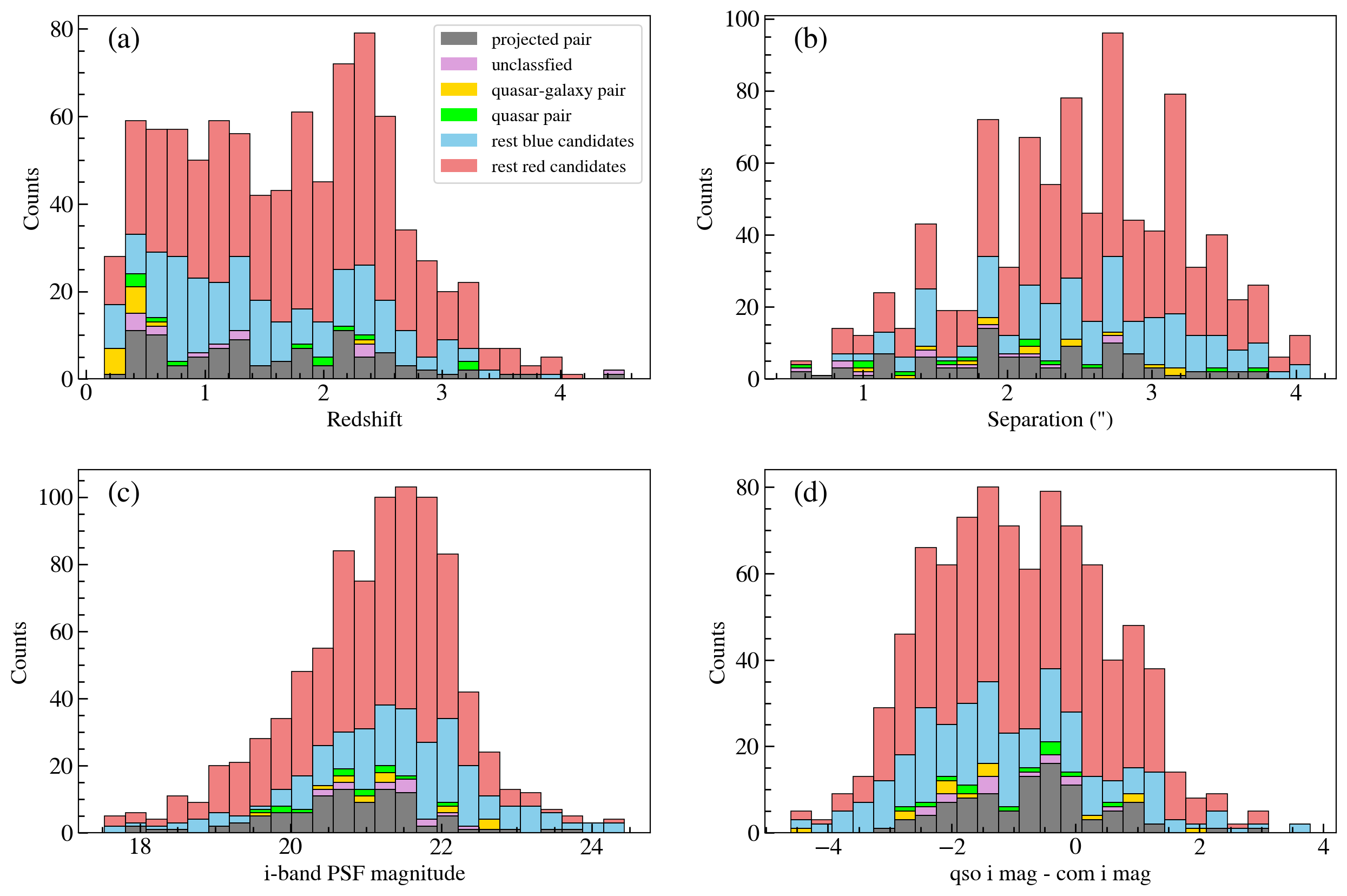}
    \caption{Classification outcomes of our dual quasar survey as a function of (a) redshift, (b) angular separation, (c) $i$-band PSF magnitude of the companion, and (d) $i$-band magnitude difference between the SDSS quasar and the companion. A total number of 122 systems have been observed among 883 dual candidates. Gray histograms denote projected pairs (96 systems where the two components are at different redshifts), pink histograms show unclassified cases (14 systems, mostly due to blending or low S/N), green and yellow histograms correspond to confirmed broad-line dual quasars (12) and quasar–galaxy pairs (14), respectively. The blue and red histograms represent unobserved candidates with companion color $g-r<1$ (212 systems) and $g-r>1$ (548 systems).}
    \label{fig:z_success}
\end{figure*}

Combining the 90 newly observed candidates in this work with 32 followed up by \citet{silverman2020dual} and \citet{tang2021optical}, we have now obtained spectroscopic data for 122 dual candidates. Among them, we confirm 12 broad-line dual quasars and 14 quasar–galaxy pairs. The overall classification status of our program is summarized in Figure~\ref{fig:z_success}. From left to right, the panels show the redshift distribution, angular separation between the two point sources in each candidate pair, the $i$-band PSF magnitude of the companion source, and the $i$-band magnitude difference between the SDSS quasar and its companion. The gray histograms represent 96 projected systems where the two components lie at different redshifts. The pink histograms denote 14 unclassified cases, mostly due to blending or insufficient S/N. The green and yellow bars correspond to the 12 confirmed quasar pairs and 14 quasar–galaxy pairs, respectively. The remainder of the sample, totaling 761 systems, remains unobserved. We further divide these by companion color into 212 “blue” candidates ($g-r<1$) and 548 “red” candidates ($g-r>1$), shown as the blue and red histograms in Figure~\ref{fig:z_success}.
\par
Based on these statistics, in this section we first attempt to estimate the dual fraction at different redshifts, and then discuss the caveats of our selection. 

\subsection{Dual fraction} \label{subsec:dual_frac}
The redshift distribution of our candidate list (Figure~\ref{fig:z_success} panel~(a)) closely follows that of the SDSS DR14 quasar catalog \citep{paris2018sloan}. Our confirmed dual quasars span $0.4<z<3.3$, while the majority (13/14) of quasar–galaxy pairs are found at $z<0.6$. This is consistent with the well-established cosmic downsizing trend in AGN evolution: luminous quasars peak at $z\sim2$–3 and decline toward lower redshift, while Seyfert-like, moderate-luminosity AGN dominate below $z\lesssim1$ \citep[e.g.,][]{fanidakis2012evolution}. These quasar–galaxy systems likely represent mergers hosting two SMBHs in different accretion states, where the companion galaxy shows a point-source morphology but lacks broad emission lines. Such cases may still harbor obscured or low-luminosity AGN, detectable through multi-wavelength diagnostics (e.g., radio jets, hard X-rays, or mid-IR dust emission; \citealt{hickox2018obscured}). A key caveat is that our selection requires one known SDSS quasar in each pair, inherently biasing the sample toward luminous broad-line AGN and making us incomplete for the low-$z$ population where most AGN are not SDSS quasars \citep{heckman2014coevolution}. 
\par
Bearing that caveat in mind, we estimate the dual fraction specifically for SDSS quasars in three redshift ranges. At $z<0.6$, the observed ratio between dual quasars and quasar–galaxy pairs (3:13) indicates that only $\sim$20\% of (major) merging systems hosting an SDSS quasar also harbor a companion with detectable broad emission lines. According to our cross-match (Section~\ref{subsection:selection}), there are 59{,}482 SDSS DR14 quasars imaged in the HSC-SSP s20a\_wide release, of which 2{,}672 lie at $z<0.6$. Among these, 113 are selected as dual candidates, comprising 47 “blue” companions ($g-r<1$) and 65 “red” companions ($g-r>1$). Our three confirmed dual quasars at $z<0.6$ (indeed, all confirmed duals in this work) fall into the blue category. They were discovered from 20 observed blue candidates, corresponding to a blue-candidate success rate of $\sim$15\%. If we conservatively assume that all 51 red candidates are not broad-line quasars and that the unobserved 27 blue candidates share the same success rate, the expected total number of dual quasars at $z<0.6$ is $\sim$7, yielding a dual fraction of 0.3\% (7/2{,}672) for SDSS quasars with separations of $0.6\arcsec$–$4\arcsec$. A more inclusive estimate comes from considering the quasar–galaxy pairs, which may host obscured or moderate-luminosity AGN. Among the observed sample, 6 such systems are identified from 20 blue candidates (30\%) and 7 from 14 red candidates (50\%). Extrapolating these rates to the unobserved candidates implies a total of $\sim$53.7 dual systems (quasar pairs $+$ quasar–galaxy pairs), corresponding to an upper limit of $\sim$2.0\% (53.7/2{,}672). We therefore report a dual fraction of 0.3\%–2.0\% for SDSS quasars at $z<0.6$ with separations of $0.6\arcsec$–$4\arcsec$. Extending the redshift range to $z<0.8$ adds two confirmed dual quasars and increases the parent SDSS$\times$HSC sample to 6{,}351 objects; following the same procedure yields a dual fraction of 0.2\%–1.2\% at $z<0.8$. These low-redshift results are consistent with previous work \citep{silverman2020dual} and with simulations predicting that only a small fraction of merging SMBHs are simultaneously luminous broad-line AGN at low $z$ \citep{steinborn2016origin,rosas2019abundances,volonteri2022dual}. 
\par
In the intermediate bin, $0.8<z<1.5$, we have not yet confirmed any dual quasars or quasar–galaxy pairs among 14{,}594 SDSS$\times$HSC quasars. Adopting the $z<0.8$ blue-candidate success rate ($5/28$) as a conservative proxy and rejecting red candidates, the implied dual fraction is 0.08\%. If we additionally assume that the fraction of quasar–galaxy pairs among blue and red candidates (6/28 and 7/84, respectively) persists in this bin and that such companions host moderate-luminosity AGN, the corresponding upper-limit fraction is 0.24\%. We therefore report a dual fraction of 0.08\%–0.24\% for SDSS quasars at $0.8<z<1.5$. 
\par
At cosmic noon ($1.5<z<3.3$), we confirm 7 dual quasars and 1 quasar–galaxy pair among 52 observed candidates. Of the 52 candidates, 46 are blue, and all confirmed systems are blue, giving a blue-candidate success rate of $\sim$15\% (7/46). Rejecting red candidates and extending this rate to the unobserved blue candidates yields an expected total of 21.3 dual quasars (or 24.3 systems including the quasar–galaxy pair), corresponding to a dual fraction of 0.06\% (21.3/36{,}527–24.3/36{,}527) for SDSS quasars at $1.5<z<3.3$. Recently, \citet{shen2023statistics} reported an integrated dual fraction of $0.062\pm0.005\%$ for SDSS DR16 quasars selected with \emph{Gaia}, at $0.3\arcsec$–$3\arcsec$ separation and $1.5<z<3.5$. Although their methodology (astrometric selection with proper-motion filtering) differs from ours, our estimates are consistent with their results at cosmic noon.
\par
Theoretical models and simulations generally predict that the dual AGN fraction increases with redshift, reaching values of a few per cent at cosmic noon, driven by the higher galaxy merger rate, elevated gas fractions, and longer quasar duty cycles at earlier cosmic times \citep[e.g.,][]{capelo2017survey, steinborn2016origin, rosas2019abundances, volonteri2022dual, chen2023properties}. In contrast, our observational results indicate an apparent decline in the dual fraction, from 0.3–2\% at $z<0.6$ to only 0.06\% at $1.5<z<3.3$.
\par
\citet{volonteri2022dual} explored the impact of observational selection effects on the redshift evolution of the simulated dual fraction. Applying selection criteria of $\log L_{\rm bol}>44$, $d<30$ kpc, $\log (M_{\rm BH,1}/M_{\odot})>7$, and $\log (M_{\rm BH,2}/M_{\odot})>7$—which best match the properties of our confirmed dual quasars (Figure~\ref{fig:BH_bol_edd})—their simulations predict a dual fraction of 1.4\% at $z<0.6$, rising to 2.5\% by $z\approx2$ and remaining approximately constant at higher redshift (see their Figure 16).
\par
The discrepancy between our observations and \citet{volonteri2022dual} emerges at $z\gtrsim0.6$, and likely has multiple contributing causes.
First, our observed dual systems are required to include at least one SDSS-selected quasar, biasing the sample toward luminous, unobscured broad-line AGN. As demonstrated by our low-redshift results, a significant population of offset or dual systems may host companions without detectable broad emission lines. Consequently, a potentially large population of obscured–unobscured and obscured–obscured dual quasars is likely missed at high redshift.
Second, a substantial fraction of the $1.5<z<3.3$ interval falls within the so-called “redshift desert”, where the lack of strong emission lines in the wavelength coverage of ground-based optical and near-infrared spectroscopy substantially hampers spectroscopic identification.
Third, our selection does not distinguish dual AGN systems from triple or more AGN systems, which were treated separately in \citet{volonteri2022dual}. We expect this effect to be minor, however, since the number density of multiple-AGN systems in the simulation is approximately one order of magnitude lower than that of dual AGNs.
Fourth, the rarity of luminous quasars in current cosmological simulations, owing to their limited volumes, leads to small-number statistics in the simulated dual-quasar population, further contributing to the apparent tension.



\subsection{Selection effects and observational limitations}
In addition to the redshift dependence and the bias to the unobscured bright sources discussed above, the candidate selection of our program is also shaped by observational constraints tied to angular separation, $i$-band PSF magnitude of the companion, and the $i$-band magnitude contrast between the pair, as shown by the remaining three panels of Figure~\ref{fig:z_success}.
\par
By construction, our candidate selection requires two distinct point sources in HSC imaging, with separations between $0.6\arcsec$ and $4\arcsec$ (Section~\ref{subsection:selection}). This lower limit is set by the seeing, below which reliable deblending becomes challenging in ground-based data. As shown in Figure~\ref{fig:z_success}, the number of selected candidates rises toward larger separations. This is partly a geometric effect: the probability of finding a projected neighbor scales with the surface density of background sources and the area of the search annulus, $\pi r^2$, so wider search radii naturally yield more chance associations. At the smallest separations ($<2\arcsec$), the available area is small, and the number of random alignments is correspondingly reduced. In addition, PSF blending at $<1\arcsec$ further suppresses completeness by making it difficult to reliably deblend companions from the bright SDSS quasar. These combined effects explain the relative paucity of selected candidates at close separations, despite theoretical expectations that the intrinsic dual fraction should increase at small physical scales during the final stages of SMBH inspiral \citep[e.g.,][]{capelo2017survey}. Yet our spectroscopic follow-ups are still fairly effective in confirming the pairs both at separations above 2\arcsec (14/70=20\%), and below (12/52=23\%). In \cite{wang2025hst}, we have tuned the selection method to push the separations of the candidates to $<0.6\arcsec$ (sub-5 kpc) with HSC, which are then followed up with HST grism spectroscopy.
\par
Our selection also required a reliable HSC $i$-band PSF measurement for the companion source. As shown in Figure~\ref{fig:z_success} panel (c), the confirmed dual quasars predominantly occupy the bright regime ($i<22.5$), while fainter companions are more often classified as quasar–galaxy pairs or remain unclassified due to insufficient S/N in the spectroscopy. The total number of selected dual candidates also decreases at the faint end. This also reflects the limitations of our source selection method. Our \texttt{GaLight}-based selection relies on the \texttt{photutils.detect\_sources} function to identify multiple point sources in HSC images \citep{ding2022galight}, which becomes increasingly inefficient when the two components differ strongly in brightness \citep{larry_bradley_2024_13989456}. In practice, this sets a contrast limit of $\Delta i \lesssim 3$ (as most of the candidates are distributed in panel (d) of Figure~\ref{fig:z_success}), beyond which faint companions are not reliably deblended from the bright SDSS quasar. As a result, this restricts us to nearly equal-luminosity pairs, thus major mergers.
\par
Taken together, these considerations highlight that our selection method and dual fraction estimates apply only to a restricted parameter space: separations of $0.6\arcsec$–$4\arcsec$, companion magnitudes of $i\lesssim22.5$, and modest luminosity contrasts. Expanding beyond this regime will require higher-resolution imaging, e.g., with JWST \citep{li2024jwst,li2025active} and Euclid \citep{ulivi2025euclid}, and deeper spectroscopy to probe fainter and more unequal pairs \citep[e.g.,][]{matsuoka2024discovery, perna2025ga, perna2026ga}. Multi-wavelength approaches, such as with X-ray, will also be essential for confirming and recovering missed obscured dual AGNs \citep[e.g.,][]{de2023x}.

\section{Conclusions} \label{sec:conclusion}
We have carried out a spectroscopic follow-up program targeting 90 dual quasar candidates selected from Subaru HSC-SSP imaging of SDSS DR14 quasars. Our selection is based on two-dimensional image decomposition and identifies PSF-like companions within $0.6\arcsec$–$4\arcsec$ ($\lesssim 30$ kpc) of an SDSS quasar. From the spectra obtained with NTT/EFOSC2, Gemini/GMOS-N, Keck/NIRES, and Subaru/FOCAS, we draw the following conclusions:

\begin{itemize}
    \item We newly confirm six broad-line dual quasars and eleven offset quasars (quasar–galaxy pairs). Taken together with our previous work, our program has now identified 12 dual quasars and 14 offset quasars. The dual quasars span $1.5 < z < 3.3$, while most quasar–galaxy pairs are found at $z < 0.6$.
    
    \item For the confirmed duals at $z>1.5$, emission-line measurements yield black hole masses of $M_{\rm BH} = 10^{8.5}$–$10^{10}\,M_\odot$ and bolometric luminosities of $L_{\rm bol} = 10^{45.5}$–$10^{47.5}\,\rm erg\,s^{-1}$. Their Eddington ratios are moderate ($\lambda_{\rm Edd} = 0.01$–0.4), indicating that while mergers can ignite both nuclei, simultaneous near-Eddington fueling is not required.
    
    \item At $z < 0.6$, our spectroscopic identifications are dominated by offset quasars, where the companion exhibits a point-like morphology but no broad emission lines. These systems likely host moderate-luminosity or obscured AGN, consistent with the downsizing trend in which Seyfert-like activity dominates at late cosmic times. The elevated fraction of offset quasars at low redshift highlights the low duty cycle of quasar activity in mergers.
    
    \item Using the full SDSS DR14Q$\times$HSC s20a\_wide quasar sample, we estimate the dual fraction of SDSS quasars within 0$\farcs$6–4\arcsec separation to be 0.2\%-1.2\% at $z<0.8$, 0.08\%-0.24\% at $0.8<z<1.5$, and 0.06\% at $1.5<z<3.3$. The lower values only consider quasar-quasar pairs, and the upper values include quasar-galaxy pairs. These observed fractions are broadly consistent with recent optical searches, but remain below theoretical expectations, which generally predict an increasing dual fraction toward high redshift.
    
    \item The apparent decline in observed dual fractions with redshift is largely attributable to selection and observation limits. Our parent sample is anchored on luminous SDSS quasars, biasing against pairs of moderate-luminosity AGN. Companions fainter than $i$-band magnitude 22 or with contrasts of $i$-band magnitude $\gtrsim 3$ are rarely detected by SExtractor deblending, and candidates with separations below $2\arcsec$ are suppressed by both geometric probability and PSF blending. As a result, our survey is most sensitive to nearly equal-luminosity pairs with intermediate separations, and provides a lower bound on the true dual AGN fraction.
\end{itemize}

In summary, our spectroscopic campaign confirms that dual quasars exist across a wide redshift range but are intrinsically rare in optical surveys anchored on SDSS quasars. Their incidence is strongly shaped by selection biases, and most low-redshift mergers appear as quasar–galaxy pairs rather than dual broad-line systems. Our results have established the incidence and properties of luminous dual quasars in the SDSS+HSC era, but a complete census of dual AGN requires deeper and higher-resolution searches. Imaging and spectroscopy with HST and JWST will extend detections to smaller separations and higher contrasts, while X-ray and mid-IR observations are essential to reveal obscured nuclei. Next-generation wide-field surveys (e.g., Rubin/LSST and Euclid) combined with high-resolution follow-up will be critical to reconcile observations with theoretical predictions of SMBH pair growth across cosmic time.

\section*{Acknowledgements}
ST and MB acknowledge funding from the Royal Society via a University Research Fellowship to MB and associated Research Fellows Enhancement Awards. Special thanks to the support astronomers at the NTT, Gemini, Keck, and Subaru telescopes and local ESO staffs who hosted the author's on-site visit to the La Silla observatory. This work cannot be done without their help. The authors sincerely thank the anonymous reviewer for helpful suggestions on this work.
\par
The Hyper Suprime-Cam (HSC) collaboration includes the astronomical communities of Japan and Taiwan, and Princeton University. The HSC instrumentation and software were developed by the National Astronomical Observatory of Japan (NAOJ), the Kavli Institute for the Physics and Mathematics of the Universe (Kavli IPMU), the University of Tokyo, the High Energy Accelerator Research Organization (KEK), the Academia Sinica Institute for Astronomy and Astrophysics in Taiwan (ASIAA), and Princeton University. Funding was contributed by the FIRST program from the Japanese Cabinet Office, the Ministry of Education, Culture, Sports, Science and Technology (MEXT), the Japan Society for the Promotion of Science (JSPS), Japan Science and Technology Agency (JST), the Toray Science Foundation, NAOJ, Kavli IPMU, KEK, ASIAA, and Princeton University. 
\par
This paper makes use of software developed for Vera C. Rubin Observatory. We thank the Rubin Observatory for making their code available as free software at http://pipelines.lsst.io/.
\par
This paper is based on data collected at the Subaru Telescope and retrieved from the HSC data archive system, which is operated by the Subaru Telescope and Astronomy Data Center (ADC) at NAOJ. Data analysis was in part carried out with the cooperation of Center for Computational Astrophysics (CfCA), NAOJ. We are honored and grateful for the opportunity of observing the Universe from Maunakea, which has the cultural, historical and natural significance in Hawaii. 
\par

\section*{Data Availability}
The ``Discovery Panels" (Section~\ref{sec:individuals}) and reduced 1D spectrum before photometric rescaling of all the observed candidates are available as MNRAS online supplementary material.
 



\bibliographystyle{mnras}
\bibliography{example} 




\appendix

\section{photometric comparison between PSF and CModel magnitudes} \label{sec:psf_vs_cmodel}
\begin{figure*}
    \includegraphics[width=0.95\textwidth]{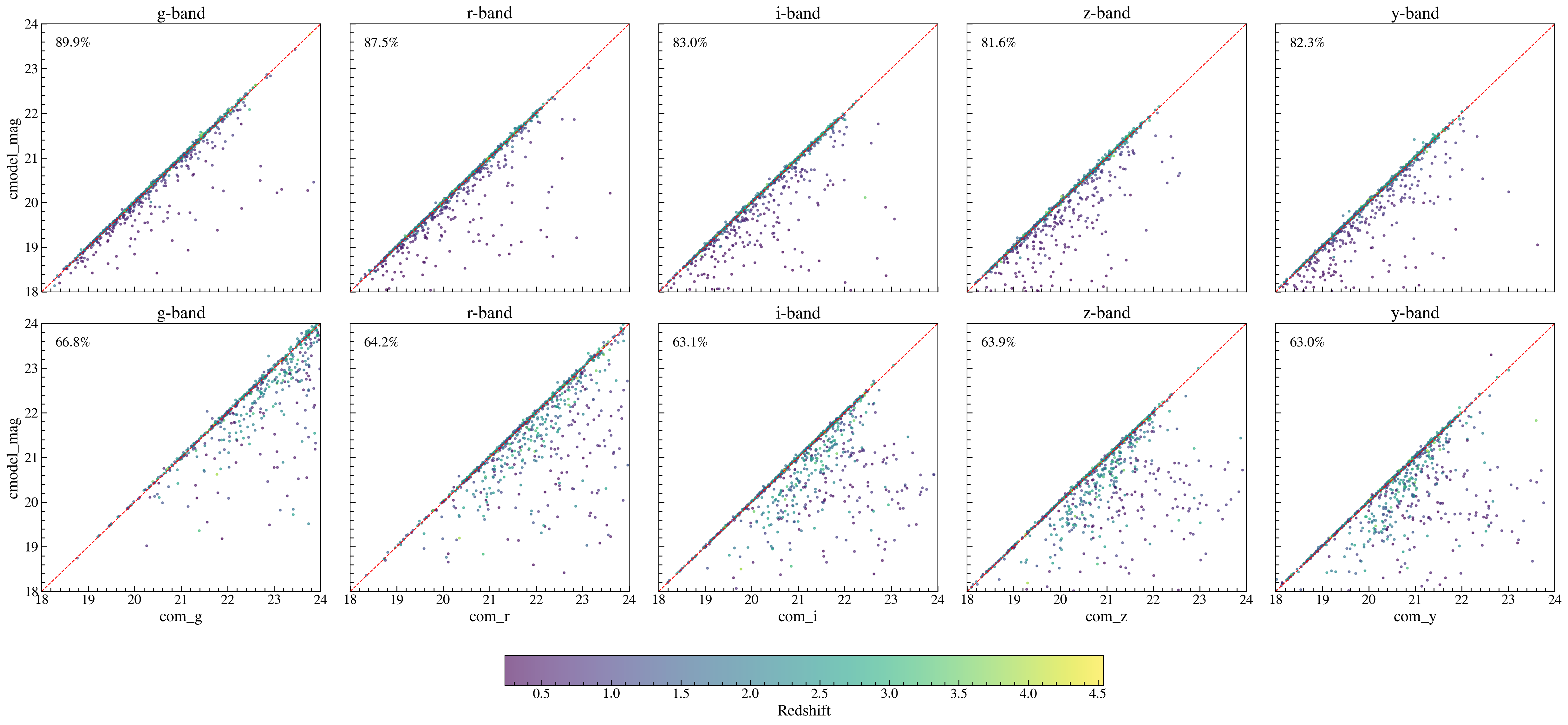}
    \caption{Comparison between \texttt{GaLight}-measured PSF magnitudes (x-axis) and HSC Pipeline CModel magnitudes (y-axis). The top row shows SDSS quasars in each candidate pair; the bottom row shows the companion sources. Each panel corresponds to one HSC band ($g,r,i,z,y$), with the red dashed line indicating the 1:1 relation. 
    Points are color-coded by the redshift of the SDSS quasar, and the percentage in the top-left of each panel gives the fraction of sources with consistent measurements ($\Delta m < 0.2$ mag).}
    \label{fig:psf_cmodel_compare}
\end{figure*}
\label{appendix:photometry}

As described in Section~\ref{sec:methods}, we used \texttt{GaLight} to measure the PSF magnitudes of candidate sources through two-dimensional image decomposition. In Figure~\ref{fig:psf_cmodel_compare}, we compare these measurements with the CModel magnitudes provided by the HSC Pipeline \citep{bosch2018hyper}. The top row shows the five-band comparison for the SDSS quasars in each candidate pair, while the bottom row shows the same for the companion sources. Red dashed lines indicate the one-to-one relation, and points are color-coded by the redshift of the SDSS quasar. On top left of each panel, the percentage indicates the fraction of sources with consistent magnitudes (agreement within $\pm 0.2$ mag) between the two methods.  
\par
For the quasars, we find good overall agreement across all bands, with consistency fractions above 80\%. The small fraction of outliers are primarily at $z<1$, where host galaxy light contributes significantly to the CModel fluxes, while our PSF magnitudes better isolate the unresolved nuclear component. Since the PSF magnitudes are used to rescale the spectra to correct for slit loss (see Section~\ref{subsec:phy_props}), this comparison validates our approach. Importantly, all of our confirmed dual quasars lie at $z>1.5$, where host-galaxy contamination is minimal, so the choice between PSF and CModel magnitudes does not affect our measurements of emission lines or black hole properties.  
\par
For the companion sources (bottom row), the agreement is substantially worse, with consistency fractions of only $\sim$63–67\% across the five bands. A notable 58 out of 883 companions are not recorded at all in the HSC Pipeline catalog, typically due to blending in pairs with $<$1\arcsec separations. This highlights the necessity of applying our own 2D image analysis techniques to select out dual candidates rather than relying exclusively on the catalog (Section~\ref{subsection:selection}). Among the measured companions, many of the most discrepant cases are associated with high-redshift quasars ($z>1.5$), suggesting that a significant fraction of these companions are foreground or background interlopers (i.e., projected pairs). We also note that the companions’ $g$- and $r$-band magnitudes are systematically fainter relative to their matched quasars, suggesting a substantial population of contaminating stars \citep{tang2021optical}. In forthcoming work, we plan to combine PSF–CModel consistency checks with more sophisticated color-based selections using the full five-band photometry, as an additional diagnostic to filter contaminants in future work.

\section{``Discovery panels" of dual and offset quasars} \label{sec:discovery_panels}
\begin{figure*}
    \includegraphics[width=0.95\textwidth]{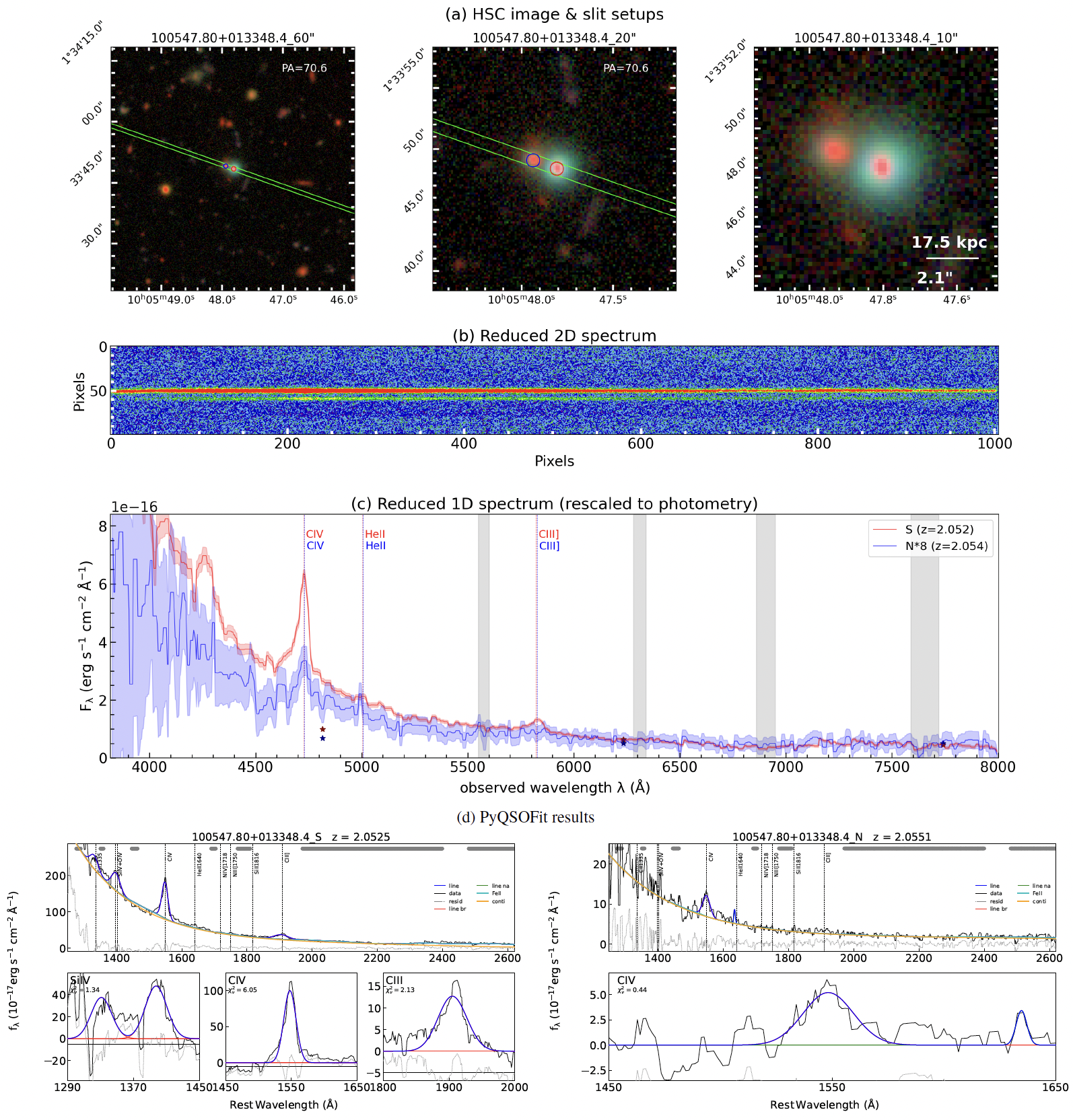}\par
    \caption{Discovery panel reporting the NTT/EFOSC2 spectrograph of SDSS J100547.80+013348.4. The bottom panels show the PyQSOFit results for both sources. This system is confirmed as a dual quasar (Section~\ref{subsubsection:J100547.80+013348.4}).}
    \label{fig:J100547.80+013348.4}
\end{figure*}

\begin{figure*}
    \includegraphics[width=0.95\textwidth]{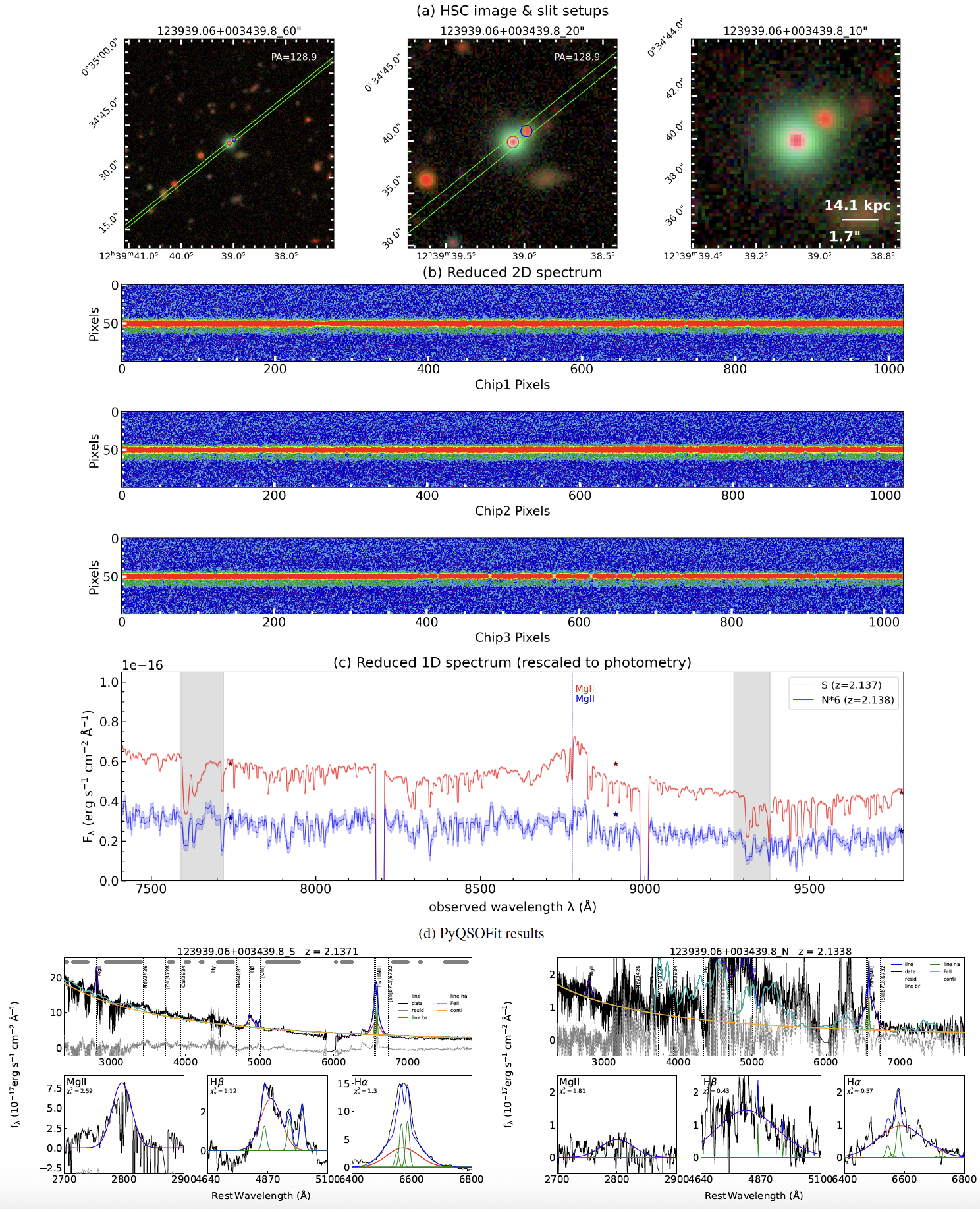}\par
    \caption{Discovery panel reporting the Gemini/GMOS spectrograph of SDSS J123939.06+003439.8. The bottom panels show the PyQSOFit results for both sources, performed on the combined spectrum including the Keck/NIRES part (Figure~\ref{fig:J123939.06+003439.8}). This system is confirmed as a dual quasar (Section~\ref{subsubsection:J123939.06+003439.8}).}
    \label{fig:J123939.06+003439.8}
\end{figure*}

\begin{figure*}
    \includegraphics[width=0.95\textwidth]{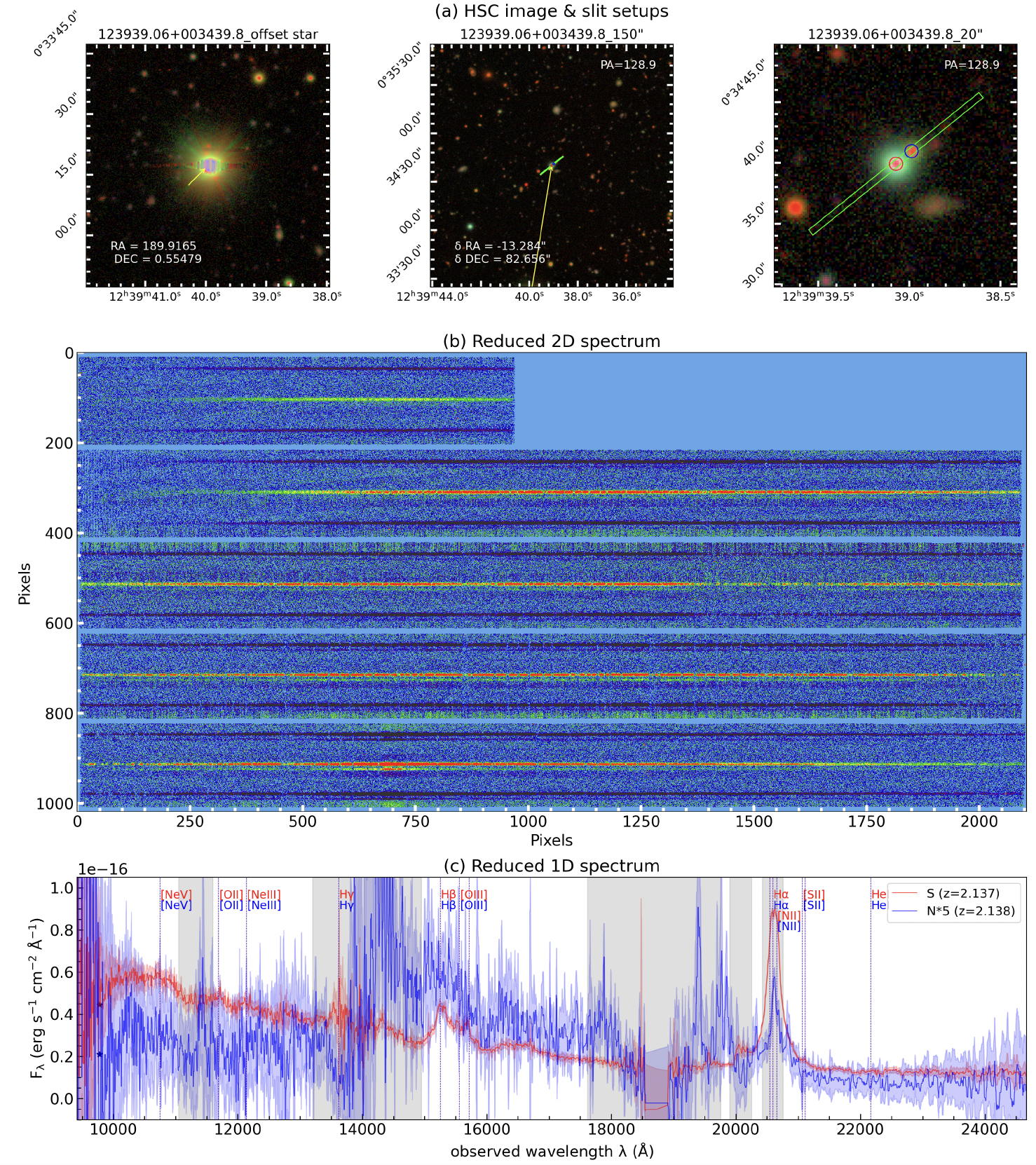}
    \caption{Discovery panel reporting the Keck/NIRES spectrograph of SDSS J123939.06+003439.8. This is the same object as in Figure~\ref{fig:J123939.06+003439.8}. For Keck/NIRES spectrograph, we need to find an offset star (left panel of a), then shift the slit to our objects (middle panel of a). The coordinate of the offset star and the offset distance are noted in these two panels.}
    \label{fig:J123939.06+003439.8_K}
\end{figure*}

\begin{figure*}
    \includegraphics[width=0.9\textwidth]{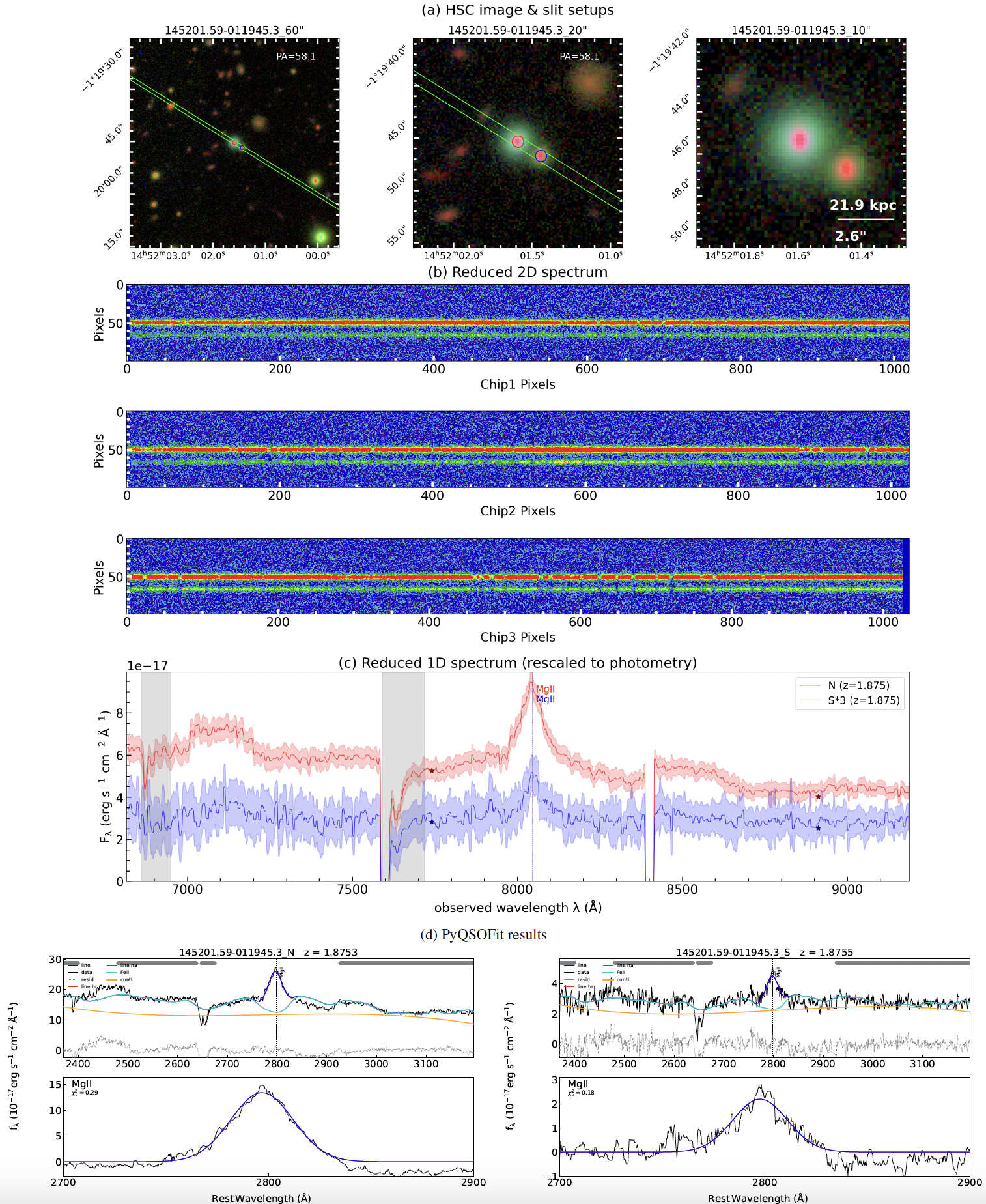}\par
    \caption{Discovery panel reporting the Gemini/GMOS spectrograph of SDSS J145201.59-011945.3. The bottom panels show the PyQSOFit results for both sources. This system is confirmed as a dual quasar (Section~\ref{subsubsection:J145201.59-011945.3}).}
    \label{fig:J145201.59-011945.3}
\end{figure*}

\begin{figure*}
    \includegraphics[width=0.95\textwidth]{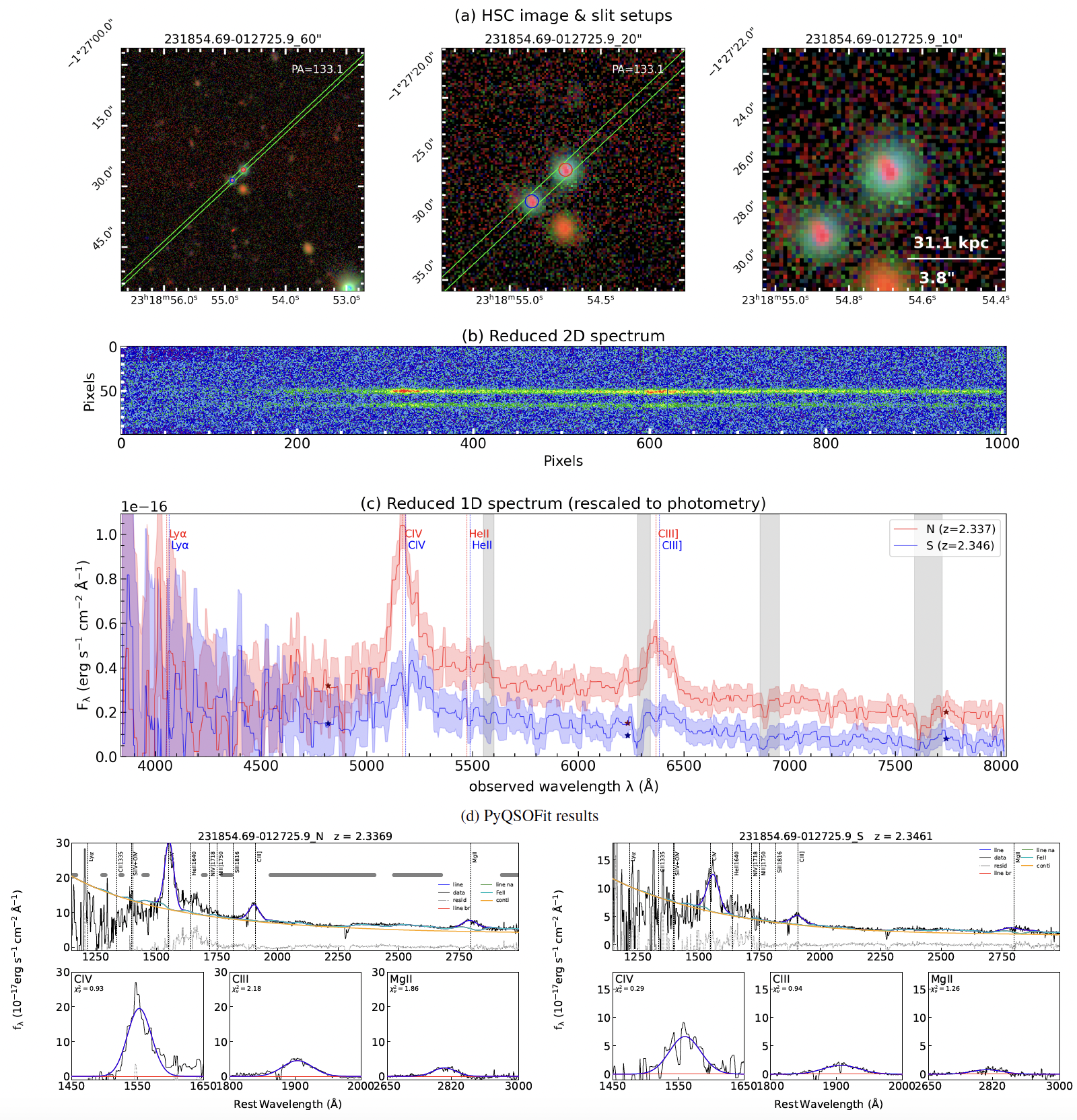}\par
    \caption{Discovery panel reporting the NTT/EFOSC2 spectrograph of SDSS J231854.69-012725.9. The bottom panels show the PyQSOFit results for both sources, performed on the combined spectrum including the Subaru/FOCAS part (Figure~\ref{fig:J231854.69-012725.9_S}). This system is confirmed as a dual quasar (Section~\ref{subsubsection:J231854.69-012725.9}).}
    \label{fig:J231854.69-012725.9}
\end{figure*}

\begin{figure*}
    \includegraphics[width=0.95\textwidth]{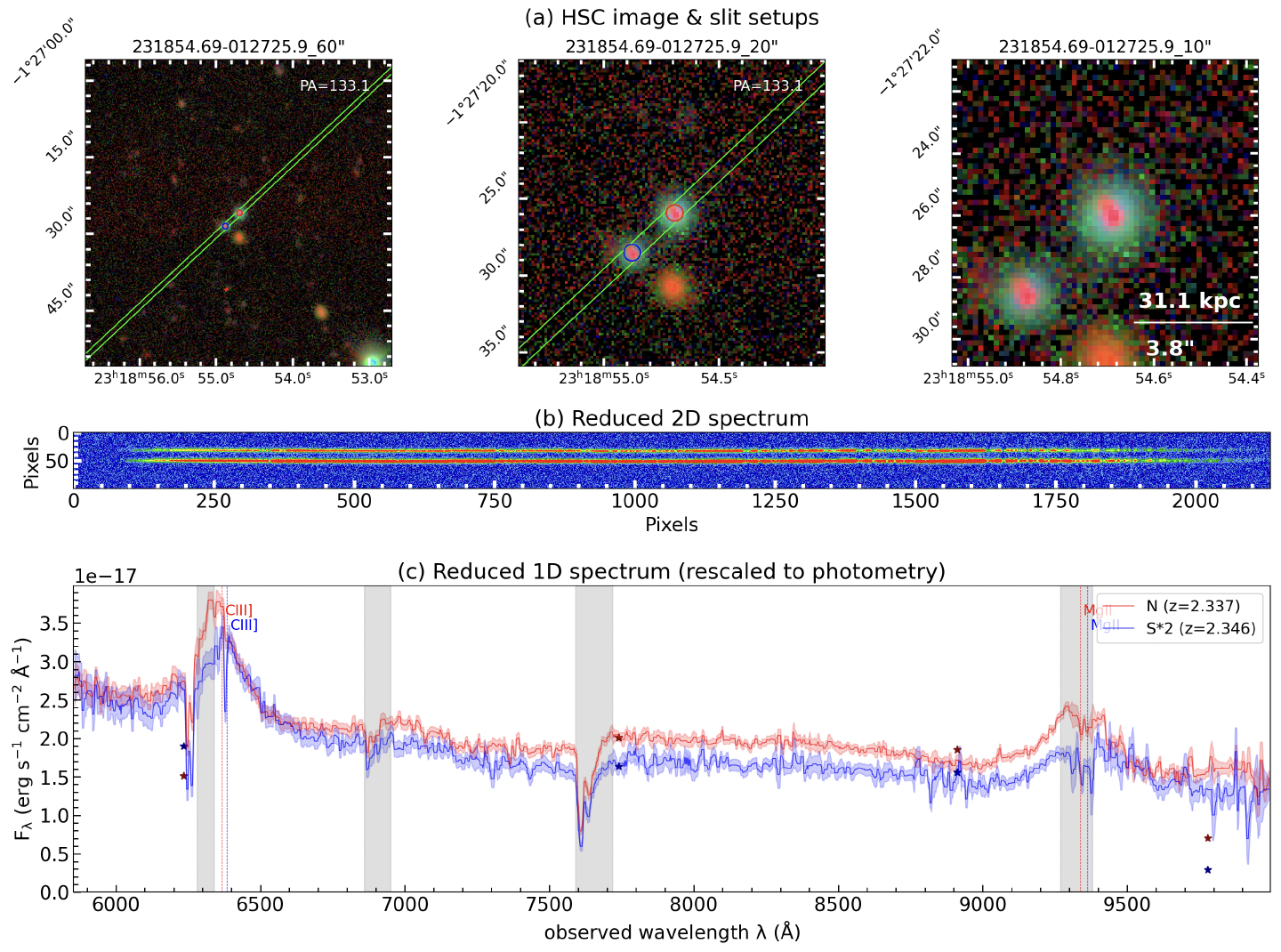}
    \caption{Discovery panel reporting the Subaru/FOCAS spectrograph of SDSS J231854.69-012725.9. This is the same source as in Figure~\ref{fig:J231854.69-012725.9}}
    \label{fig:J231854.69-012725.9_S}
\end{figure*}

\begin{figure*}
    \includegraphics[width=0.9\textwidth]{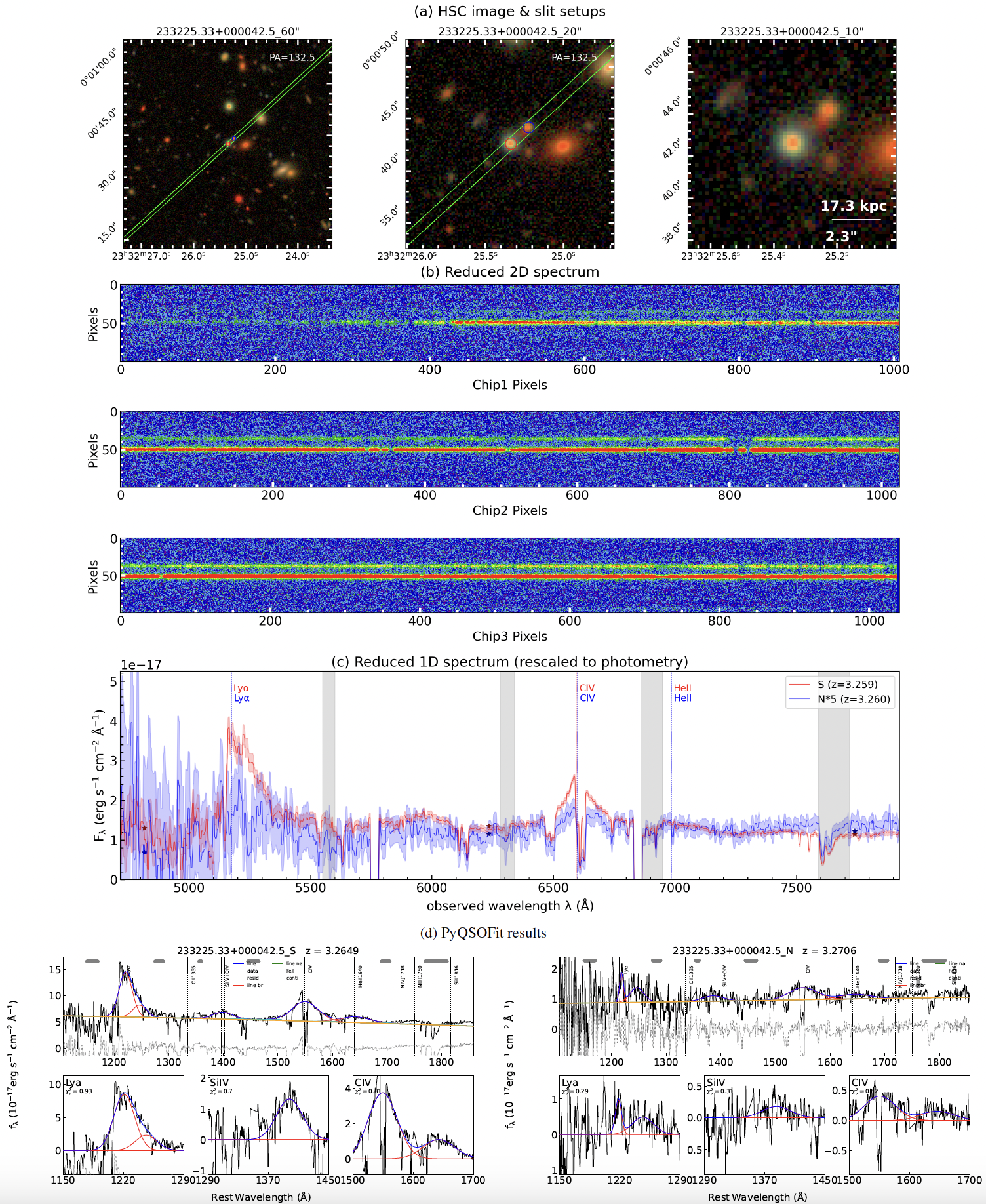}\par
    \caption{Discovery panel reporting the NTT/EFOSC2 spectrograph of SDSS J233225.33+000042.5. The bottom panels show the PyQSOFit results for both sources. This system is confirmed as a dual quasar (Section~\ref{subsubsection:J233225.33+000042.5}).}
    \label{fig:J233225.33+000042.5}
\end{figure*}

\begin{figure*}
\centering
    \includegraphics[width=0.95\textwidth]{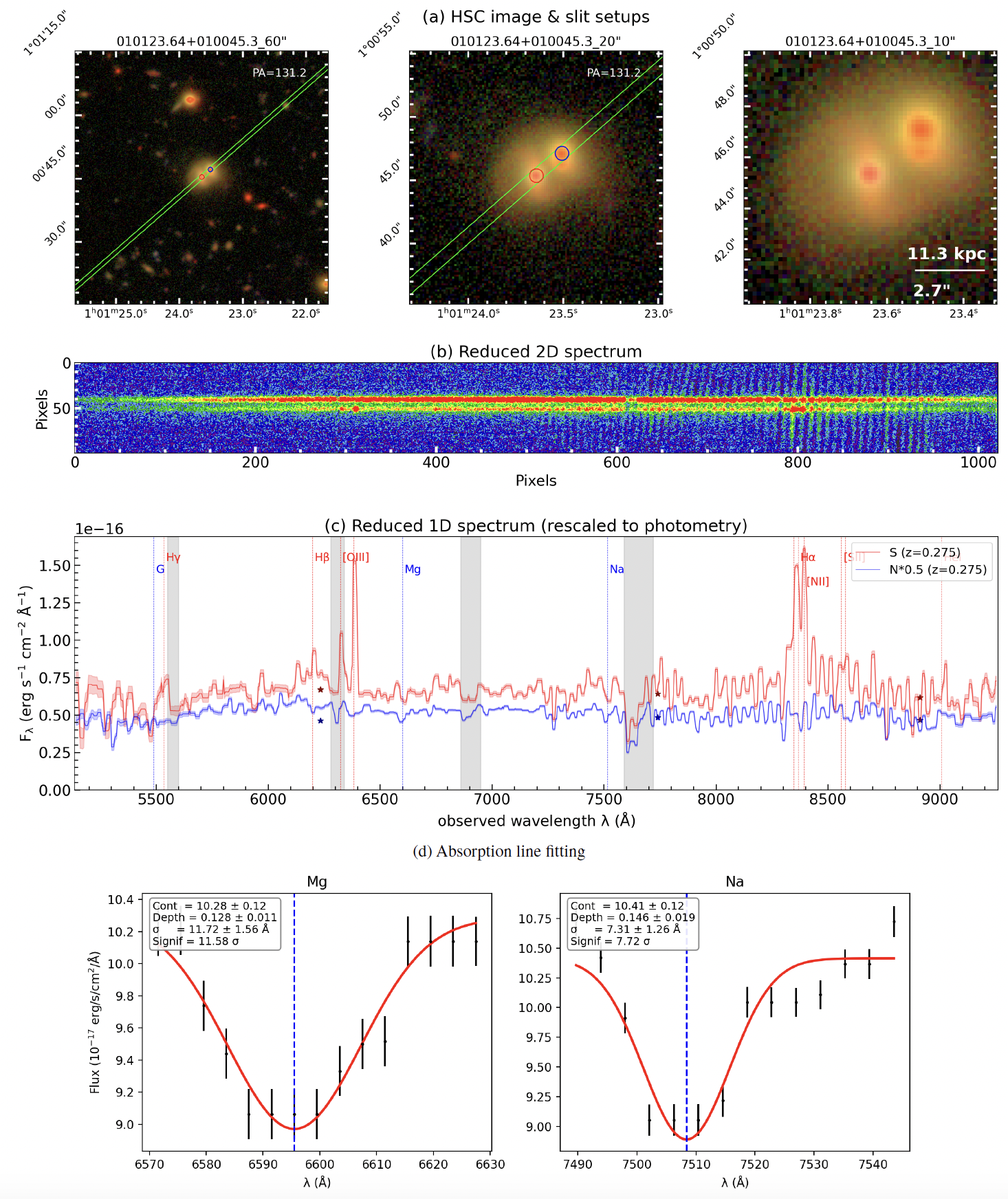}\par
    \caption{Discovery panel reporting the NTT/EFOSC2 spectrograph of SDSS J010123.64+010045.3. This system is confirmed as a quasar-QG pair (Section~\ref{subsubsection:J010123.64+010045.3}). The bottom two panels show the fitting of Mg and Na absorption lines of the companion's spectrum. For the absorption fitting plots in general, the window is centered at the expected position of the absorption lines at the redshift given in the 1D spectrum panel. the data are plotted as the black points with 1$\sigma$ flux error. Note that the flux level might be different from what is plotted in panel (c) in some cases, this is because we rescaled the spectrum in panel (c) for visibility. The rescaling factor is noted in the labels of panel (c). The fitting is performed on the original spectrum before rescaling. We apply double Gaussian model to the Ca H\&K and single Gaussian to the other lines with \texttt{scipy.opt.curve\_fit}. The best-fit models are plotted as the red curve. The line centers are marked with the vertical blue dashed line. The best-fit parameters are noted on top left of each panel, which include: (1) The continuum level; (2) The depth of the absorption, defined as the distance of the peak from the continuum floor; (3) The width of the line, which is the $\sigma$ of the Gaussian profile; (4) Significance of the line, defined as the depth divided by the uncertainty of the depth. The same format applies for the rest of the samples with absorption line fitting panels.}
    \label{fig:J010123.64+010045.3}
\end{figure*}

\begin{figure*}
\centering
    \includegraphics[width=0.95\textwidth]{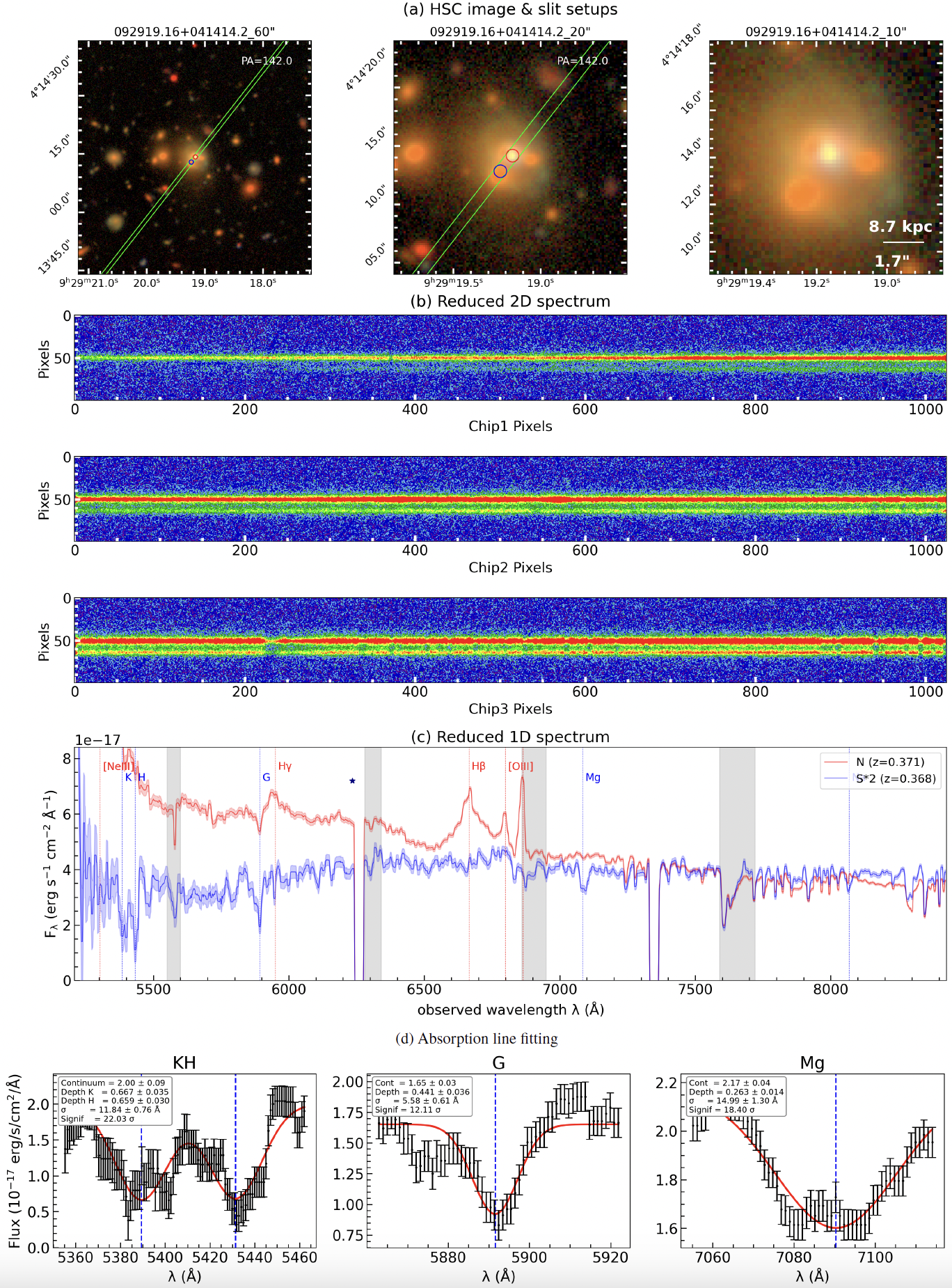}\par
    \caption{Discovery panel reporting the Gemini/GMOS spectrograph of SDSS J092919.16+041414.2, covering the quasar and the lower-left companion. This system is confirmed as a quasar-QG-QG triple system (Section~\ref{subsubsection:J092919.16+041414.2}). The bottom panels show the fitting of the Ca H\&K, G-band, and Mg absorption lines of the companion's spectrum. }
    \label{fig:J092919.16+041414.2A}
\end{figure*}

\begin{figure*}
\centering
    \includegraphics[width=0.95\textwidth]{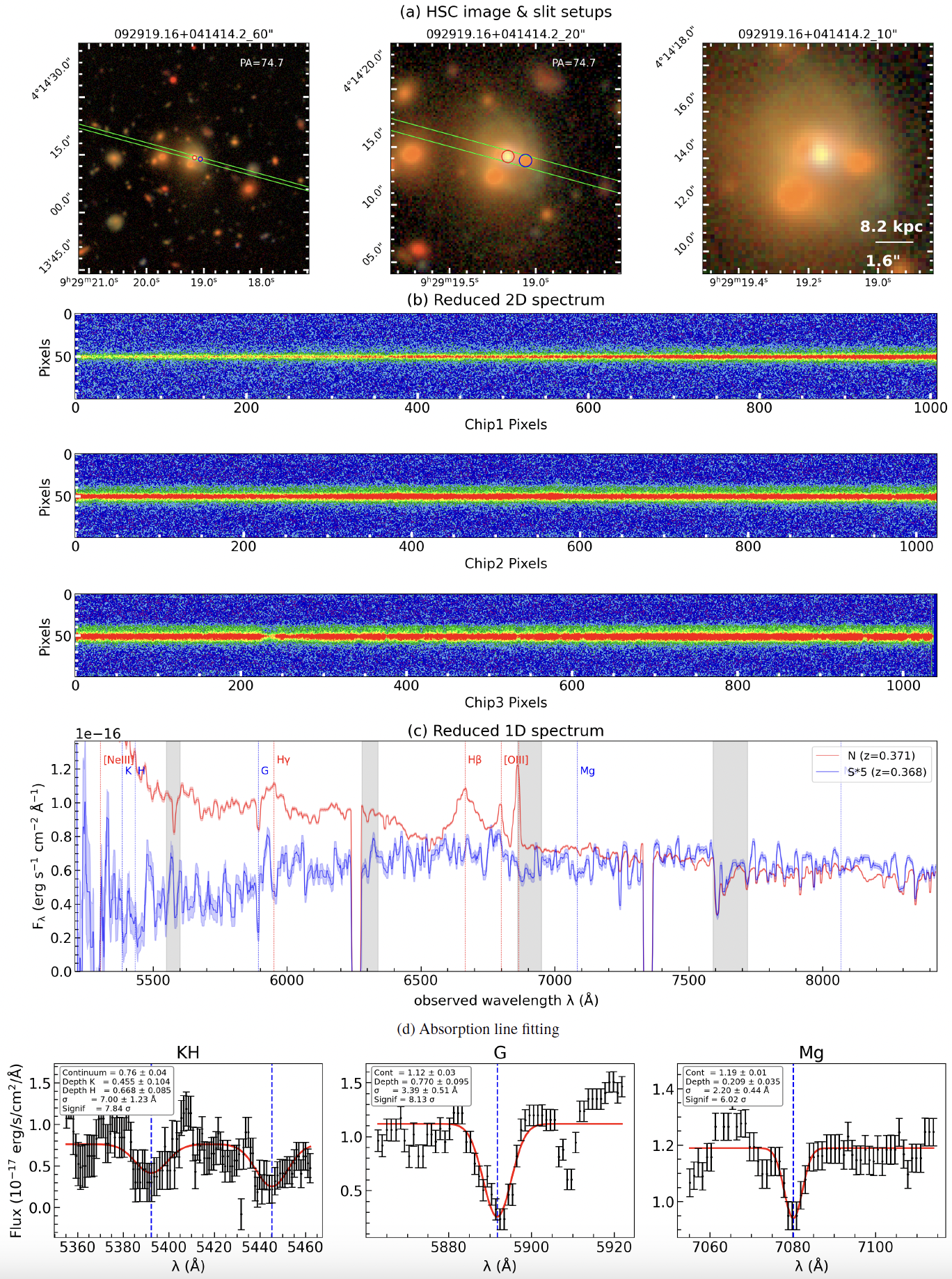}\par
    \caption{Discovery panel reporting the Gemini/GMOS spectrograph of SDSS J092919.16+041414.2, covering the quasar and the lower-left companion. This is the same system as in Figure\ref{fig:J092919.16+041414.2A}, but covering different companions. We observed this system twice instead of once with the slit covering both of the companions for two reasons: (1) To better deblend the quasar emission from the companions with the method described in Section~\ref{subsubsec:spec_reduce}; (2) To cover the ``arc" structure above the quasar. The bottom panels show the fitting of the Ca H\&K, G-band, and Mg absorption lines of the companion's spectrum.}
    \label{fig:J092919.16+041414.2B}
\end{figure*}

\begin{figure*}
\centering
    \includegraphics[width=0.9\textwidth]{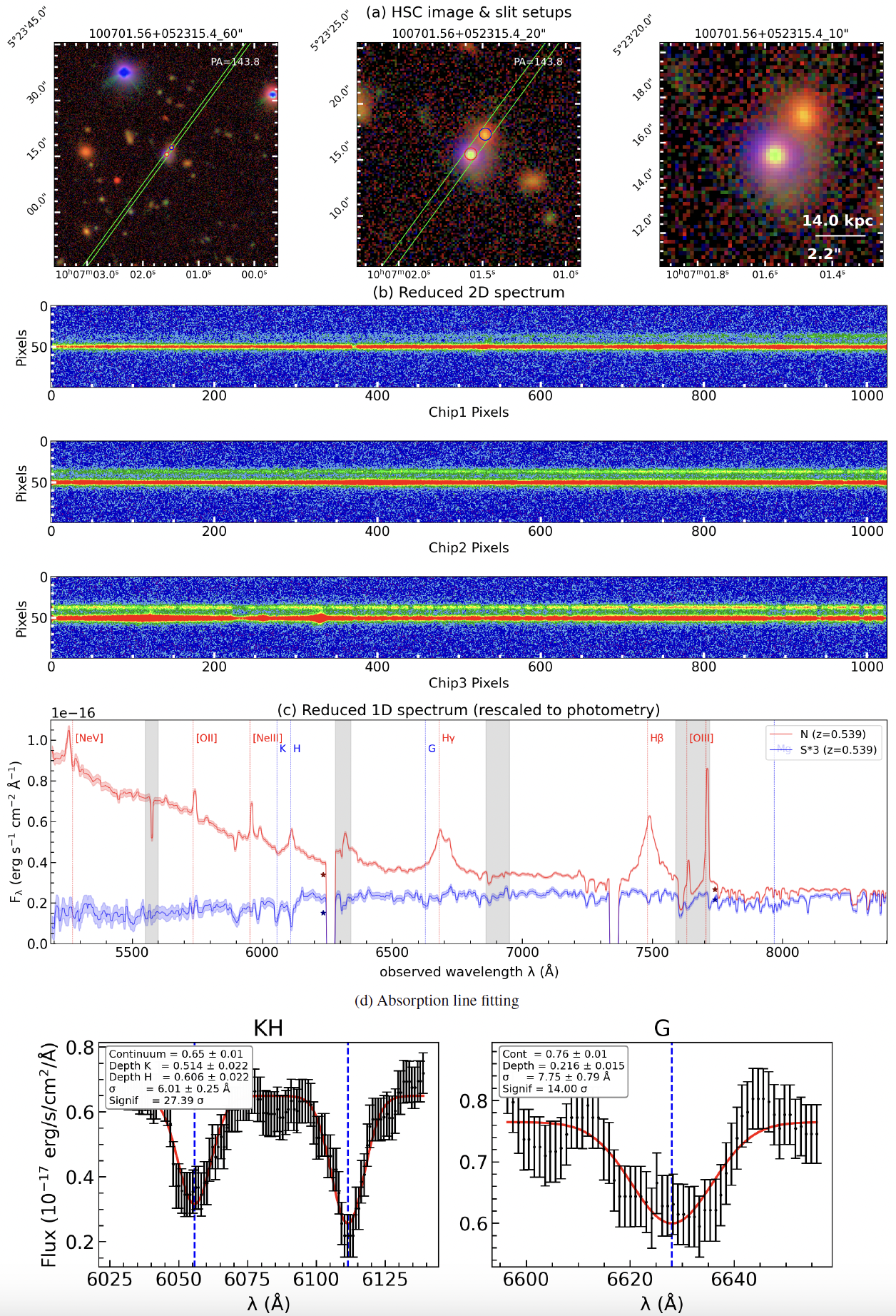}\par
    \caption{Discovery panel reporting the Gemini/GMOS spectrograph of SDSS J100701.56+052315.4. This system is confirmed as a quasar-QG pair (Section~\ref{subsubsection:J100701.56+052315.4}). The bottom panels show the fitting of the Ca H\&K and G-band absorption lines of the companion's spectrum.}
    \label{fig:J100701.56+052315.4}
\end{figure*}

\begin{figure*}
    \includegraphics[width=0.95\textwidth]{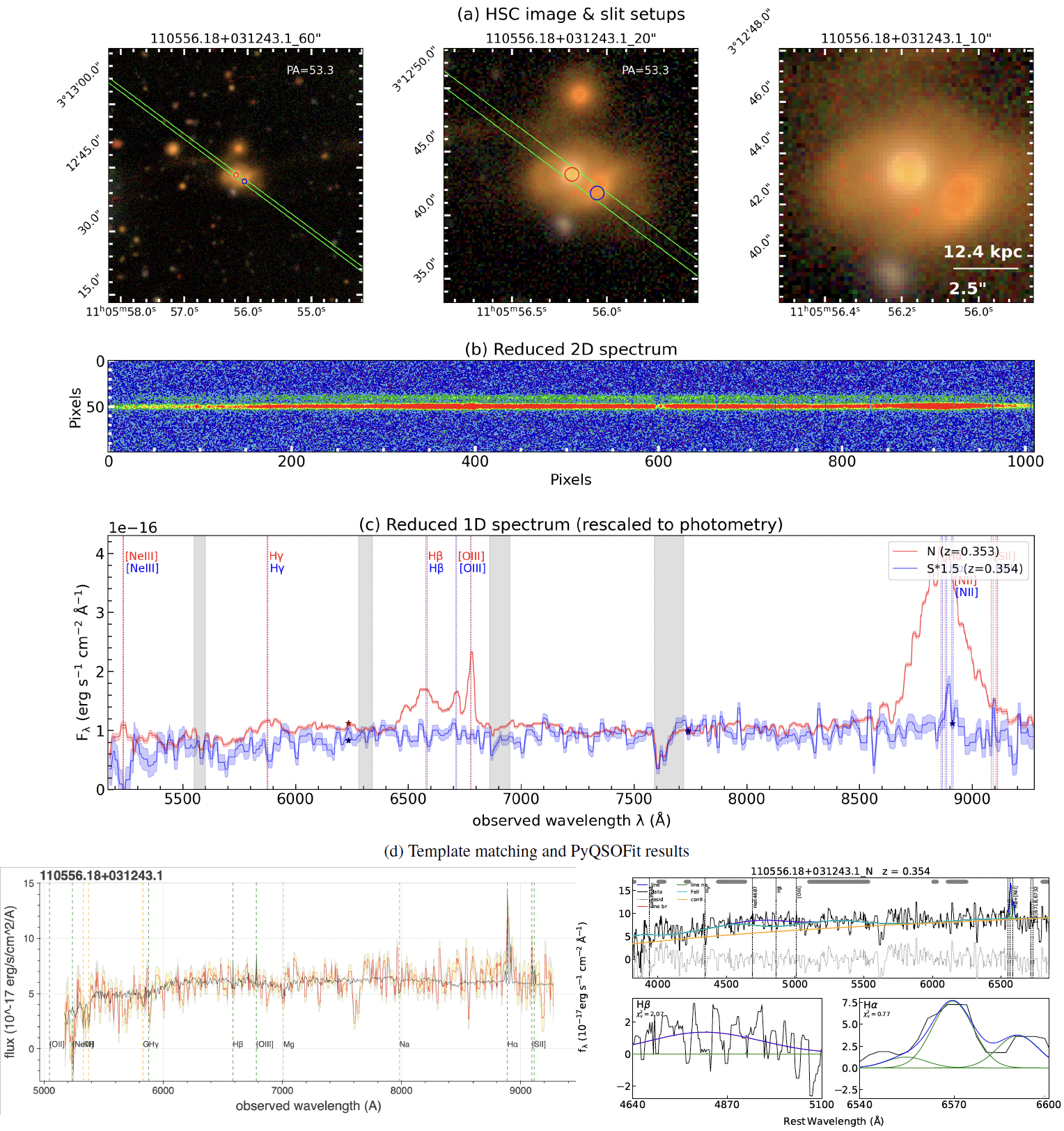}\par
    \caption{Discovery panel reporting the NTT/EFOSC2 spectrograph of SDSS J110556.18+031243.1. This system is confirmed as a quasar-SFG/QG pair (Section~\ref{subsubsection:J110556.18+031243.1}). The bottom left panel shows the comparison between the spectrum of the companion (red curve) and a SDSS galaxy template at z=0.354. The bottom right panel shows the PyQSOFit results of the companion's spectrum.}
    \label{fig:J110556.18+031243.1}
\end{figure*}

\begin{figure*}
    \includegraphics[width=0.95\textwidth]{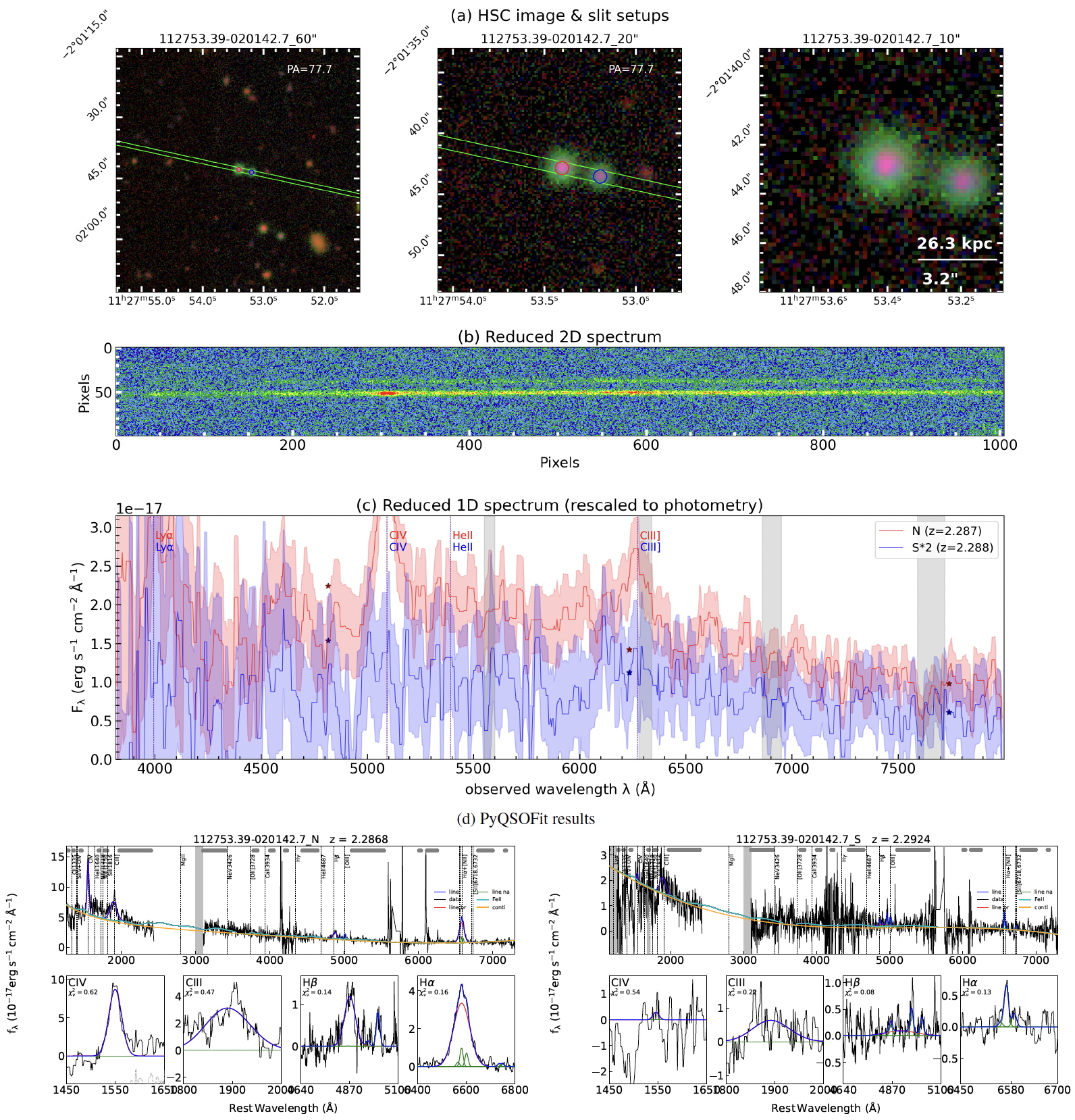}\par
    \caption{Discovery panel reporting the NTT/EFOSC2 spectrograph of SDSS J112753.39-020142.7. This system is confirmed as a quasar-SFG pair (Section~\ref{subsubsection:J112753.39-020142.7}). The bottom panels show the PyQSOFit results on the combined spectrum including the Keck/NIRES part (Figure~\ref{fig:J112753.39-020142.7_K}), Figure~\ref{fig:J112753.39-020142.7_K}) of both the quasar and the companion.}
    \label{fig:J112753.39-020142.7}
\end{figure*}

\begin{figure*}
    \includegraphics[width=0.95\textwidth]{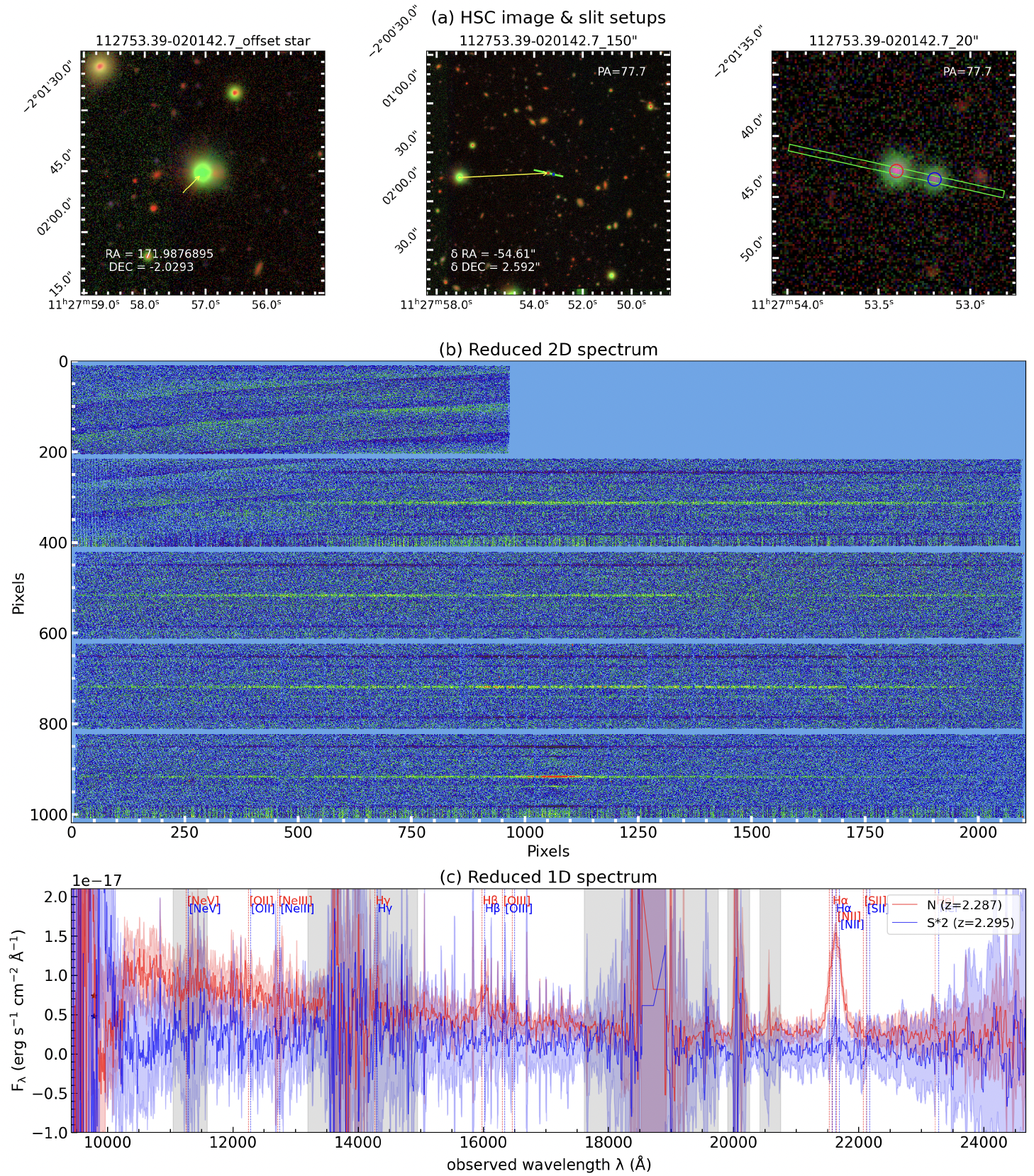}
    \caption{Discovery panel reporting the Keck/NIRES spectrograph of J112753.39-020142.7. This is the same system as in Figure~\ref{fig:J112753.39-020142.7}.}
    \label{fig:J112753.39-020142.7_K}
\end{figure*}

\begin{figure*}
    \includegraphics[width=0.95\textwidth]{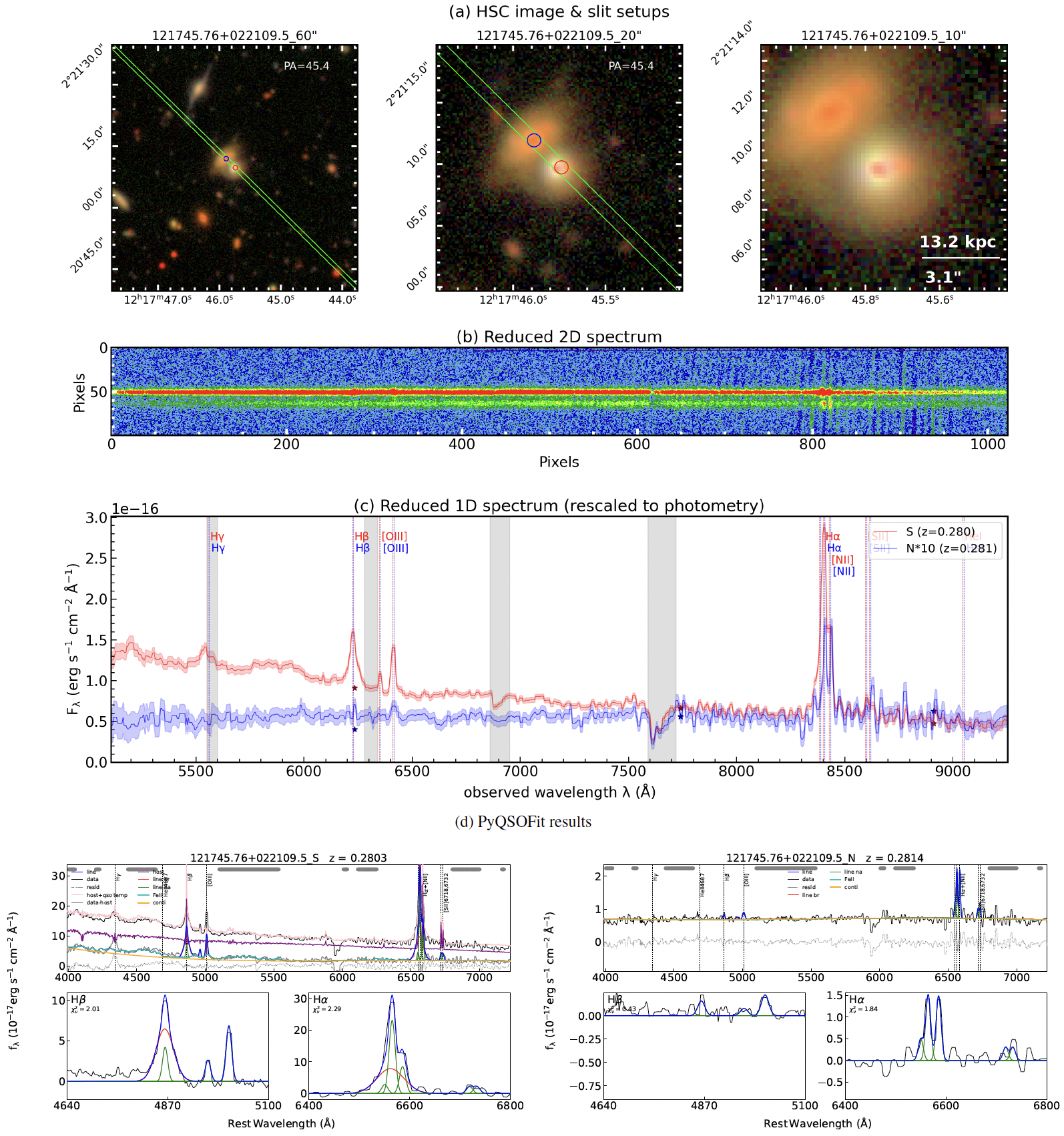}\par
    \caption{Discovery panel reporting the NTT/EFOSC2 spectrograph of SDSS J121745.76+022109.5. This system is confirmed as a quasar-SFG pair (Section~\ref{subsubsection:J121745.76+022109.5}). The bottom panels show the PyQSOFit results for both the quasar and the companion.}
    \label{fig:J121745.76+022109.5}
\end{figure*}

\begin{figure*}
    \includegraphics[width=0.9\textwidth]{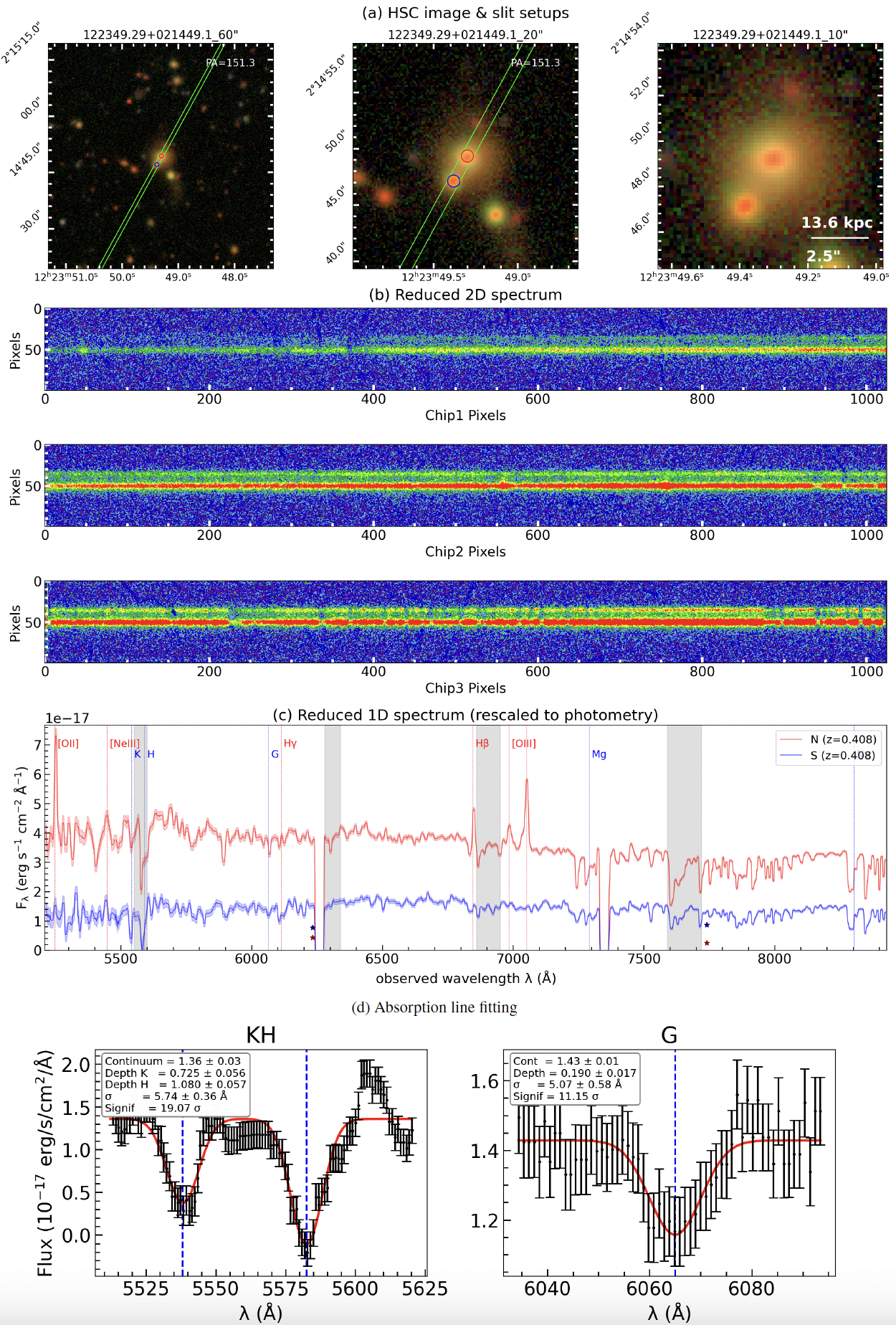}\par
    \caption{Discovery panel reporting the Gemini/GMOS spectrograph of SDSS J122349.29+021449.1. This system is confirmed as a quasar-QG pair (Section~\ref{subsubsection:J122349.29+021449.1}). The bottom panels show the fittings of the Ca H\&K and G-band absorption lines of the companion's spectrum.}
    \label{fig:J122349.29+021449.1}
\end{figure*}

\begin{figure*}
    \includegraphics[width=0.9\textwidth]{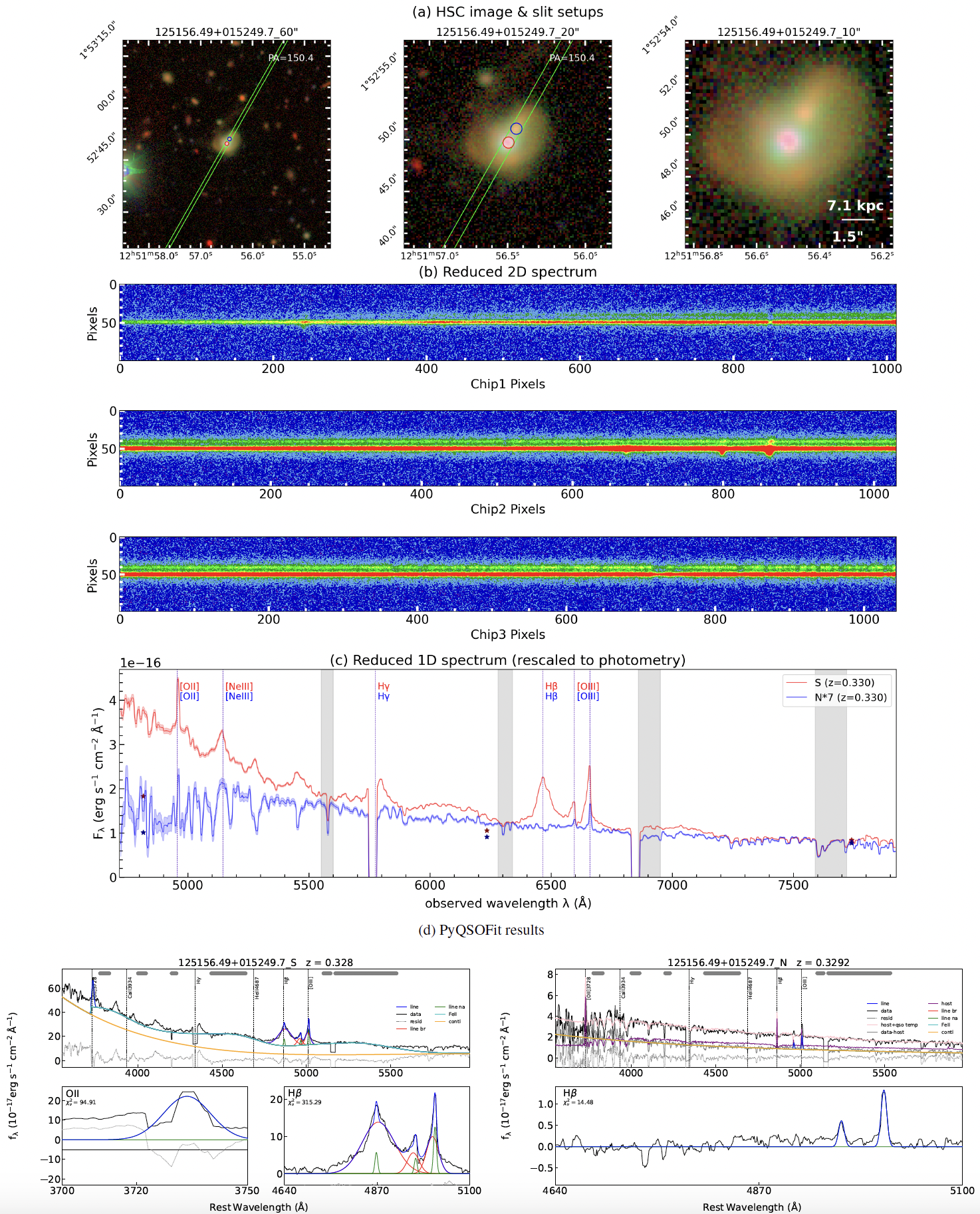}\par
    \caption{Discovery panel reporting the Gemini/GMOS spectrograph of SDSS J125156.49+015249.7. This system is confirmed as a quasar-SFG pair (Section~\ref{subsubsection:J125156.49+015249.7}). The bottom panels show the PyQSOFit results of both the quasar and the companion.}
    \label{fig:J125156.49+015249.7}
\end{figure*}

\begin{figure*}
    \includegraphics[width=0.9\textwidth]{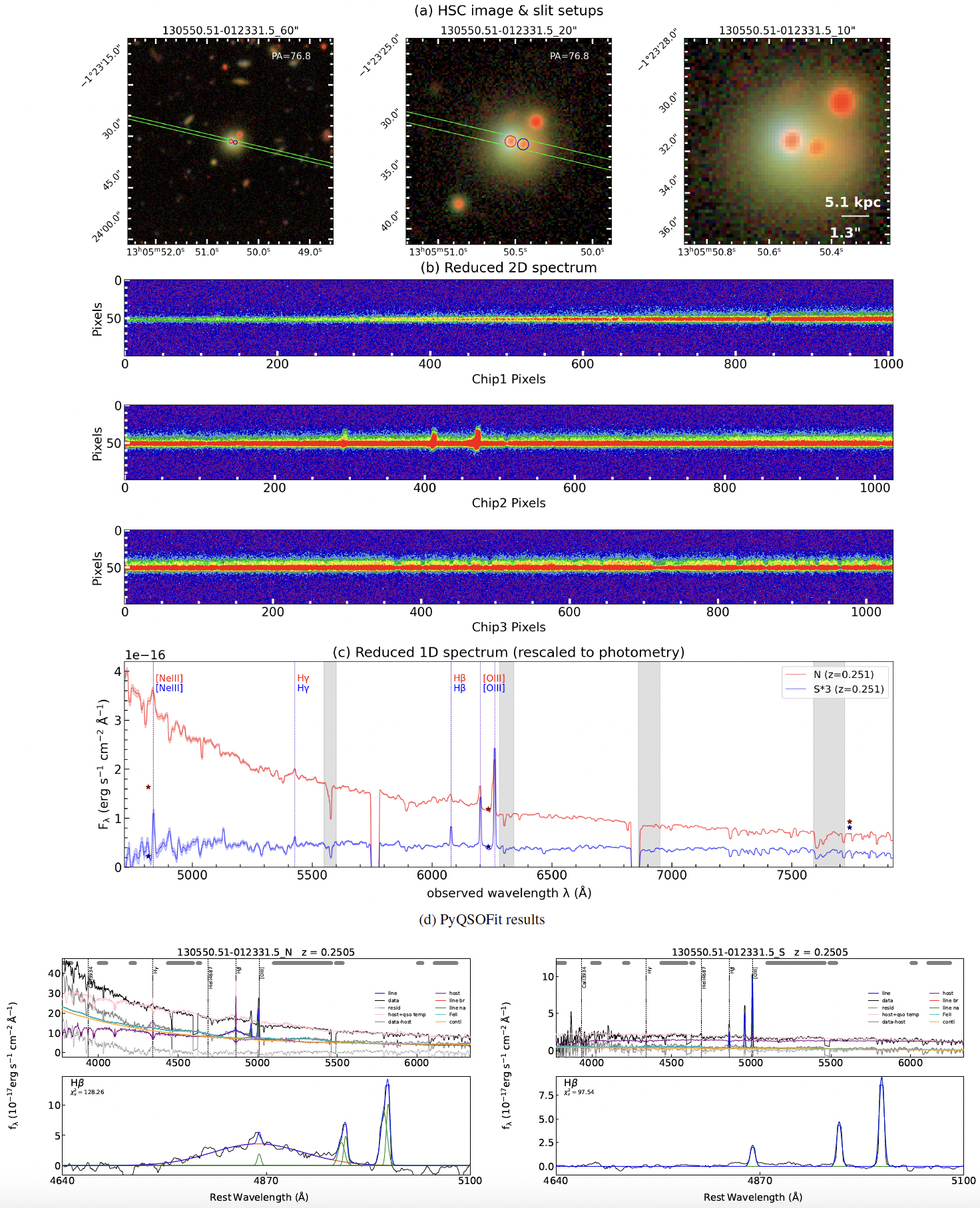}\par
    \caption{Discovery panel reporting the Gemini/GMOS spectrograph of SDSS J130550.51-012331.5. This system is confirmed as a quasar-SFG pair (Section~\ref{subsubsection:J130550.51-012331.5}). The bottom panels show the PyQSOFit results of both the quasar and the companion.}
    \label{fig:J130550.51-012331.5}
\end{figure*}

\begin{figure*}
    \includegraphics[width=0.95\textwidth]{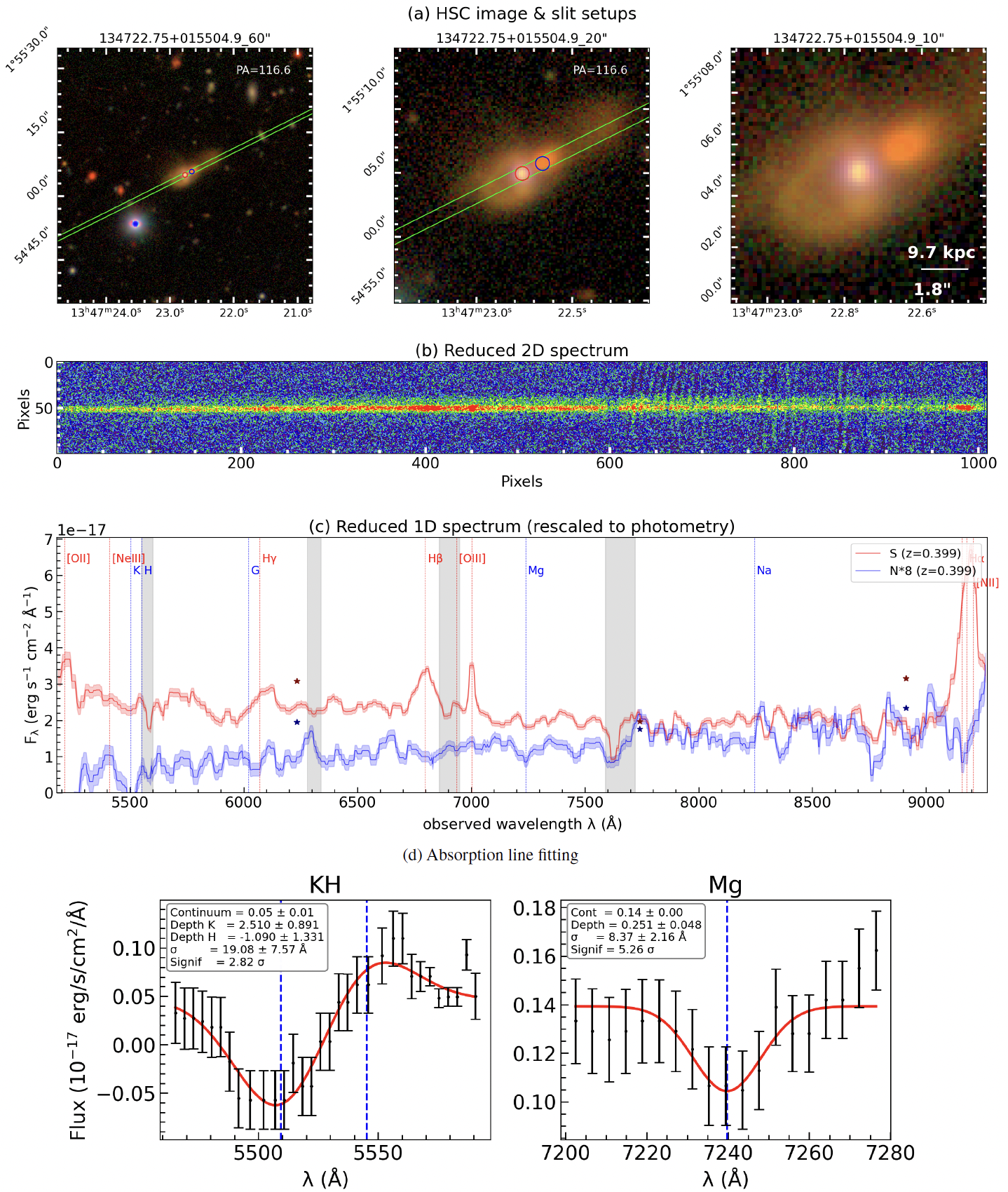}\par
    \caption{Discovery panel reporting the NTT/EFOSC2 spectrograph of SDSS J134722.75+015504.9. This system is considered as a tentative quasar-QG pair (Section~\ref{subsubsection:J134722.75+015504.9}), given that the companion spectrum is relatively faint and the absorption lines are not very significant. The bottom panels show the fitting of the Ca H\&K and Mg absorption lines of the companion's spectrum.}
    \label{fig:J134722.75+015504.9}
\end{figure*}

\begin{figure*}
    \includegraphics[width=0.9\textwidth]{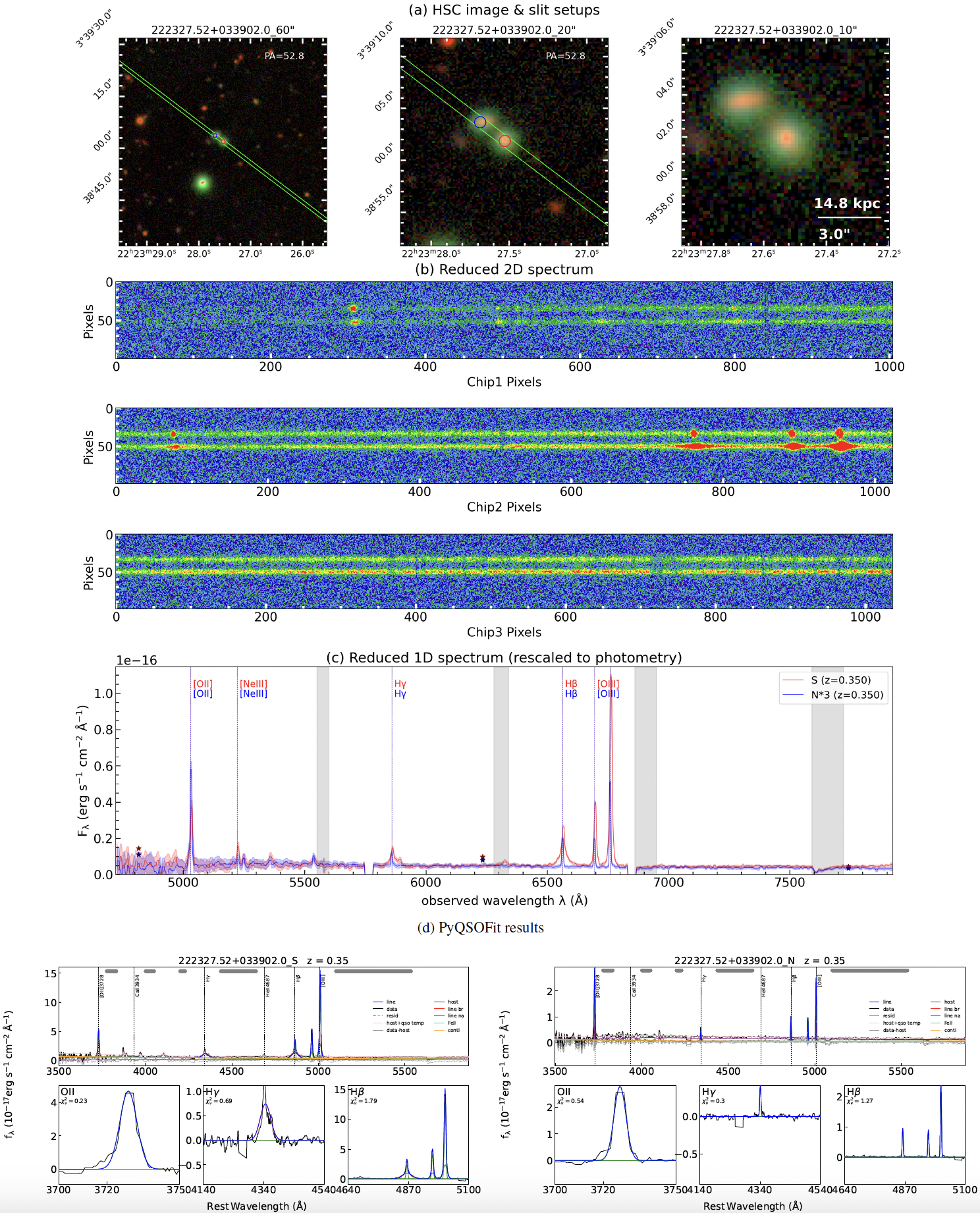}\par
    \caption{Discovery panel reporting the Gemini/GMOS spectrograph of SDSS J222327.52+033902.0. This system is confirmed as a quasar-SFG pair (Section~\ref{subsubsection:J222327.52+033902.0}). The bottom panels show the pyQSOFit results of both the quasar and the companion.}
    \label{fig:J222327.52+033902.0}
\end{figure*}

Here we present the ``Discovery panels" of dual and offset quasars as described in Section~\ref{sec:individuals}. The format of the panels is similar to Figure~\ref{fig:J000129.98+001911.3}. Figure~\ref{fig:J100547.80+013348.4}-\ref{fig:J233225.33+000042.5} present the dual quasars, and Figure~\ref{fig:J010123.64+010045.3}-\ref{fig:J222327.52+033902.0} present the offset quasars. All these ``Discovery panels", including the other observed sources are available via the MNRAS online supplementary material. 

\bsp	
\label{lastpage}
\end{document}